\documentclass[a4paper,11pt]{article}
\pdfoutput=1 
\usepackage{jheppub} 
\usepackage{amsmath,amssymb}
\usepackage{latexsym}
\usepackage{slashed}

\newcommand{\A}{{\cal A}}
\newcommand{\B}{{\cal B}}

\font\mybb=msbm10 at 12pt
\def\bb#1{\hbox{\mybb#1}}

\def\bR {\bb{R}}

\def\bC {\bb{C}}

\def\al{\alpha}
\def\be{\beta}
\def\ga{\gamma}
\def\de{\delta}
\def\ep{\epsilon}

\def\la{\lambda}

\def\si{\sigma}
\def\om{\omega}

\def\De{\Delta}

\def\Om{\Omega}

\def\bal{\bar{\alpha}}
\def\bbe{\bar{\beta}}
\def\bga{\bar{\gamma}}

\def\cF{{\cal{F}}}

\title{Spinorial geometry, off-shell Killing spinor identities and higher 
derivative 5D supergravities}
\author[a]{Federico Bonetti,}
\author[b,c]{Dietmar Klemm,}
\author[d]{Wafic A. Sabra}
\author[e,f]{and Peter Sloane.}
\affiliation[a]{Department of Physics and Astronomy,\\ Johns Hopkins University, 3400 North Charles Street, Baltimore, MD 21218, USA.}
\affiliation[b]{Universit\`a di Milano, Dipartimento di Fisica, \\
 Via Celoria 16, 20133 Milano, Italy.}
 \affiliation[c]{INFN, Sezione di Milano, Via Celoria 16, 20133 Milano, Italy.}
\affiliation[d]{Centre for Advanced Mathematical Sciences and Physics Department,\\
American University of Beirut, Lebanon.}
\affiliation[e]{Facultdad de Ciencias en F\'isica y Matem\'aticas, Universidad Autonoma de Chiapas (FCFM-UNACH), Ciudad 
Universitaria UNACH, \\
Carretera Emiliano Zapata Km. 4, 
Real del Bosque (Ter\'an),
Tuxtla Guti\'errez, \\ Chiapas,
C. P. 29050, M\'exico.}
\affiliation[f]{Mesoamerican Centre for Theoretical Physics, Universidad Autonoma de Chiapas (MCTP-UNACH), Ciudad 
Universitaria UNACH, \\
Carretera Emiliano Zapata Km. 4, 
Real del Bosque (Ter\'an),
Tuxtla Guti\'errez, \\ Chiapas,
C. P. 29050, M\'exico.}
\emailAdd{fdrc.bonetti@gmail.com}
\emailAdd{dietmar.klemm@mi.infn.it}
\emailAdd{ws00@aub.edu.lb}
\emailAdd{psloane@mctp.mx}

\abstract{			Killing spinor identities relate components of 
equations of motion to each other for supersymmetric backgrounds. The only input 
required is the field content and the supersymmetry transformations of the 
fields, as long as an on-shell supersymmetrization of the action without 
additional fields exists. If we consider off-shell supersymmetry it is clear 
that the same relations will occur between components of the equations of 
motion independently of the specific action considered, in particular the 
Killing spinor identities can be derived for arbitrary, including higher derivative, 
supergravities, with a specified matter content.
We give the Killing spinor identities for 
five-dimensional $\mathcal{N}=2$ ungauged supergravities coupled to Abelian vector multiplets, and then using spinorial geometry techniques so that we have explicit 
representatives for the spinors, we discuss the particular case 
of the time-like class of solutions to theories with perturbative corrections at 
the four derivative level. We also discuss the maximally supersymmetric solutions in the general off-shell case.}

\keywords{Supergravity Models, Flux compactifications, Supersymmetry and 
Duality, Black Holes in String Theory}
\preprint{IFUM-1060-FT}		
\begin{document}
\maketitle
\flushbottom

\section{Introduction}
In recent years much technology has been developed in order to complete the 
important task of classifying the supersymmetric solutions of supergravity 
theories.
In this paper we would like to point out the utility of the combination of two 
of these pieces of technology, the so called spinorial geometry approach introduced in \cite{Gillard:2004xq} and the 
Killing spinor identities \cite{Kallosh:1993wx,Bellorin:2005hy}, particularly in the context of classifying the supersymmetric solutions of off-shell 
supergravities, including in the presence of higher derivative terms.

The spinorial geometry approach is to represent the space of spinors using 
differential forms and use the Spin$(d-1,1)$ gauge freedom of the Killing spinor 
equations. The backgrounds that solve the Killing spinor equations for the representative spinors of each orbit of 
Spin$(d-1,1)$ in the spinor space are then related by a local Lorentz 
tranformation to the solution for any other spinor in that orbit.
An oscillator basis for the gamma-matrices then facilitates the reduction of the Killing 
spinor equations to linear systems for the spin connection and fields. To investigate solutions with more than the minimal amount of supersymmetry one may then use
the isotropy group of the first Killing spinor to simplify the second, a process that may be repeated until the common isotopy subgroup of the Killing spinors reduces to the trivial group.

In \cite{Kallosh:1993wx,Bellorin:2005hy} the Killing spinor identities were 
derived which relate components of the equations of motion of supergravity 
theories for backgrounds which preserve some proportion of the supersymmetry. 
The derivation does not require that the supersymmetric action is specified, just 
that the action is supersymmetric under the given supersymmetry variations of 
the fields. In \cite{Meessen:2007ef}
the Killing spinor identities were used in the off-shell $\mathcal{N}=2$ $d=5$ 
superconformal theory to show that the maximally supersymmetric vacua of the 
two derivative theory are the vacua of arbitrarily higher derivative corrected 
theories, up to a generalization of the very special geometry condition. However 
in that work the compensating multiplet was taken to be an on-shell 
hyper-multiplet. We generalize the results of \cite{Meessen:2007ef} to 
the case of an off-shell compensator, extending the results of that 
work to arbitrary higher derivative terms involving the compensating multiplet, 
an example of which is the Ricci scalar squared invariant constructed in \cite{Ozkan:2013nwa}.
The previously constucted Weyl tensor squared invariant \cite{Hanaki:2006pj} is independent of the compensator. Our analysis also extends that of \cite{Meessen:2007ef} to include the gauged case, and thus AdS$_5$ vacua.
We will also be interested in what the Killing spinor identities have to say about 
solutions with less supersymmetry. The spinorial geometry techniques allow us to 
use our simple representatives to show which of the (components of the) 
equations of motion are automatically satisfied for supersymmetric solutions.
 
We will use the Killing spinor identities in 
order to study curvature-squared corrections to $\mathcal{N}=2$, $D=5$ ungauged 
supergravity coupled to an arbitrary number of Abelian vector multiplets. In 
particular we will focus our attention on a gravitational Chern-Simons term of 
the form
$A \wedge \mathrm{tr}(R \wedge R)$
where $R$ denotes the curvature 2-form \cite{Hanaki:2006pj}, and a Ricci scalar squared term \cite{Ozkan:2013nwa}. 

We will use the off-shell superconformal formalism on which there is an extensive literature. We will use mostly the conventions of \cite{Hanaki:2006pj,Fujita:2001kv,Kugo:2000hn,Kugo:2000af}.
The very helpful appendix B in \cite{Ozkan:2013nwa} provides a map from the conventions of \cite{Bergshoeff:2001hc,Bergshoeff:2002qk,Bergshoeff:2004kh,Coomans:2012cf,Ozkan:2013uk} to those we use.
Earlier work on off-shell Poincar\'e supergravity can be found in \cite{Zucker:1999ej}. There is also an extensive literature on off-shell superconformal gravity in five dimensions in superspace, see \cite{Kuzenko:2006mv,Kuzenko:2007cj,Kuzenko:2007hu,Kuzenko:2008wr,Kuzenko:2013rna,Kuzenko:2015lca,Kuzenko:2014eqa} and particularly \cite{Butter:2014xxa}, which contains the superspace contruction of the invariants we consider here amongst much else.
In appendix \ref{constructaction} we 
summarize the construction of supermultiplets whose supersymmetry algebra 
closes without any reference to the equations of motion. These supermultiplets 
can then be used to obtain supersymmetric actions with derivatives of arbitrary 
order without making the supersymmetry transformations of the fields any more 
complicated. Another advantage of the off-shell formalism is the disentanglement 
of kinematic properties (e.g. BPS conditions) from dynamic properties (e.g. 
equations of motion).
The off-shell formulation greatly restricts ambiguities arising from 
field redefinitions, such as
\begin{equation}
g_{\mu\nu}' = g_{\mu\nu} + a R g_{\mu\nu} + b R_{\mu\nu} + \dots  \ ,
\end{equation}
which plague higher-derivative theories in the on-shell formalism. In fact, the 
supersymmetry algebra is not invariant under such transformations, even though 
the on-shell Lagrangian may be. 

We shall be interested in the ungauged $\mathcal{N}=2$, $D=5$ supergravities, 
and so we will appropriately gauge fix the superconformal theory similiarly to 
\cite{Hanaki:2006pj}, see also \cite{Castro:2008ne}, however we will use an 
off-shell compensating linear multiplet, as in \cite{Ozkan:2013nwa}.
This allows us to be sure that our results will hold even on the addition of invariants 
formed from the compensating multiplet. 

The supersymmetric solutions of the minimal ungauged two derivative theory were classified in \cite{Gauntlett:2002nw} and the generalisation to a coupling to arbitrarily many Abelian vector multiplets was reported in \cite{Gutowski:2004yv,Gutowski:2005id}.
The supersymmetric solutions of higher derivative theory have been considered before. In, for example, \cite{Guica:2005ig,Bena:2005ae,Castro:2007hc,Castro:2007sd,Castro:2007ci}
a variety of ansatz were considered, whilst in \cite{Castro:2008ne} the classification of the supersymmetric solutions was presented, following the two derivative analysis of \cite{Gauntlett:2002nw}.
We will reanalyze these results making use of the Killing spinor identities, and give the full equations of motion that remain to be solved in a compact form, for the time-like class. We will show that the Ricci squared invariant does not contribute to any of the equations of motion either in the time-like or null classes of supersymmetric solutions, and so that this classification is valid also in the presence of this invariant.
The supersymmetric near-horizon geometries of this theory were classified, up to the existence of non-constant solutions of a non-linear vortex equation in \cite{Gutowski:2011nk}, assuming that the horizon is Killing with respect to the Killing vector coming from the Killing spinor bilinear. If such solutions exist, they fall outside the classification of \cite{Kunduri:2007vf}, are half supersymmetric and may admit scalar hair. In \cite{Manton:2012fv} it was shown that this equation does indeed admit some non-constant solutions.
It would be particularly interesting to construct explicitly such near-horizon geometries and the corresponding full black hole solutions, or, on the other hand, to extend the uniqueness theorem of \cite{Gutowski:2004bj} under some regularity assumptions. This work, when combined with the results of \cite{Gutowski:2011nk,Manton:2012fv} offers some necessary ingredients to pursue this.

The structure of the paper is as follows: in section \ref{ksigen} we review the 
derivation of the Killing spinor identities 
\cite{Kallosh:1993wx,Bellorin:2005hy} and fix our conventions. In section 
\ref{ksisn2d5} we derive the particular Killing spinor identities for off-shell 
$\mathcal{N}=2$, $d=5$ supergravity with Abelian vector multiplets. In section \ref{halftl} we then review the classsification of 
solutions of the Killing spinor equations at order $\alpha'$ in the time-like class for particular 
four derivative corrections to the two derivative action and the implications of 
the Killing spinor identities for the equations of motion of these solutions. This classification is also valid for 
any off-shell $\mathcal{N}=2$, $d=5$ theory constructed using the standard-Weyl 
gravitational multiplet and with the same matter content if we consistently 
truncate all of the SU$(2)$ triplet fields, the scalar $N$ and the vector $P_\mu$.\footnote{Note that this 
immediately excludes the gauged case, as it is the field $V_\mu^{\mathbf{ij}}$ 
that enters into the gauge covariant derivatives and is set to a combination of 
physical vector multiplets through its equation of motion.}
In section \ref{maxtime-like} we consider the maximally supersymmetric cases in the time-like class and we reproduce the classification 
of \cite{Gauntlett:2002nw,Fiol:2003yq}, 
which is simplified considerably by using the spinorial geometry techniques.
In \cite{Gauntlett:2002nw} a number of maximally supersymmetric solutions were found in the time-like class that were conjectured to be isometric to the near-horizon geometry of the BMPV black hole, and were indeed later shown to be so in \cite{Fiol:2003yq}. Here we obtain this result directly by analysing the Killing spinor equations. 
In section \ref{nullriccisquared} we show that the Ricci squared invariant does not contribute to the equations of motion for the null class of solutions, in a simple calculation using the Killing spinor identities, without going into the details of the resulting geometry.
In section \ref{allordersmax} we extend Meessen's argument \cite{Meessen:2007ef} to include an off-shell compensator in the construction, using the untruncated version of the off-shell theory, necessarily also considering the gauged case.
In appendix \ref{spinors} we give the necessary information on the description of the 
spinors of this theory in terms of forms, and find representatives for each 
orbit of Spin$(4,1)$ on the space of spinors.
We introduce a basis \eqref{tlbasis}  adapted to the case of 
time-like spinors, and use it to derive linear systems 
from the Killing spinor equations for a generic spinor in appendix \ref{tlsystem}.
In appendix \ref{ksis} we give the linear systems for the 
Killing spinor identities in the time-like (\ref{ksitime-like}) and null 
(\ref{ksinull}) bases, the latter using an adapted basis detailed in \eqref{nullbasis}. 

\section{Off-shell Killing spinor identities}\label{ksigen}
We now recall the general derivation of the Killing spinor identities \cite{Kallosh:1993wx,Bellorin:2005hy,Meessen:2007ef} and fix 
our conventions.
Let $S[\phi_b, \phi_f]$ be any supergravity action, constructed in terms of 
bosonic fields $\phi_b$ and fermionic fields $\phi_f$. Let us further assume 
$S[\phi_b, \phi_f]$ is the spacetime integral of a Lagrangian density:
\begin{equation}
S[\phi_b, \phi_f] = \int d^dx \sqrt{g} \mathcal{L}[\phi_b, \phi_f] \; .
\end{equation}
The invariance under supersymmetry transformations of the action can be written
\begin{eqnarray}
 0=\delta_Q S[\phi_b, \phi_f] =
 \int d^dx \sqrt{g} \left\{ \mathcal{L}_b[\phi_b, \phi_f] \delta_Q 
\phi_b[\phi_b, \phi_f] + \mathcal{L}_f[\phi_b, \phi_f] \delta_Q \phi_f [\phi_b, 
\phi_f] \right\} \; ,
\end{eqnarray}
where $\delta_Q$ denotes a local supersymmetry transformation of arbitrary 
parameter, subscripts $b,f$ denote functional derivative with respect to 
$\phi_b, \phi_f$ respectively, and a sum over fields is understood.

Next consider a second variation of the action functional by varying $\delta_Q 
S[\phi_b, \phi_f]$ with respect to fermionic fields only. Since $\delta_Q 
S[\phi_b, \phi_f]$ is identically zero for arbitrary $\phi_b, \phi_f$, we have
\begin{equation}
\delta_Q S[\phi_b, \phi_f + \delta_F \phi_f] = 0 \; ,
\end{equation}
and we set the fermions to zero after the variation. Hence we get
\begin{eqnarray}
 \lefteqn{\left. \delta_F \delta_Q S \right|_{\phi_f=0} =0}  \\
&=& \int d^dx \sqrt{|g|} \left[ (\delta_F \mathcal{L}_b)( \delta_Q\phi_b ) + 
\mathcal{L}_b (\delta_F \delta_Q \phi_b)  + (\delta_F \mathcal{L}_f)( 
\delta_Q\phi_f ) + \mathcal{L}_f (\delta_F \delta_Q \phi_f) 
\phantom{\frac{1}{1}}\right]_{\phi_f=0} \; . \nonumber
\end{eqnarray}
Since $\delta_Q \phi_b$ and $\mathcal{L}_f$ are odd in fermions we are left 
with 
\begin{equation}
 \int d^dx \sqrt{|g|} \left[ ( \mathcal{L}_b (\delta_F \delta_Q \phi_b) + 
(\delta_F \mathcal{L}_f)( \delta_Q\phi_f ) \right]_{\phi_f=0} =0\; .
\end{equation}

Calculating $(\delta_F \mathcal{L}_f)_{\phi_f=0}$ requires knowledge of the 
entire Lagrangian, not only its bosonic truncation. However if we restrict 
ourselves to supersymmetry transformations having Killing spinors as 
parameters, 
$\delta_K$, we have
\begin{equation}
(\delta_K \phi_f)_{\phi_f = 0} = 0 \; .
\end{equation}
Note that
\begin{equation}
\mathcal{L}_b := \frac{1}{\sqrt{|g|}} \frac{\delta S[\phi_b,\phi_f] }{\delta 
\phi_b}= \frac{1}{\sqrt{|g|}} \frac{\delta S_B[\phi_b] }{\delta \phi_b} + 
\frac{1}{\sqrt{|g|}} \frac{\delta S_F[\phi_b,\phi_f] }{\delta \phi_b} \;,
\end{equation}
where the last term vanishes if $\phi_f=0$. We are thus led to define
\begin{equation}
\mathcal{E}_b := \frac{1}{\sqrt{|g|}} \frac{\delta S_B[\phi_b] }{\delta \phi_b} 
\; ,
\end{equation}
so that bosonic equations of motion take the form
\begin{equation}
\mathcal{E}_b=0 \; .
\end{equation}
Thus the Killing spinor identities may be written as
\begin{equation}
\int d^dx \sqrt{|g|} \; \mathcal{E}_b (\delta_F \delta_K \phi_b)_{\phi_f=0} = 0 
\; .
\end{equation}

We will now derive the Killing spinor identities 
 for off-shell $\mathcal{N}=2$, $D=5$ 
supergravity, which have been discussed in \cite{Meessen:2007ef}. We discuss 
the construction of such superconformal theories in appendix 
\ref{superconformaltheory} and their gauge fixing to Poincar\'e supergravity in 
appendix \ref{gaugefix}. What we need are the off-shell supersymmetry variations for 
the bosonic field content, and
we record the relevant terms for our discussion here for ease of reference:
\begin{align}
\delta e^a_\mu &= -2i \bar{\epsilon} \gamma^a \psi_\mu\ ,\nonumber\\
\delta v_{ab} &= - \tfrac{1}{8}i \bar{\epsilon} \gamma_{ab} \chi +\cdots \ , \nonumber\\
\delta D &=-\tfrac{1}{3}i \bar{\epsilon}\gamma^{\mu\nu} \chi v_{\mu\nu} 
-i\bar{\epsilon}\gamma^\mu \nabla_\mu \chi  
+i\bar{\epsilon}^{\mathbf{i}}\gamma^\mu V_{\mathbf{ij}\mu}\chi^{\mathbf{j}}-\tfrac{i}{6}\bar{\epsilon}^{\mathbf{i}}(\slashed{E}+N)L_{\mathbf{ij}}\chi^{
\mathbf{j}} 
+\tfrac{i}{3}\bar{\epsilon}^{\mathbf{i}}\gamma^a{V'}_{a\mathbf{ij}}\chi^{\mathbf
{j}}+ \cdots \ ,\nonumber\\
\delta V^{\mathbf{ij}}_\mu &= 
-\tfrac{i}{4}\bar{\epsilon}^{(\mathbf{i}}\gamma_\mu \chi^{\mathbf{j})} +
\cdots \ ,\nonumber\\
\delta A^I_\mu & = -2i \bar{\epsilon} \gamma_\mu \Omega^I + \cdots \ ,\nonumber\\
\delta M^I &= 2i \bar{\epsilon} \Omega^I \ ,\nonumber\\
\delta Y^{I\mathbf{ij}} &= 2i\bar{\epsilon}^{(\mathbf{i}}\gamma^a \nabla_a 
\Omega^{\mathbf{j})I}
-2i\bar{\epsilon}^{(\mathbf{i}}\gamma^a 
V_{a\phantom{\mathbf{j})}\mathbf{k}}^{\phantom{a}\mathbf{j})} 
\Omega^{\mathbf{k}I}
-\tfrac{2i}{3}V_a^{\mathbf{k}(\mathbf{i}}\bar{\epsilon}_{\mathbf{k}}
\gamma_a\Omega^{\mathbf{j})I}
-\tfrac{i}{3}\bar{\epsilon}^{(\mathbf{i}}\gamma_{ab}v^{ab}\Omega^{\mathbf{j})I} 
-\tfrac{i}{4}\bar{\epsilon}^{(\mathbf{i}}\chi^{\mathbf{j})}M^I \ ,\nonumber\\
 \delta N &= \tfrac{i}{2}L_{\mathbf{ij}}\bar{\epsilon}^{\mathbf{i}}\chi^{\mathbf{j}}
\ . 
\end{align}

In the above we have supressed terms involving the gravitino, and in particular have not listed the variation of the auxiliary vector $P_a$ as it only involves the gravitino. This is due to our 
taking the strategy of solving the equations of motion of all other fields 
before turning to solve the Einstein equation. Because of this the only term 
involving the gravitino that will not lead to a term involving an equation of 
motion of a bosonic field that we have solved will come from the vielbien 
variation. As to be expected from the complexity of the Einstein equation of 
higher derivative theories and the ubiquity of the gravitino in the 
supersymmetry transformations, if we keep these terms we may obtain long 
expressions for the components of the Einstein equation in terms of components 
of the other equations of motion and the fields. However as long as we keep in 
mind that our gravitino Killing spinor identity is only valid after solving the 
other equations of motion, we may proceed by ignoring the gravitino terms in the 
above variations, greatly simplifying the derivation.
So if we set 
$\mathcal{E}(e)^\mu_a := \tfrac{1}{\sqrt{|g|}} \tfrac{\delta S}{ \delta 
e^a_\mu}$
,
we get 
\begin{equation}
\left. \mathcal{E}(e)_a^\mu \gamma^a \epsilon^{\mathbf{i}} \right|_{\text{other 
bosons
on-shell}} = 0 \; .
\end{equation}
To proceed we will need one more ingredient, the gravitino variation which reads
\begin{eqnarray}
\delta \psi^{\mathbf{i}}_\mu &=& \nabla_\mu \epsilon^{\mathbf{i}} + \tfrac{1}{2} \gamma_{\mu ab} 
v^{ab} \epsilon^{\mathbf{i}} - \tfrac{1}{3} \gamma_\mu \gamma_{ab} v^{ab} \nonumber\\
&&+ V^{\mathbf{ij}}_{\mu}\epsilon_{\mathbf{j}}+  
\tfrac{1}{6}\gamma_\mu(\slashed{P}+N)L^{\mathbf{ij}}\epsilon_{\mathbf{j}}- 
\tfrac{1}{3}\gamma_\mu\gamma^a{V'}_a^{\mathbf{ij}}\epsilon_{\mathbf{j}} 
=0\ ,
\end{eqnarray}
where $V_\mu^{\mathbf{ij}}=V_\mu L^{\mathbf{ij}} + {V'}^{\mathbf{ij}}_{\mu}$ so that ${V'}^{\mathbf{ij}}_{\mu} L_{\mathbf{ij}}=0$, since $L^2:=L_{\mathbf{ij}}L^{\mathbf{ij}}=1$ from the gauge fixing of the superconformal theory down to the super-Poincar\'e theory, which is discussed in section \ref{gaugefix}. 
We define the same splitting for any SU(2) symmetric field $A^{\mathbf{ij}}$, in particular we define $A^{\mathbf{ij}}=AL^{\mathbf{ij}} + {A'}^{\mathbf{ij}}$ so that ${A'}^{\mathbf{ij}}L_{\mathbf{ij}}=0$.
It will be useful to derive the following identity for SU(2) symmetric fields.
Consider two such fields $A^{\mathbf{ij}},B^{\mathbf{ij}}$. We may easily show that
\begin{equation}
 2A^{[\mathbf{i|k}}B^{\mathbf{|j}]}_{\mathbf{k}} = A_{\mathbf{kl}}B^{\mathbf{kl}}\epsilon^{\mathbf{ij}} = (AB + {A'}_{\mathbf{kl}}{B'}^{\mathbf{kl}})\epsilon^{\mathbf{ij}} \ .
\end{equation}
We also note the identity
\begin{equation}
 L_{\mathbf{ij}}A^{\mathbf{ik}}B^{\mathbf{j}}_{\mathbf{k}} = L_{\mathbf{ij}}{A'}^{\mathbf{ik}}{B'}^{\mathbf{j}}_{\mathbf{k}} \ ,
\end{equation}
which clearly vanishes for $A=B$.

Let us now write the KSI associated to a variation of gauginos. We set
\begin{equation}
\mathcal{E}(A)^\mu_I := \frac{1}{\sqrt{|g|}} \frac{\delta S}{ \delta A_\mu^I} \;
, \qquad \mathcal{E}(M)_I := \frac{1}{\sqrt{|g|}} \frac{\delta S}{ \delta M^I}
\; , \qquad \mathcal{E}(Y)_{I\mathbf{ij}} := \frac{1}{\sqrt{|g|}}\frac{\delta 
S}{ \delta
Y^{I\mathbf{ij}}} \; ,
\end{equation}
and have therefore
\begin{eqnarray}
0 &=& \int d^dx \sqrt{|g|} \left[ \mathcal{E}(A)^\mu_I \left( -2 i
\bar{\epsilon}^{\mathbf{i}} \gamma_\mu \right) + \mathcal{E}(M)_I (2i 
\bar{\epsilon}^{\mathbf{i}}) +
\mathcal{E}(Y)_{I\mathbf{jk}}(2i\bar{\epsilon}^{\mathbf{j}})\gamma^aV_a^{\mathbf
{ki}} \right. \\
&& \left.+ 
\tfrac{2i}{3}\mathcal{E}(Y)_{I\mathbf{k}}^{\mathbf{i}}V_a^{\mathbf{jk}}\bar{
\epsilon}_{\mathbf{j}}\gamma_a -
\mathcal{E}(Y)_{I}^{\mathbf{ij}}
(\tfrac{i}{3}\bar{\epsilon}_{\mathbf{j}}\gamma^{ab}v_{ab}) 
\right]\delta\Omega^I_{\mathbf{i}}
 + \mathcal{E}(Y)_{I}^{\mathbf{ij}}
(2i\bar{\epsilon}_{\mathbf{j}}\gamma^a) \nabla_a \delta \Omega^I_{\mathbf{i}} 
\; .\nonumber
\end{eqnarray}
Integrating by parts and using the fact that the gravitino Killing spinor 
equation implies
\begin{equation}
\gamma^a\nabla_a \epsilon^{\mathbf{i}} = 
\tfrac{5}{6}(v\cdot\gamma)\epsilon^{\mathbf{i}} - \gamma^aV_aL^{\mathbf{ij}}\epsilon_{\mathbf{j}}
+\tfrac{2}{3}{V'}^{a\mathbf{ij}}\gamma_a\epsilon_{\mathbf{j}} - \tfrac{5}{6}(\slashed{P}+N)L^{\mathbf{ij}}\epsilon_{\mathbf{j}} \ ,
\label{gravdotgamma}
\end{equation}
we obtain
\begin{eqnarray}
0&=&\left[ \mathcal{E}(A)^\mu_I \gamma_\mu  - \mathcal{E}(M)_I + \tfrac{5}{12}\mathcal{E}(Y)(\slashed{P}+2\slashed{V} +N)\right] 
\epsilon^{\mathbf{i}}  \label{gauginoksigen}\\ 
&+&\left[\left(\nabla^a{\mathcal{E}(Y)_{I}}^{\mathbf{ij}}
\right)\gamma_a - \tfrac{5}{6}\mathcal{E}({Y'})_I^{\mathbf{ik}}(\slashed{P} +2\slashed{V} + N )L^{\mathbf{j}}_{\mathbf{k}}
-{\mathcal{E}(Y)_I}^{\mathbf{ij}}\slashed{v} \right]\epsilon_{\mathbf{j}}
 \; . \nonumber
\end{eqnarray}
Next we consider the KSI associated with the auxiliary fermion. We define
\begin{eqnarray}
\mathcal{E}(v)^{ab} &:=& \frac{1}{\sqrt{|g|}} \frac{\delta S}{ \delta v_{ab}} \ ,
\qquad 
\mathcal{E}(D) := \frac{1}{\sqrt{|g|}} \frac{\delta S}{ \delta D} \ ,\qquad
\mathcal{E}(N) := \frac{1}{\sqrt{|g|}} \frac{\delta S}{ \delta N} \ ,\qquad
\nonumber\\
\mathcal{E}(P)^a &:=& \frac{1}{\sqrt{|g|}} \frac{\delta S}{ \delta P_a}\ , \qquad
\mathcal{E}(V)^\mu_{\mathbf{ij}} := \frac{1}{\sqrt{|g|}} \frac{\delta 
S}{V_\mu^{\mathbf{ij}}}\ ,
\end{eqnarray}
and thus obtain 
\begin{eqnarray}
 0&=&\int d^5 x \sqrt{|g|} \left[ -\tfrac{i}{8} \mathcal{E}(v)^{ab}
\bar{\epsilon}^{\mathbf{i}} \gamma_{ab} 
-i\mathcal{E}(D)\bar{\epsilon}^{\mathbf{j}}\gamma_aV^{a}L^{\mathbf{i}}_{\mathbf{j}} 
-\tfrac{i}{3} \mathcal{E}(D) v^{ab} \bar{\epsilon}^{\mathbf{i}} \gamma_{ab} 
\right.\nonumber\\
&&\left. +\tfrac{i}{6}\mathcal{E}(D)\bar{\epsilon}^{\mathbf{j}}(\slashed{P}+N)L^{\mathbf{i}}_{\mathbf{j}}
-\mathcal{E}(D)\tfrac{4i}{3}\bar{\epsilon}^{\mathbf{j}}{V'}^{\mathbf{i}}_{a\mathbf{
j}}\gamma^a
+\tfrac{i}{4}\mathcal{E}(V)^{\mu 
\mathbf	{i}}_{\phantom{\mu\mathbf{i}}\mathbf{j}}
\bar{\epsilon}^{\mathbf{j}}\gamma_\mu 
+\tfrac{i}{4}\mathcal{E}(Y)^{\mathbf{i}}_{I\mathbf{j}}\bar{\epsilon}^{\mathbf{j}
}M^I\right.\nonumber\\
&&\left. 
 -\tfrac{i}{2}\mathcal{E}(N)L^{\mathbf{i}}_{\mathbf{j}}\right] \delta \chi_{\mathbf{i}} 
+ \left[ -i\bar{\epsilon}\mathcal{E}(D) \gamma^\mu \right] \nabla_\mu \delta 
\chi \; .
\end{eqnarray}
Integrating
the last term by parts, discarding the total derivative and making use of
the gravitino Killing spinor equation
we obtain 
\begin{eqnarray}
0&=&\left[ \tfrac{1}{8} \mathcal{E}(v)^{ab} + \tfrac{1}{2} \mathcal{E}(D) v^{ab}
\right] \gamma_{ab} \epsilon^{\mathbf{i}} + \nabla^a \mathcal{E}(D) \gamma_a 
\epsilon^{\mathbf{i}} 
-\tfrac{1}{4}\mathcal{E}(V)_a^{ \mathbf{ij}}\gamma^a\epsilon_{\mathbf{j}} 
-\tfrac{1}{4}\mathcal{E}(Y)_{I}^{\mathbf{ij}}M^I\epsilon_{\mathbf{j}} \nonumber\\
&&+2\mathcal{E}(D){V'}^{\mathbf{ij}}_a\gamma^a\epsilon_{\mathbf{j}} +\tfrac{1}{2}\mathcal{E}(N)L^{\mathbf{ij}}\epsilon_{\mathbf{j}}
-\mathcal{E}(D)(\slashed{P}+N))L^{\mathbf{ij}}\epsilon_{\mathbf{j}}
 \; . \label{auxfermksigen}
\end{eqnarray}

In order to use these equations we need either to solve explicitly for the 
Killing spinors or better to find representatives for them for different 
(classes of) solutions. Our strategy will be to expand the Killing spinor 
identities in suitable bases for their solution using the spinorial geometry 
techniques.
It is especially easy to solve these system as we have already reduced the system to equations that 
are algebraic in the Killing spinors, using the gravitino Killing spinor 
equation.

In the two derivative ungauged on-shell theory with Abelian vectors all 
supersymmetric solutions (locally) preserve four or eight supersymmetries. 
However this is no longer a priori true in the off-shell theory unless the 
auxiliary SU$(2)$ fields vanish. Because of this it is possible that a number of new 
features arise in the off-shell case in theories with suitably complicated 
actions which are normally associated with higher dimensional or gauged 
supergravities.  
Note that the Killing spinor identities derived above will be valid for 
supersymmetric solutions with the appropriate number of Killing spinors, i.e. 
spinors which satisfy all of the Killing  spinor equations. This is due to the implicit sum over fields.

\section{N=2, d=5 ungauged supergravity with four derivative 
corrections}\label{ksisn2d5}
 
We review the construction of the superconformal Lagrangian in appendix 
\ref{superconformaltheory}, and the gauge fixing to Poincar\'e supergravity in 
\ref{gaugefix}. We do not break the R-symmetry down to global U(1), which could be achieved by choosing a particular value for $L^{\mathbf{ij}}$. 

Now we will specialize to a particular consistent truncation that is sufficient to study first order perturbative string theory corrections. In particular we remove
terms in $\mathcal{L}_4$ that do not contribute to 
linear order in $\alpha'$ using the two derivative equations of motion for the 
auxiliary fields.
In particular note that since $V_\mu^{\mathbf{ij}},Y^{I\mathbf{ij}},N,P_\mu$ have trivial equations of motion at the two derivative level one can write for example
$V_\mu^{\mathbf{ij}}=\mathcal{O}(\alpha')$. However the corrections to these equation of motion are themselves of order $\alpha'$ so in fact
\begin{equation}
 {V}^{\mathbf{ij}}=\mathcal{O}{(\alpha')^2}\ , \qquad Y^{I\mathbf{ij}}=\mathcal{O}{(\alpha')^2}\ ,\qquad N= \mathcal{O}{(\alpha')^2}\ ,\qquad P_\mu=\mathcal{O}{(\alpha')^2}\ .
\end{equation}
Due to this we may truncate them from the action and the supersymmetry tranformations when studying the perturbatively corrected four derivative theory at first order and to all orders in the consistent truncation.
In \cite{Castro:2008ne, Meessen:2007ef} only higher derivative terms independent of the compensator were considered, and the above statement follows for the fields ${V}^{\mathbf{ij}}, Y^{I\mathbf{ij}}$ as they could only couple to each other in the action, and have trivial equations of motion at two derivative level.
However in invariants involving the compensator, one must check that these fields are in fact higher order, as they could appear contracted with $L^{\mathbf{ij}}$. Clearly the order of the fields $N$ and $P_\mu$ must also be checked. However an inspection of the Ricci scalar squared superconformal invariant \eqref{ricci_inv}, assures us that these fields are in fact $\mathcal{O}(\alpha'^2)$. We would like to emphasize, however that this may not be the case with all invariants involving the compensating multiplet, and must be checked.

The resulting Lagrangian of $R^2$ corrected $\mathcal{N}=2$, $D=5$ ungauged 
Poicar\'e supergravity coupled to Abelian vector multiplets is given by
\begin{equation}
\mathcal{L} = \mathcal{L}_2 + \mathcal{L}_4 \; .\label{2plus4Lagrangian}
\end{equation}
At two derivative level we have
\begin{align}
\mathcal{L}_2 = \mathcal{L}_V + 2\mathcal{L}_L &= \tfrac{1}{2} D(\mathcal{N}-1) 
- \tfrac{1}{4} R(\mathcal{N}+3) + v^2 (3\mathcal{N}+1) + 2 \mathcal{N}_I v^{ab} 
F^I_{ab} + \nonumber\\
& + \mathcal{N}_{IJ} \left( \tfrac{1}{4} F^I_{ab}F^{Iab} - \tfrac{1}{2} 
\nabla_a M^I \nabla^a M^J  \right) + \tfrac{1}{24} c_{IJK} e^{-1} 
\epsilon^{abcde} A^I_a F^J_{bc} F^K_{de} \; . 
\end{align}
where the Levi-Civita symbol is denoted by $\epsilon^{abcde}$. Note the sign 
of the scalar kinetic term which corrects that in eqn.(78) of \cite{Castro:2008ne}.

As far as the four derivative Lagrangian is concerned we will take 
$\mathcal{L}_4=\mathcal{L}_{C^2} + \mathcal{L}_{R_s^2}$, where
\begin{align}
\mathcal{L}_{C^2} &= \tfrac{c_{2I}}{24} \left\{ \tfrac{1}{16} e^{-1} 
\epsilon^{abcde} A^I_a C_{bcfg} C_{de}^{\phantom{de}fg} + \tfrac{1}{8} M^I 
C^{abcd} C_{abcd} +\right. \nonumber\\
& + \tfrac{1}{12} M^I D^2 + \tfrac{1}{6} D v^{ab} F^I_{ab} + \tfrac{1}{3} M^I 
C_{abcd} v^{ab} v^{cd} + \tfrac{1}{2} C_{abcd}F^{Iab} v^{cd} + \nonumber\\
&+ \tfrac{8}{3} M^I v_{ab} \nabla^b \nabla_c v^{ac} -\tfrac{16}{9}M^I 
v^{ab}v_{bc}R_a^{\phantom{a}c} - \tfrac{2}{9} M^I v^2 R +\nonumber\\
& + \tfrac{4}{3} M^I \nabla_a v_{bc} \nabla^a v^{bc} +\tfrac{4}{3} M^I \nabla_a 
v_{bc} \nabla^b v^{ca} +\nonumber\\
& - \tfrac{2}{3} M^I e^{-1}\epsilon^{abcde}v_{ab} v_{cd} \nabla^f v_{ef} + 
\tfrac{2}{3} e^{-1}\epsilon^{abcde} F^I_{ab} v_{cf} \nabla^f v_{de} +\nonumber\\
&+ \epsilon^{abcde} F^I_{ab} v_{cf} \nabla_d v_{e}^{\phantom{e}f} -\tfrac{4}{3} 
F^I_{ab} v^{ac} v_{cd} v^{db} - \tfrac{1}{3} F^I_{ab}v^{ab} v_{cd} v^{cd} +\nonumber\\
&\left. + 4 M^I v_{ab} v^{bc} v_{cd} v^{da} - M^I v_{ab} v^{ab} v_{cd} v^{cd} 
\phantom{\tfrac{1}{1}}\right\} \; ,
\end{align}
where $C$ denotes the Weyl tensor and we are using the conventions
 ${R_{\mu\nu\sigma}}^{\rho}= -2\partial_{[\mu}\Gamma^\rho_{\nu]\sigma} + 
2\Gamma^\tau_{[\mu|\sigma}\Gamma^\rho_{\tau|\nu]}$,
$R_{\mu\nu}={R_{\mu\rho\nu}}^{\rho}$ and
\begin{equation}
 C_{\mu\nu\sigma\rho}=R_{\mu\nu\sigma\rho} - \tfrac{2}{3}(g_{\mu[\sigma 
}R_{\rho]\nu} - g_{\nu[\sigma}R_{\rho]\mu}) + 
\tfrac{1}{6}Rg_{\mu[\sigma}g_{\rho]\nu} \;,
\end{equation}
which are different to the conventions in \cite{Hanaki:2006pj}. In \ref{eoms} 
we give the contributions to the equations of motion for this contribution to 
the action, which are quite involved.

For the Ricci tensor squared contribution one finds
\begin{align}
 e^{-1}\mathcal{L}_{R^2_s} =& \mathcal{E} (\tfrac{2}{3}D-\tfrac{4}{3}v^2 + R)^2 \ ,  
\end{align}
where we have absorbed a factor into the definition of $\mathcal{E}=e_IM^I$ and we also provide the contributions to the equations of motion in appendix \ref{eoms}, which are rather simpler.

In order to solve the Killing spinor equations to order $(\alpha')$ or to all orders in a consistent truncation, we may
remove the same fields from the Killing spinor equations and identities 
which now read
\begin{align}
\nabla_\mu \epsilon^{\mathbf{i}} + \left[ \tfrac{1}{2} \gamma_{\mu ab} 
v^{ab} - \tfrac{1}{3} \gamma_\mu \gamma_{ab} v^{ab}\right] \epsilon^{\mathbf{i}} =0 \ ,\nonumber\\
\left[-\frac{1}{4} F^I_{ab} \gamma^{ab} - \frac{1}{2} 
\gamma^\mu \partial_\mu M^I - \frac{1}{3}M^I v^{ab} \gamma_{ab} 
\right]\epsilon^{\mathbf{i}} = 0\ ,\nonumber\\
\left[D - \frac{8}{3} v^2 + \left(2\nabla_b v^{ba} - \frac{2}{3} 
\epsilon^{abcde} v_{bc} v_{de}\right)
\gamma_a + \epsilon^{abcde}\gamma_{ab}\nabla_c v_{de}\right] \epsilon^{\mathbf{i}} = 0\ , \nonumber\\
 \mathcal{E}(e)_\mu^a\gamma_a\epsilon^{\mathbf{i}} =0\ ,\nonumber\\
 \left[ \mathcal{E}(A)^\mu_I \gamma_\mu  - \mathcal{E}(M)_I\right] \epsilon^{\mathbf{i}} =0 \ , \nonumber\\
  \left[ \tfrac{1}{8} \mathcal{E}(v)^{ab} + \tfrac{1}{2} \mathcal{E}(D) v^{ab}\right]
 \gamma_{ab} \epsilon^{\mathbf{i}} =0\ .
\end{align}
In appendix \ref{tlsystem} we give the linear systems associated to the Killing spinor equations in a time-like basis, whilst 
for the Killing spinor identities we present the linear systems in the time-like and null bases in appendices \ref{ksitime-like} and \ref{ksinull}, respectively. These bases are adapted to the 
time-like and null orbits of Spin(4,1) on the space of spinors which can be 
found in appendix \ref{spinors}. In the next two sections we shall use these systems 
to analyse the equations of motion of the truncated theory, which is sufficient 
to study the order $\alpha'$ four derivative corrections to the ungauged theory.

In the interests of completeness we give the full form of the KSI for the 
gravitino for this truncation, which we calculate using the full supersymmetry transformations in 
\cite{Fujita:2001kv} to be
\begin{eqnarray}
\mathcal{E}(e)^\mu _a (2 \bar{\epsilon}\gamma^a ) &=& \mathcal{E}(A)^\mu _I (2 
M^I\bar{\epsilon}) + \mathcal{E}(v)^{ab} ( \tfrac 12  v_{a b}\bar{\epsilon} 
\gamma^\mu   
- \tfrac 12  {v_{a}}^\mu \bar{\epsilon}\gamma_b + \tfrac 34  \nabla_b 
\bar{\epsilon} {\gamma_{a}}^\mu ) 
\nonumber\\
&+& \mathcal{E}(v)^{a\mu }({v_a}^b \bar{\epsilon} \gamma_b + \tfrac 32  
\nabla_a\bar{\epsilon} - \tfrac 34  \nabla_b \bar{\epsilon} {\gamma_a}^b) + 
\nabla_a\mathcal{E}(v)^{a \mu }( \tfrac 32  \bar{\epsilon})   + 
\nabla^b\mathcal{E}(v)^{a \mu }( - \tfrac 34  \bar{\epsilon} \gamma_{ab}) 
\nonumber\\
&+& \mathcal{E}(D)\left(
4 \bar{\epsilon} \nabla_bv^{b \mu } - 2 \epsilon^{\mu 
defg}\bar{\epsilon}v_{de}v_{fg} + (D-\tfrac{2}{3}v^2)\bar{\epsilon}\gamma^\mu  
+\tfrac{22}{3} v_{a b}v^{\mu  b}\bar{\epsilon}\gamma^a
\right.\nonumber\\
 &-&\left.2{\epsilon_d}^{efgh}v_{ef}v_{g h}\bar{\epsilon}\gamma^{\mu d} - 2 
\nabla^\mu v_{a b} \bar{\epsilon} \gamma^{a b} -  4\nabla^av_{b 
a}\bar{\epsilon} \gamma^{b\mu }
-4\nabla_av_{\mu  b}\bar{\epsilon}\gamma^{b a} \right.\nonumber\\&+& \left. 12 
\nabla^a( {v_{a}}^\mu \bar{\epsilon}) - 4\nabla^a(v^{\mu 
b}\bar{\epsilon}\gamma_{ab}) + 4\nabla^a (v_{a b}\bar{\epsilon}\gamma^{\mu 
b})\right)   
+\nabla_b\mathcal{E}(v)^{ab}(\tfrac 34  \bar{\epsilon} {\gamma_a}^\mu )\nonumber\\
&+&4\nabla^a\mathcal{E}(D)( 3{v_{a}}^\mu \bar{\epsilon} - v^{\mu 
b}\bar{\epsilon}\gamma_{ab} + v_{ab}\bar{\epsilon}\gamma^{\mu b}) \ .
\label{fullgravitinoKSI}
\end{eqnarray}
We can then write this in terms of the variation with respect to the metric 
using
\begin{equation}
\tfrac{\delta S[e^a_\mu, v_{ab},D, A^I_\mu, M^I]}{\delta e^a_\lambda} = -2 
g^{\lambda(\mu} e^{\nu)}_a \tfrac{\delta S[g_{\mu\nu}, v_{\mu\nu} ,D, A^I_\mu, 
M^I]}{\delta g^{\mu\nu}} 
 - 2 v_{ab} e^{b}_{[\mu} \delta^{\lambda}_{\nu]}\tfrac{ \delta S[g_{\mu\nu}, 
v_{\mu\nu},D, A^I_\mu, M^I ]}{\delta v^{\mu\nu}} \ .
\end{equation}
We will not find this expression particularly enlightening in what follows.

\section{Half supersymmetric time-like solutions}\label{halftl}
In the section we shall analyse the supersymmetry conditions arising from the
existence of one time-like Killing spinor and reproduce the results of 
\cite{Castro:2008ne}, which we will add to in the next section by examining the 
Killing spinor identitities and equations of motion of the theory considered 
there with the addition of the Ricci scalar squared invariant.

\subsection{Killing spinor equations and geometric constraints}
Let us turn first to solving the Killing spinor equations. We shall see 
that demanding one supersymmetry leads to 4 out of the
8 possible supersymmetries being preserved.
It is convenient to work in the oscillator basis defined in (\ref{tlbasis}),
whose action on the basis elements is recorded in table \ref{tab:gamma}. The
Killing spinor equations have been expanded in this basis to yield the linear
system in appendix \ref{tlsystem}. For the representative of the $\text{SU}(2)$ 
orbit of
$\text{Spin}(1,4)$ we may always choose (cf.~ eq. \ref{SU(2)-rep1})\footnote{As
discussed in appendix \ref{spinors}, there are two different representatives, 
one for each of the
different $\text{SU}(2)$ orbits, which are related by a $\text{Pin}$ 
transformation. The results for the representative of the other $\text{SU}(2)$ 
orbit are closely related to what we shall find for the
representative we consider here, and we shall summarize the results in section 
\ref{secondorbit}.}
\begin{equation}
\epsilon = (\epsilon^1,\epsilon^2) = (e^{\phi}1,-ie^{\phi}e^{12})\ . 
\label{SU(2)-rep1'}
\end{equation}
Inspecting the linear system in appendix \ref{tlsystem} it is easy to see that
the two components of the spinor yield equivalent conditions. Now consider the
spinor $\eta=(\eta^1,\eta^2)=(-ie^{\phi}e^{12},-e^{\phi}\;1)$. This is clearly
linearly independent from $\epsilon$, however it yields an equivalent linear
system, thus the system preserves at least two supersymmetries. In
fact the system preserves half of the supersymmetry, as the spinors
$\chi=(i\epsilon^1,-i\epsilon^2)=(ie^{\phi}1,-e^{\phi}e^{12})$ and
$\zeta=(i\eta^1,-i\eta^2)=(e^{\phi}e^{12},ie^{\phi}1)$ also yield identical
systems. To summarize, demanding the existence of one (time-like) supersymmetry
implies that the solution is half supersymmetric and it is sufficient to
solve the Killing spinor equations of the first component of that spinor.

From the gravitino eqns.~\eqref{lin-sys-time-grav} we obtain
\begin{eqnarray}
\partial_0 \phi = 0\ , \qquad \om_{\al,12}=0\ , \qquad
v_{0\al}=-\tfrac{3}{2}\partial_\al \phi =-\tfrac{3}{4}\om_{0,0\al}
= -\tfrac{3}{2}{\om_{\al\ga}}^\ga = - 
\tfrac{3}{2}\om_{\bbe,12}{\epsilon^{\bbe}}_\al\ , \nonumber \\
v_{\al\be}=-\tfrac{3}{2}\om_{0,\al\be}=-\tfrac{3}{2}\om_{\al,0\be}\ , \qquad
v_{1\bar2}=-\tfrac{1}{2}\om_{1,0\bar2}=\tfrac{1}{2}\om_{\bar2,01}\ , \nonumber 
\\
{v_\gamma}^\gamma=-\tfrac{3}{2}{\om_{0,\gamma}}^\gamma=-\tfrac{3}{2}{\om_{\ga,0}
}^\ga\ , \qquad
2v_{1\bar1} - v_{2\bar2}=-\tfrac{3}{2}\om_{1,0\bar1}\ , \qquad
v_{1\bar1} - 2v_{2\bar2}=\tfrac{3}{2}\om_{2,0\bar2}\ . 
\end{eqnarray}
where $\epsilon_{\alpha\beta}$ is antisymmetric with $\epsilon_{12}=1$.
 From this we can easily read off the geometric constraints
\begin{eqnarray}
\partial_0 \phi = \om_{\al,12} &=& 0\ , \label{integrable}\\
\om_{(i,|0|j)}&=&0\ , \label{impkill1}\\
{\om_{0,\gamma}}^\gamma&=&{\om_{\ga,0}}^\ga\ , \label{tsstrace} \\
\om_{0,\al\be}&=&\om_{\al,0\be}\ , \label{tss02} \\
2\partial_\al \phi =\om_{0,0\al}
&=& 2{\om_{\al\ga}}^\ga = 2\om_{\bbe,12}{\epsilon^{\bbe}}_\al\ . 
\label{impkill2}
\end{eqnarray}
Consider next the one-form bilinear $V=e^{2\phi}e^0$ constructed from the 
spinor \eqref{SU(2)-rep1'}.
$V$ is clearly time-like and it is easy to show that  (\ref{impkill1}) and the 
first equation in (\ref{impkill2})
imply that it is Killing. We can thus introduce coordinates $t,x^m$ such that
\begin{equation}
V=\frac{\partial}{\partial t} \; ,
\end{equation}
as a vector. The metric takes the form
\begin{equation}
ds^2 = e^{4\phi}(dt+\Om)^2 - e^{-2\phi}\hat g_{mn}dx^mdx^n\ ,
\end{equation}
and we may adapt a frame such that $ds_5^2=(e^0)^2 - ds^2_4= (e^0)^2 - 
\hat{\eta}_{ij}e^ie^j$,
\begin{equation}
e^0 = e^{2\phi}(dt + \Om)\ , \qquad e^i = e^{-\phi}\hat{e}_n^i dx^n\ ,
\end{equation}
where $\hat{\eta}_{ij}$ denotes the flat euclidean metric, $\hat{e}^i$ is a 
vierbein for $\hat g$ and
$\phi, \om$ and $e^i$ are independent of $t$.
Next consider the torsion free condition for the f\"unfbein $e^A$,
\begin{equation}
de^A + \om_{B,\phantom{A}C}^{\phantom{B,}A} e^B \wedge e^C=0\ .
\end{equation}
In particular setting $A=i$ and considering the part with either of $B,C=0$ we 
find conditions compatible
with the constraints (\ref{tsstrace}) and (\ref{tss02}),
but in addition this implies that the trace free $(1,1)$ part of 
$\om_{0,ij}=\om_{i,0j}$ must also be satisfied.
It is convenient to introduce the two form $G$,
\begin{equation}
G = e^{2\phi} d\Omega\ .
\end{equation}
Then the components of the five-dimensional spin connection are
\begin{displaymath}
\omega_{0,0i} = 2 e^\phi \hat{\nabla}_i\phi\ , \qquad
\omega_{0,ij}  = \omega_{i,0j}  = -\tfrac{1}{2} G_{ij}\ , \qquad
\omega_{i,jk}  = -e^{\phi} \left( \hat{\omega}_{i,jk} - 2 {\hat{\eta}}_{i[j} 
\hat{\nabla}_{k]} \phi \right)\ ,
\end{displaymath}
where hats refer to four-dimensional quantities and we note that all components 
are determined 
in terms of the base space. We can see that this means 
\eqref{impkill1}-\eqref{tss02} and the first
equality in \eqref{impkill2} are satisfied, and it remains to interpret 
\eqref{integrable} and the remainder
of \eqref{impkill2}. Examining the first of these we see that 
$\om_{\alpha,12}=0$
implies that the $(3,0) + (0,3)$ part of the connection vanishes, and thus the 
complex structure is integrable.
The remaining conditions can also be expressed in terms of the Gray-Hervella 
classification for an
$\text{SU}(2)$ structure manifold, and it can be seen that the manifold is in 
the special Hermitian class \cite{Gillard:2005ic}.
We will not pursue this here, as we shall show instead that the base space is 
hyper-K\"ahler, i.e. we will describe it instead via its integrable Sp$(1)(\cong$SU(2)) structure. 
We can now write $v$ as
\begin{eqnarray}
v&=&v_{0\al} e^0 \wedge e^{\al} + v_{0\bal} e^0 \wedge e^{\bal} +
 \frac{1}{2}\left(v_{\al\be} e^{\al} \wedge e^{\be}  + v_{\bar{\al}\bar{\be}} 
e^{\bar{\al}} \wedge e^{\bar{\be}} \right) \nonumber\\
&\phantom{=}& + \de_{\al\bbe}{v_{\gamma}}^{\gamma}e^{\alpha}\wedge 
e^{\bar\beta} +
\left( v_{\al\bbe} - \de_{\al\bbe} {v_\gamma}^{\gamma} \right) e^{\al}\wedge 
e^{\bbe}\ ,
\end{eqnarray}
where the $(1,1)$ piece with respect to the complex structure has been split 
into its traceful and traceless parts.
It is convenient instead to decompose the spatial part of $v$ into selfdual, 
$v^+$, and antiselfdual, $v^-$, parts.
Note that the nonzero components of the decomposition of a two-form $\alpha$ in 
the oscillator basis are
\begin{eqnarray*}
\alpha^{(+)}_{1\overline{1}} &=& \tfrac{1}{2} (\alpha_{1\overline{1}} - 
\alpha_{2\overline{2}})\ , \qquad
\alpha^{(+)}_{1\overline{2}} = \alpha_{1\overline{2}}\ , \qquad
\alpha^{(+)}_{\overline{1}2} = \alpha_{\overline{1}2}\ , \qquad
\alpha^{(+)}_{2\overline{2}} = - \tfrac{1}{2}( \alpha_{1\overline{1}} - 
\alpha_{2\overline{2}} )\ , \\
\alpha^{(-)}_{1\overline{1}} &=& \tfrac{1}{2} (\alpha_{1\overline{1}} + 
\alpha_{2\overline{2}})\ , \qquad
\alpha^{(-)}_{12} = \alpha_{12}\ , \qquad
\alpha^{(-)}_{\overline{1} \overline{2}} = \alpha_{\overline{1} \overline{2}}\ 
, \qquad
\alpha^{(-)}_{2\overline{2}} = \tfrac{1}{2} (\alpha_{1\overline{1}} + 
\alpha_{2\overline{2}})\ ,
\end{eqnarray*}
so that with respect to the complex structure $\alpha^+$ is the trace-free 
$(1,1)$ part,
whilst $\al^-$ is the $(2,0) + (0,2)$ part and the trace. We observe that we 
may thus write
\begin{equation}
 v^{(+)}_{ij} = \tfrac{1}{4} G^{(+)}_{ij}\ , \qquad
 v^{(-)}_{ij} = \tfrac{3}{4} G^{(-)}_{ij}\ ,
\end{equation}
so $v$ is given by
\begin{equation}
v = -\tfrac{3}{2} e^0 \wedge d\phi + \tfrac{1}{4} G^{(+)} + \tfrac{3}{4} G^{(-)}
= \tfrac{3}{4} de^0 - \tfrac{1}{2} G^{(+)}\ .
\end{equation}
The two-form bilinears of the spinor \eqref{SU(2)-rep1'} are
\begin{eqnarray}
X^{(1)} &=& -e^{2\phi} (e^1 \wedge e^2 + e^{\overline{1}} \wedge 
e^{\overline{2}})\ , \nonumber\\
X^{(2)} &=& -ie^{2\phi} (e^1 \wedge e^2 - e^{\overline{1}} \wedge 
e^{\overline{2}})\ , \nonumber\\
X^{(3)} &=& -ie^{2\phi} (e^1 \wedge e^{\overline{1}} + e^{2} \wedge 
e^{\overline{2}})\ .
\end{eqnarray}
Notice that the constraints on the connection imply that they are closed, since 
$dX^{(i)} = 0$
is equivalent to demanding
\begin{eqnarray}
2 \nabla_0 \phi &=& (\omega_{1,0\overline{1}} + \omega_{2,0\overline{2}}) - 
(\omega_{0,1\overline{1}} +
\omega_{0,2\overline{2}}) =  \omega_{1,0\overline{1}} + 
\omega_{\overline{1},01} = \omega_{2,0\overline{2}} + \omega_{\overline{2},02}\ 
, \nonumber\\
\omega_{0,12} &=& \omega_{1,02}\ , \qquad
\omega_{1,02} + \omega_{2,01} = 0\ , \qquad
\omega_{\alpha,12} = 0\ , \nonumber\\
\nabla_1 \phi &=& \omega_{1,1\overline{1}} + \omega_{1,2\overline{2}} = 
\omega_{\overline{2},12}\ , \qquad
\nabla_2 \phi = \omega_{2,1\overline{1}} + \omega_{2,2\overline{2}} = 
-\omega_{\overline{1},12}\ ,
\end{eqnarray}
which are all implied by the gravitino Killing spinor equation. Defining
\begin{equation}
\mathcal{X}^{(i)i}_{\phantom{(i)i}j} := \hat{\eta}^{ik} \hat{X}^{(i)}_{kj}\ , 
\label{calX}
\end{equation}
such that $\hat{X}_{ij}^{(i)}$ are the components with respect to the vierbein 
$\hat{e}^i$,
\begin{equation}
\tfrac{1}{2} X^{(i)}_{ij} e^i \wedge e^j = \tfrac{1}{2} (X^{(i)}_{ij} 
e^{-2\phi}) \hat{e}^i \wedge \hat{e}^j = \tfrac{1}{2} \hat{X}^{(i)}_{ij} 
\hat{e}^i \wedge \hat{e}^j\ ,
\end{equation}
we find that the $\mathcal{X}^{(i)}$ obey the algebra of the imaginary unit 
quaternions,
\begin{equation}
\mathcal{X}^{(i)} \mathcal{X}^{(j)} = -\delta_{ij} \mathbb{I} + \epsilon_{ijk} 
\mathcal{X}^{(k)}\ . \label{unit-quat}
\end{equation}
This defines an almost quaternionic structure on the base space. If they are 
covariantly constant they define an
integrable hypercomplex structure on the base, so we examine
\begin{equation}
\hat{\nabla} \mathcal{X}^{(i)} = 0\ , \qquad i = 1,2,3\ , \label{nablaX=0}
\end{equation}
which is equivalent to demanding
\begin{align*}
& \hat{\omega}_{\alpha1\overline{1}} + \hat{\omega}_{\alpha2\overline{2}}  = 0\ 
, &
& \hat{\omega}_{\alpha12} = 0\ , &
& \hat{\omega}_{\alpha \overline{1}\overline{2}}= 0\ ,
\end{align*}
which are again implied by the gravitino Killing spinor equation. We thus 
conclude the base space is
hyper-K\"ahler. Note that the spin connection and the curvature two-form on the 
base are selfdual,
$\hat{\om}_{i,jk}^{(-)}=\hat{R}_{ij}^{(-)}=0$.

We turn next to the gaugini equations. For our representative, the linear 
system \eqref{lin-sys-time-gaug}
boils down to
\begin{equation}
\partial_0 M^I = {\cF^{I\al}}_\al = \cF^I_{\bar1 \bar2}= 0\ , \qquad 
\partial_{\bal}M^I = 4\cF^I_{0\bal}\ .
\end{equation}
Thus we have
\begin{equation}
\partial_0 M^I = 0\ , \qquad
F^I_{0i} = -\tfrac{4}{3} M^I v_{0i} + \nabla_i M^I\ , \qquad
F^{I(-)}_{ij} = -\tfrac{4}{3} M^I v^{(-)}_{ij}\ .
\end{equation}
We can eliminate $v$ to find
\begin{eqnarray}
F^I &=& e^{-2\phi} e^0 \wedge d(M^I e^{2\phi}) - M^I G^{(-)} + F^{I(+)} 
\nonumber \\
&=& -d(M^I e^0) + M^I G^{(+)} + F^{I(+)}\ ,
\end{eqnarray}
where the selfdual part of $F$ is undetermined. Note that
\begin{equation}
V \lrcorner F^I = d(M^I e^{2\phi})\ ,
\end{equation}
which, together with the Bianchi identity, implies that the Lie derivative of 
$F^I$ along $V$ is zero,
\begin{equation}
\mathcal{L}_{V} F^I = d(V \lrcorner F^I) + V \lrcorner dF^I = 0\ ,
\end{equation}
and thus $F^I$, including its undetermined part, is independent of $t$. Since 
\begin{equation}
dF^I = dM^I \wedge G^{(+)} + M^I dG^{(+)} + dF^{I(+)}\ ,
\end{equation}
the undetermined part of the field strength satisfies
\begin{equation}
dF^{I(+)} = -dM^I \wedge G^{(+)} - M^I dG^{(+)}\ .
\end{equation}
Let us introduce the selfdual two-form
\begin{equation}
\Theta^{I(+)} := M^I G^{(+)} + F^{I(+)}\ ,
\end{equation}	
so imposing the Bianchi identity for $F^I$ is equivalent to demanding 
\begin{equation}
d \Theta^{I(+)} = 0\ .
\end{equation}

We now turn to the auxiliary fermion Killing spinor equation. Next we wish to 
substitute for $v$ in terms of $\hat{G}$ and
$\phi$. Carefully evaluating the covariant derivative of $v$ we obtain
\begin{eqnarray}
 \nabla_0 v_{0i} &=& 2 e^{3\phi} \hat{v}_{il} \hat{\nabla}^l \phi + \frac{1}{2} 
e^{3\phi} \hat{G}_{il} \hat{v}^{(0)l} \ ,\qquad
 \nabla_0 v_{ij} = 4 e^{2\phi} \hat{v}^{(0)}_{[i} \hat{\nabla}_{j]} \phi + 
e^{4\phi} \hat{v}_{[i|l} \hat{G}_{j]}^{\phantom{m}l}\nonumber\ ,  \\
 \nabla_k v_{0i} &=& e^{2\phi} \hat{\nabla}_k \hat{v}^{(0)}_i + e^{2\phi} 
\hat{v}^{(0)}_k \hat{\nabla}_i \phi + e^{2\phi} \hat{v}^{(0)}_i \hat{\nabla}_k 
\phi - e^{2\phi} \hat{\eta}_{ik} \hat{v}^{(0)}_l \hat{\nabla}^l \phi - 
\frac{1}{2} e^{4\phi} \hat{v}_{il} \hat{G}_{k}^{\phantom{m}l} \ ,\label{nablav}\\
 \nabla_k v_{ij} &=& e^{3\phi} \hat{\nabla}_k \hat{v}_{ij} + 2 e^{3\phi} 
\hat{v}_{ij} \hat{\nabla}_{k} \phi + 2 e^{3\phi} \hat{v}_{[i|k} 
\hat{\nabla}_{j]} \phi + 2 e^{3\phi} \hat{\eta}_{[i|k} \hat{v}_{j]l} 
\hat{\nabla}^l\phi + e^{3\phi} \hat{v}^{(0)}_{[i} \hat{G}_{j]k} \nonumber\ .
\end{eqnarray}
 Using this the expressions defined in \eqref{defs-aux} become
\begin{eqnarray}
\mathcal{A} &=& D - \tfrac{3}{2} e^{4\phi} \hat{G}^{(-)}\cdot\hat{G}^{(-)}  - 
\tfrac{1}{2} e^{4\phi }\hat{G}^{(+)}\cdot\hat{G}^{(+)}  - 3 e^{2\phi} 
\hat{\nabla}^2 \phi +18 e^{2\phi} (\hat{\nabla}\phi \cdot\hat{\nabla}\phi)\ , 
\nonumber \\
\mathcal{A}^i &=& 3 e^{3\phi} \left[\tfrac{1}{2} \hat{\nabla}_j \hat{G}^{(+)ji} 
-  \tfrac{1}{2} \hat{\nabla}_j \hat{G}^{(-)ji} - \hat{G}^{(+)ji} \hat{\nabla}_j 
\phi + \hat{G}^{(-)ji} \hat{\nabla}_j \phi \right]\ , \nonumber \\
\mathcal{A}^{ij} &=& 0\ .
\end{eqnarray}
Recall that in four dimensions for a two-form $\alpha$ we have the identity
\begin{equation}
\hat{\nabla}_{j}  \alpha^{ji} =\left( *d*\alpha \right)^i\ ,
\end{equation}
so $\A^i$ is proportional to the Hodge dual of the 3 form $d\left( e^{-2\phi} G 
\right)$, but
$G=e^{2\phi} d\Omega$, and hence $\mathcal{A}^i = 0$. Using this together with 
$\mathcal{A}^{ij}=0$
in the linear system \eqref{lin-sys-time-aux'}, one sees that the latter is 
satisfied iff $\mathcal{A}=0$.
Thus the only additional condition arising from the auxilary fermion equation 
is an expression for $D$,
\begin{equation}
D = \tfrac{3}{2} e^{4\phi} \hat{G}^{(-)} \cdot \hat{G}^{(-)} + \tfrac{1}{2} 
e^{4\phi }\hat{G}^{(+)} \cdot \hat{G}^{(+)} + 3 e^{2\phi} \hat{\nabla}^2 \phi - 
18 e^{2\phi} (\hat{\nabla}\phi)^2\ .
\end{equation}

\subsection{Killing spinor identities and equations of motion}

Here we will examine the equations of motion using the Killing spinor 
identities in the time-like basis,
given in section \ref{ksitime-like} for the representative \eqref{SU(2)-rep1'}. 
We obtain
\begin{eqnarray} 
\mathcal{E}(A)_I^0 - \mathcal{E}(M)_I = 0\ , \qquad
\mathcal{E}(A)_I^i &=& 0\ , \nonumber \\
\left( 
\tfrac{1}{4}\mathcal{E}(v)+\mathcal{E}(D)v\right)^\alpha_{\phantom{\alpha}\alpha
} + \nabla^0\mathcal{E}(D)&=&0\ , \nonumber \\
\left( \tfrac{1}{4}\mathcal{E}(v)+\mathcal{E}(D)v\right)^{0i} - \nabla^i 
\mathcal{E}(D) &=& 0\ , \nonumber \\
\left( \tfrac{1}{4}\mathcal{E}(v)+\mathcal{E}(D)v\right)^{12} = 0\ , \qquad
\mathcal{E}(e)_a^\mu &=& 0\ . \label{halfbpstlksi}
\end{eqnarray}
Note that as the KSI are a consequence of the off-shell supersymmetry, these 
are valid for all higher order corrections that can be added to the theory with 
the same field content, i.e. for any consistent truncation in which the SU(2) 
triplet fields in addition to $N$ and $P_\mu$ are set to zero. 
 In particular for any such corrected action, including the one under 
consideration, it is sufficient to impose the equations of motion
\begin{equation}
\mathcal{E}(D) = 0\ , \qquad \mathcal{E}(v)^{(+)ij} = 0\ , \qquad 
\mathcal{E}(M)_I = 0\ .
\end{equation}

Consider the contribution to the equation of motion coming from the Ricci scalar squared action. Looking at the equations of motion coming from this invariant, we see that the contribution to the gauge field equation of motion vanishes. But we know from the Killing spinor identities that 
$\mathcal{E}(A)_I^0 = \mathcal{E}(M)_I$. Looking at the scalar equation we read off the identity
\begin{equation}
 R=\tfrac{4}{3}v^2-\tfrac{2}{3}D^2\ ,
\end{equation}
where these quantities are all defined on the full five dimensional space.
Using the conditions we have found on the geometry and the expressions for the auxiliary fields we can verify this identity directly.
Turning to the contributions from this density to the other equations of motion, we see that they vanish identically for any supersymmetric background in the time-like class.

The equation of motion for $D$ is therefore given by
\begin{align}
0 &= \tfrac{1}{2}(\mathcal{N} - 1) + \tfrac{c_{2I}}{48} e^{2\phi} 
\left[\tfrac{1}{4} e^{2\phi} M^I \left(\tfrac{1}{3}
\hat{G}^{(+)} \cdot \hat{G}^{(+)} + \hat{G}^{(-)} \cdot 
\hat{G}^{(-)}\right)\right. \nonumber\\
&+ \left.\tfrac{1}{12} e^{2\phi} \hat{G}^{(+)} \cdot \hat{\Theta}^{(+)I}  + M^I 
\hat{\nabla}^2 \phi +
\hat{\nabla}\phi \cdot \hat{\nabla}M^I - 4 M^I 
\hat{\nabla}\phi\cdot\hat{\nabla}\phi\right]\ . \label{dsimplified}
\end{align}
The $M^I$ equation is more involved, but using \eqref{nablav}, and the various 
identities we have collected in appendix \ref{tlidentities}, we find
\begin{align}
0 &=e^{4\phi} \left[ \tfrac{1}{4} c_{IJK} \hat{\Theta}^{(+)J} \cdot 
\hat{\Theta}^{(+)K} - \hat{\nabla}^2\left( e^{-2\phi} \mathcal{N}_I \right) 
\right] + \nonumber\\
&+ \tfrac{c_{2I}}{24} e^{4\phi} \left\{ \hat{\nabla}^2\left( 
3\hat{\nabla}\phi\cdot\hat{\nabla}\phi - \tfrac{1}{12}e^{2\phi}\hat{G}_{(+)}^2 
- \tfrac{1}{4}e^{2\phi}\hat{G}_{(-)}^2   \right)  + \tfrac{1}{8} \hat{R}_{ijkl} 
\hat{R}^{ijkl} \right\}\ . \label{scalarsimplified}
\end{align}
This computation has been checked in Mathematica using the package xAct 
\cite{Martin-Garcia:2002xa, Portugal:1998qi}, and the two equations above are 
in agreement with \cite{Castro:2008ne}.

Finally, after a very long calculation and making extensive use of the 
identities in appendix \ref{tlidentities} we find the equation of motion for 
$v$ yields
\begin{eqnarray}
0=&&-4e^{2\phi}\hat{G}^{(+)}_{ij} + 2e^{2\phi}\mathcal{N}_I 
\hat{\Theta}^{I(+)}_{ij} \nonumber\\&&+ \tfrac{c_{2I}}{24}\left\{ 
\tfrac{1}{2}e^{6\phi} \left(\tfrac{1}{3}\hat{G}_{(+)}^2 + \hat{G}_{(-)}^2 
\right)\hat{\Theta}^{(+)I}_{ij}
- \tfrac{1}{3}e^{4\phi}\left(M^I \hat{G}^{(+)}_{kl} + 2\hat{\Theta}^{I(+)}_{kl} 
\right){\hat{R}_{ij}}^{\phantom{ij}kl} \right. \label{vsimplified}\\
&&\left. +e^{4\phi}\hat{\nabla}^2\left[  M^I ( \hat{G}^{(-)}_{ij} - \tfrac{1}{3} 
\hat{G}^{(+)}_{ij}) \right]
-\tfrac{1}{6}e^{-2\phi}\hat{\nabla}^2 [e^{6\phi} \hat{\Theta}^{I(+)}_{ij}]
-4e^{4\phi}\hat{\nabla}_{[i}\hat{\nabla}_k [ M^I \hat{G}^{(-)k}_{\phantom{(-)k}j]} ] 
\right\}  \, ,\nonumber
\end{eqnarray}
where we have substituted for $\mathcal{N}$ using the equation of motion for 
$D$. To obtain this we found it useful to consider the equation
\begin{equation}
 \mathcal{E}(v)_{ab} + 4k\mathcal{E}(D)v_{ab}=0 \; .
\end{equation}
We have checked the KSI for this equation explicitly and indeed the electric component and the anti-self-dual component automatically vanishes for $k=1$, so that these parts of the $\mathcal{E}(v)_{ab}$ are automatic up to solving $\mathcal{E}(D)$.
It is then sufficient to solve the self-dual part and taking $k=9$ gives the equation above. 
This equation 
was not given in full generality in \cite{Castro:2008ne}, where the equation of motion was contracted 
with $\hat{G}^+$.
Note that the covariant derivatives on the last term commute, and that whilst 
$\hat{\Theta}^I$ is harmonic with respect to the form Laplacian, it 
is not harmonic with respect to the connection Laplacian and 
instead obeys \eqref{laplaciantheta}. Finally note that this equation is selfdual as the antiselfdual part of the last term and the manifestly antiselfdual term $\nabla^2M^I\hat{G}^{(-)}_{ij}$ cancel using the identity \eqref{dK23}.

\subsection{Towards general black hole solutions}

In this section we shall comment briefly on solving the remaining equations of motion, in the case that the solution is a single centre black hole with a regular horizon.
In \cite{Gutowski:2011nk} a systematic analysis of the possible supersymmetric near horizon geommetries of the five dimensional theory inculding the truncated Weyl-squared invariant was performed,
assuming a regular compact horizon, regular fields and that the horizon is Killing with respect to the Killing vector assocated to the Killing spinor bilinear.
In the case of horizon topology $S^3$ it was found that the geometry may be squashed if a certain vortex like equation admits non-constant solutions.
Whether there exist squashed solutions or not, following the analysis of the two derivative case in \cite{Gutowski:2004bj}, it was demonstated that for a supersymmetric black hole the geometry may be written as a U(1) fibration of $\mathbb{R}^4$, and the $\hat{\Theta}^I$ must vanish under some regularity assumptions.
So to investigate the supersymmetric black hole solutions with regular horizons one may always take $\hat{R}_{ijkl}=\hat{\Theta}^I=0$. This means that \eqref{scalarsimplified} may be solved for a set of harmonic functions on $\mathbb{R}^4$ which we label $H_I$
\begin{equation}
e^{2\phi}H_I +    \mathcal{N}_I  
= \tfrac{c_{2I}}{24}  \left\{ 3e^{2\phi}(\hat{\nabla}\phi)^2 - \tfrac{1}{12}e^{4\phi} \hat{G}_{(+)}^2 - \tfrac{1}{4}e^{4\phi}\hat{G}_{(-)}^2    \right\}\ . \label{integratedscalarBH}
\end{equation}
Contracting this with the scalars and using it in \eqref{dsimplified} we find
\begin{equation}
 e^{-2\phi}(1-4\mathcal{N})=H_IM^I+\tfrac{c_{2I}}{24}\left\{ M^I(\hat{\nabla}^2\phi +(\hat{\nabla}\phi)^2) - \hat{\nabla}\phi\cdot\hat{\nabla} M^I \right\} \ .
\end{equation}
The $v$ equation also simplifies to yield
\begin{eqnarray}
0=&&-4e^{2\phi}\hat{G}^{(+)}_{ij} + \tfrac{c_{2I}}{24}\left\{ 
 e^{4\phi}\hat{\nabla}^2\left[  M^I ( \hat{G}^{(-)}_{ij} - \tfrac{1}{3} 
\hat{G}^{(+)}_{ij}) \right]
-4e^{4\phi}\hat{\nabla}_{[i}\hat{\nabla}_k [ M^I \hat{G}^{(-)k}_{\phantom{(-)k}j]} ] 
\right\}  \, ,
\end{eqnarray}
We note that at two derivative level $\hat{G}^{+}$ vanishes, and can thus be dropped from the correction terms to the equations of motion to order $\alpha'$.
Making this assumption the above further simplifies to give an expression for $\hat{G}^{+}$ in terms of second derivatives of $M^I$ and $\phi$, and $d\omega^{-}$. Note that the Laplacian of $M^I\hat{G}^{(-)}$ only occurs to cancel the antiselfdual part of $dK^-$, where $dK^-$ is defined as in \eqref{dK23}, with $\alpha=M^I\hat{G}$.
One would perhaps expect that $\hat{G}^+$ will only be non-zero in the case that the horizon is squashed, corresponding to the loss of two commuting rotational isometries. It would be especialy interesting to investigate this further, and also to use the analysis of \cite{Gutowski:2011nk,Dunajski:2016rtx} to investigate the black ring solutions, and we hope to report on these issues at a later date.
\subsection{The second time-like representative}\label{secondorbit}

As is discussed in appendix \ref{spinors} there is a second orbit with isotropy 
group $\text{SU}(2)$
in the space of spinors. This is related to the first orbit by a $\text{Pin}$ 
transformation that is not in
$\text{Spin}$, which is thus associated to a reflection, rather than a proper 
Lorentz rotation of the frame.
In this section we will briefly give the solution to the Killing spinor 
equations for a representative of this orbit, which are of course very similar 
and which may be read off from the general linear system presented in
appendix \ref{tlsystem}.

The first component is given by $\epsilon^1=e^\phi e^1$, and again inspecting 
the linear system we see that if
it is satified for this component of the spinor, then it is automatically 
satisfied for the second component
$\epsilon^2$, and indeed for the four linearly independent spinors with first 
components
$\epsilon^1,\epsilon^2,i\epsilon^1,i\epsilon^2$.
The one-form bilinear of the representive is the same as in the case of the 
first orbit, and the associated time-like vector field is again Killing so we 
may adapt the same coordinates. 
The non-zero components of the spin connection are antiselfdual, 
$\hat{\omega}_{i,jk}^{(+)}=0$ and thus
$\hat{R}_{ij}^{(+)}=0$. The two-forms associated to this representative are 
different, and are now selfdual,
\begin{eqnarray}
X^{(1)} &=&-e^{2\phi} (e^1 \wedge e^{\bar{2}} + e^{\bar{1}} \wedge e^2)\ , 
\nonumber\\
X^{(2)} &=& +i e^{2\phi} (e^1 \wedge e^{\bar{2}} - e^{\bar{1}} \wedge e^2)\ , 
\nonumber\\
X^{(3)} &=& +i e^{2\phi} (e^1 \wedge e^{\bar{1}} - e^{2} \wedge e^{\bar{2}})\ .
\end{eqnarray}
They are closed, and induce endomorphisms $\mathcal{X}^{(i)}$ on the base 
space, defined by \eqref{calX}.
The $\mathcal{X}^{(i)}$ satisfy \eqref{unit-quat} and \eqref{nablaX=0}, so one 
has again
an integrable quaternionic structure, and thus the base is hyper-K\"ahler.
The gaugino equation \eqref{lin-sys-time-gaug} gives us an expression for $F^I$,
\begin{eqnarray}
F^I &=& -e^{-2\phi} e^0 \wedge d(M^I e^{2\phi}) + M^I G^{(+)} + F^{I(-)} 
\nonumber \\
&=& d(M^I e^0) - M^I G^{(+)} + F^{I(-)}\ ,
\end{eqnarray}
where now it is the antiselfdual part of the flux which is undetermined.
Thus we define the closed form
\begin{equation}
\Theta^{I(-)} := F^{I(-)} - M^I G^{(-)} \; .
\end{equation}
and again, using the Bianchi identity, this is independent of $t$.

From the auxilary fermion equation we just get the same expression for $D$, 
after interchanging $\hat{G}^{\pm}$.
\begin{equation}
D = \tfrac{1}{2} e^{4\phi} \hat{G}^{(-)} \cdot \hat{G}^{(-)} + \tfrac{3}{2} 
e^{4\phi }\hat{G}^{(+)} \cdot \hat{G}^{(+)} + 3 e^{2\phi} \hat{\nabla}^2 \phi - 
18 e^{2\phi} (\hat{\nabla}\phi)^2 \;. 
\end{equation}
In this case the independent EOM's are
\begin{equation}
\mathcal{E}(D) = 0 ,\qquad
\mathcal{E}(M)_I = 0 , \qquad
\mathcal{E}(v)^{(-)ij} = 0 \; .
\end{equation}
The first equation gives
\begin{align}
0 &= \tfrac{1}{2}(\mathcal{N} - 1) + \tfrac{c_{2I}}{24}\tfrac{1}{2} e^{2\phi} \left[
\tfrac{1}{4} e^{2\phi}  M^I \left[ \hat{G}^{(+)} \cdot \hat{G}^{(+)} + 
\tfrac{1}{3} \hat{G}^{(-)} \cdot \hat{G}^{(-)} \right]  \right. \nonumber\\
& \left.- \tfrac{1}{12} 
e^{2\phi}  \hat{G}^{(-)} \cdot \hat{\Theta}^{(-)I} + 
 M^I \hat{\nabla}^2 \phi +  \hat{\nabla}\phi \cdot \hat{\nabla}M^I - 4  M^I 
\hat{\nabla}\phi \cdot \hat{\nabla} \phi \right] \; ,
\end{align}
whilst the second equation reads
\begin{align}
0 &=e^{4\phi} \left[ \tfrac{1}{4} c_{IJK} \hat{\Theta}^{(-)J} \cdot 
\hat{\Theta}^{(-)K} - \hat{\nabla}^2\left( e^{-2\phi} \mathcal{N}_I \right) 
\right]  \nonumber\\
&+ \tfrac{c_{2I}}{24} e^{4\phi} \left\{ \hat{\nabla}^2\left( 
3\hat{\nabla}\phi\cdot\hat{\nabla}\phi - \tfrac{1}{12}e^{2\phi}\hat{G}_{(-)}^2 
- \tfrac{1}{4}e^{2\phi}\hat{G}_{(+)}^2   \right)  + \tfrac{1}{8} \hat{R}_{ijkl} 
\hat{R}^{ijkl} \right\}\ .
\end{align}
The auxiliary two form equation of motion is
\begin{eqnarray}
0=&&-4e^{2\phi}\hat{G}^{(-)}_{ij} + 2e^{2\phi}\mathcal{N}_I 
\hat{\Theta}^{I(-)}_{ij} \nonumber\\&&+ \tfrac{c_{2I}}{24}\left\{ 
\tfrac{1}{2}e^{6\phi} \left(\tfrac{1}{3}\hat{G}_{(-)}^2 + \hat{G}_{(+)}^2 
\right)\hat{\Theta}^{(-)I}_{ij}
- \tfrac{1}{3}e^{4\phi}\left(M^I \hat{G}^{(-)}_{kl} + 2\hat{\Theta}^{I(-)}_{kl} 
\right){\hat{R}_{ij}}^{\phantom{ij}kl} \right. \\
&&\left. +e^{4\phi}\hat{\nabla}^2\left[  M^I ( \hat{G}^{(+)}_{ij} - \tfrac{1}{3} 
\hat{G}^{(-)}_{ij}) \right]
-\tfrac{1}{6}e^{-2\phi}\hat{\nabla}^2 [e^{6\phi} \hat{\Theta}^{I(-)}_{ij}]
-4e^{4\phi}\hat{\nabla}_{[i}\hat{\nabla}_k [ M^I \hat{G}^{(+)k}_{\phantom{(+)k}j]} ] 
\right\}  \, ,\nonumber
\end{eqnarray}
which is antiselfdual.

\section{Maximal time-like supersymmetry}\label{maxtime-like}
In the consistent trunaction we are considering it is clear that we need only demand two linearly independent Killing spinors to impose maximal supersymmetry. We include this derivation here,
as it is rather more direct than that presented in \cite{Gauntlett:2002nw}, which left some solutions only conjecturally isometric to the near horizon BMPV geometry, and these conjectures were subsequently proven in \cite{Fiol:2003yq}.
\subsection{Killing spinor equations and geometric constraints}
In the previous section we have only imposed the existence of 
one time-like Killing spinor, so we wish to choose a second Killing spinor. 
Decomposing $\De_C$ under $\text{SU}(2)$ we find
\begin{equation}
\De_C= \mathbb{C}\,\langle1,e^{12}\rangle + \mathbb{C}\,\langle 
e^1,e^2\rangle\,.
\end{equation}
Note that for linear independence the second spinor must have a component in
$\mathbb{C}\,\langle e^1,e^2\rangle$, since we have seen that the spinors 
implied by the existence
of one spinor span $\mathbb{C}\,\langle 1,e^{12}\rangle$. Now notice that we 
may act with the
residual $\text{SU}(2)$ gauge symmetry to write the spinor as
\begin{equation}
\xi^1 = \lambda \; 1 + \sigma e^{12} + e^\chi e^1\,,
\end{equation}
where $\chi$ is real.
So choosing this as the first component of a symplectic Majorana spinor we have
\begin{equation}
\xi=(\lambda \; 1 + \sigma e^{12} + e^\chi e^1,  i\sigma^* \; 1 -  i \la^* 
e^{12} + i e^\chi e^2 )\,.
\end{equation}
Recall that the linear system is equivalent under the symplectic Majorana 
conjugate, in fact it yields the
(dual of the) complex conjugate system. Thus not only 
is it sufficient to consider the
Killing spinor equations for the first component of $\xi$, but this implies 
that the linearly independent spinor
$(\xi^2,\xi^1)$ is also Killing. Now note that $(i\xi^1,-i\xi^2)$ and 
$(i\xi^2,-i\xi^1)$ are also linearly
independent and their linear systems are equivalent to the system from $\xi^1$. 
Finally we note that the
sigma group \cite{Gran:2007fu} of the plane of parallel spinors of the half-supersymmetric 
solution,
$\Sigma(\mathcal{P})=\textrm{Stab}(\mathcal{P})/ \textrm{Stab}(\epsilon,\eta, 
\chi,\zeta)$, is a rigid
$\text{SU}(2)$, where $\mathcal{P}=\mathbb{C}\,\langle 
e^{\phi}1,e^{\phi}e^{12}\rangle$, due to the supersymmetry enhancement found in the previous section. 
So to summarize, by demanding the existence of one time-like Killing spinor 
$\epsilon$ we saw that this implied
the existence of another three linearly independent Killing spinors, and when 
demanding the existence of
one more linearly independent to these we have maximal supersymmetry. 

First let us consider the gravitino equation. The linear system 
\eqref{lin-sys-time-grav} for $\xi^1$ yields
\begin{eqnarray}
 \sqrt{2}\partial_0 \lambda 
 -e^\chi \left( \omega_{0,01} - \tfrac{4}{3}v_{01}
\right)  &=& 0\,,\\
 \partial_0 \chi
 -\left( \tfrac{1}{2} \left( \omega_{0,1\bar{1}}
-\omega_{0,2\bar{2}}  \right) - \tfrac{1}{3} \left( v_{1\bar{1}} - v_{2\bar{2}}
\right)  \right) &=&0\,,\\
 \omega_{0,1\bar{2}} -\tfrac{2}{3}v_{1\bar{2}} = 0\,, \qquad
 e^\chi \left( \omega_{0,0\bar{2}} - \tfrac{4}{3}v_{0\bar{2}}
\right) 
 +\sqrt{2} \partial_0 \sigma &=&0\,, \\
 \sqrt{2}\partial_\alpha \lambda -\sqrt{2}\la \partial_\alpha \phi 
 - e^\chi \left( \omega_{\alpha,01} + 2\delta_{\alpha2}v_{12}
\right) &=&0\,, \label{second2}\\
 -\partial_\alpha\chi + \left(
\tfrac{1}{2}\left( \omega_{\alpha,1\bar{1}} - \omega_{\alpha,2\bar{2}} \right) +
\tfrac{1}{3}\delta_{1\alpha}v_{01} + \delta_{\alpha2}v_{02} \right) &=&0\,, 
\label{third2}\\
   \omega_{\alpha,1\bar{2}} -
\tfrac{2}{3}\delta_{\alpha 2}v_{01}  
 = 0\,, \qquad
\left( \om_{\bal, 1\bar2} +
\tfrac{2}{3}\de_{\bal\bar1}v_{0\bar2} \right) &=&0\,, \label{first1}
\end{eqnarray}
\begin{eqnarray}
 e^\chi \left( \omega_{\alpha, 0 \bar{2}} -
\tfrac{2}{3}\delta_{\alpha1}v_{1\bar{2}} - \tfrac{2}{3}\delta_{\alpha 2} \left(
v_{1\bar{1}} +2v_{2\bar{2}}  \right) \right)
+\sqrt{2}\partial_\alpha \sigma - \si \sqrt{2}\partial_\alpha \phi 
\label{fourth2}
&=&0\,, \\
 \sqrt{2}\partial_{\bar{\alpha}}\lambda -\sqrt{2}\la\partial_{\bar{\alpha}}\phi
-  e^\chi\left( \omega_{\bar{\alpha},01} +
\tfrac{2}{3}\delta_{\bal \bar1}\left( 2v_{1\bar1} + v_{2\bar2} \right) +
\tfrac{2}{3}\de_{\bal \bar 2}v_{1\bar2} \right) \label{fourth1}
&=&0\,,\\
-\partial_{\bal}\chi
+\left( \tfrac{1}{2}\left(\om_{\bal, 1\bar1}-\om_{\bal,
2\bar2} \right) + \de_{\bal\bar1}v_{0\bar1}
+\tfrac{1}{3}\de_{\bal\bar2}v_{0\bar2}\right) \label{third1}
&=&0\,, \\
 e^\chi \left( \om_{\bal, 0\bar2} - 2\de_{\bal \bar1}v_{\bar1 \bar2} \right)
+\sqrt2\partial_{\bal}\si - \sqrt2 \si \partial_{\bal}\phi 
&=&0\,.\label{second1}
\end{eqnarray}
The first four equations give
\begin{equation}
\sqrt{2}\partial_0 \la = 4 e^{\phi +\chi}\hat{\nabla}_1\phi\,, \qquad
-\sqrt{2}\partial_0 \si = 4 e^{\phi +\chi}\hat{\nabla}_{\bar2}\phi\,, \qquad
\partial_0 \chi = G^{(+)} =0\,. \label{d0f}
\end{equation}
From (\ref{first1}) and \eqref{third2}, \eqref{third1} we obtain respectively
\begin{eqnarray}
\hat{\om}_{1,1\bar2} = \hat{\om}_{2,\bar{1}2} = 0\,, \qquad
\hat{\om}_{1,\bar{1}2} = -2\hat{\nabla}_2 \phi\,, \qquad \hat{\om}_{2,1\bar2} = 
2\hat{\nabla}_1 \phi\,,\nonumber\\
\hat{\om}_{1,1\bar1}-\hat{\om}_{1,2\bar2}= 2\hat{\nabla}_1\phi\,, \qquad
\hat{\om}_{2,1\bar1}-\hat{\om}_{2,2\bar2}=-2\hat{\nabla}_2\phi\,, \qquad
d\phi=-d\chi\,. \label{phi-is-chi}
\end{eqnarray}
From (\ref{second2}), (\ref{second1}) we get
\begin{eqnarray}
\hat{\nabla}_1 (e^{-\phi} \la)=0\,, \qquad
\hat{\nabla}_2 (e^{-\phi} \si^*)=0\,, \nonumber\\
\sqrt{2}e^{\chi}\hat{G}^{(-)}_{12}=\hat{\nabla}_1(\si^* 
e^{-\phi})=\hat{\nabla}_2(\la e^{-\phi})\,, \label{gminus1}
\end{eqnarray}
and finally (\ref{fourth2}) and (\ref{fourth1}) give
\begin{eqnarray}
\hat{\nabla}_1(e^{-\phi}\si)=0\,, \qquad
\hat{\nabla}_2(e^{-\phi}\la^*)=0\,, \nonumber\\
\sqrt{2}e^\chi \hat{G}^{(-)}_{1\bar1}=\hat{\nabla}_1(e^{-\phi}\la^*)= 
\hat{\nabla}_2(e^{-\phi}\si)\,. \label{gminus2}
\end{eqnarray}
The gaugini equations \eqref{lin-sys-time-gaug} boil down to
\begin{equation}
\nabla_A M^I = {\mathcal{F}}^I =0\,,
\end{equation}
so
\begin{equation}
F^I=2M^Ie^0\wedge d\phi  -M^IG^{(-)}\,.
\end{equation}
The Bianchi identity for $F^I$ is therefore satisfied,
\begin{equation}
dF^I=2M^I de^0\wedge d\phi - M^I dG^{(-)}=0\,.
\end{equation}
We can write the auxilary fermion equation as
\begin{equation}
(\B+\B^i\gamma_i) e^1=0\,, \label{maxtl-aux-1}
\end{equation}
since $e^\chi$ is non-zero. Consider first the $\B^i$ part, substituting 
$\A^i=0$ one gets
\begin{eqnarray}
\B^i&=& -4\epsilon^{ijkl}\nabla_j 
v_{kl}=-6\epsilon^{ijkl}e^{3\phi}\hat{\nabla}_j (e^{-2\phi}G^{(-)}_{kl}) =0\,.
\label{maxtlbi}
\end{eqnarray}
Thus the condition remaining from (\ref{maxtl-aux-1}) becomes simply $\B=0$, 
which yields
\begin{equation}
0= 6e^{2\phi} \left( \hat{\nabla}^i\hat{\nabla}_i\phi -2\hat{\nabla}^i\phi 
\hat{\nabla}_i\phi \right)=6e^{4\phi} \hat{\nabla}^i\hat{\nabla}_i 
e^{-2\phi}\,. \label{maxtlb0}
\end{equation}
Thus $H=e^{-2\phi}$ is harmonic on the base, whilst the expression for the 
auxiliary scalar $D$ becomes
\begin{equation}
D=\tfrac{3}{2}e^{4\phi}(\hat{G}^{(-)})^2  -12\hat{\nabla}^i \phi \hat{\nabla}_i 
\phi\,.\label{auxhat}
\end{equation}
We note that as $\widehat{d\Omega}=e^{-2\phi}\hat{G}^{(-)}$ is a closed 
anti-selfdual two-form, it can be written as a constant linear combination of 
the hyper-K\"ahler two-forms on the base. As they are covariantly constant with 
respect to the $\hat{\nabla}$ connection,  so is $\widehat{d\Omega}$.
We can calculate $(\hat{G}^{(-)})^2$  from (\ref{gminus1}), (\ref{gminus2}) to 
get
\begin{eqnarray}
(\hat{G}^{(-)})^2&=&  
\textrm{Re}(\lambda)^2\hat{\nabla}_i\phi\hat{\nabla}^i\phi -2 
\textrm{Re}(\lambda)\hat{\nabla}_i\phi \hat{\nabla}^i \textrm{Re}(\lambda) + 
\hat{\nabla}_i\textrm{Re}(\lambda)\hat{\nabla}^i\textrm{Re}(\lambda) \ ,\nonumber\\
&=& \textrm{Im}(\lambda)^2\hat{\nabla}_i\phi\hat{\nabla}^i\phi -2 
\textrm{Im}(\lambda)\hat{\nabla}_i\phi \hat{\nabla}^i \textrm{Im}(\lambda) + 
\hat{\nabla}_i\textrm{Im}(\lambda)\hat{\nabla}^i\textrm{Im}(\lambda)\,,
\label{g-as-lambda}
\end{eqnarray}
with similar expressions involving $\sigma$, where we have used the last 
equation of (\ref{phi-is-chi}) to see
that $e^{2(\phi+\chi)}$ is just some positive constant, and moreover we can 
always rescale the spinor $\xi$ such that  $e^{(\phi+\chi)}=1/4$ .

The connection 1-forms $\hat{\omega}$ are completely determined and to compute 
the curvature two-form,  it is convenient to write
\begin{eqnarray}
\hat{\omega}_1&=&\hat{\nabla}_1 \phi \left[M,\bar{M}\right] + 
2\hat{\nabla}_2\phi M\,, \qquad
\hat{\omega}_{\bar1} = -\hat{\nabla}_{\bar1} \phi \left[M,\bar{M}\right] + 
2\hat{\nabla}_{\bar2}\phi\bar{M}\,,
\nonumber\\
\hat{\omega}_2&=&-\hat{\nabla}_2 \phi \left[M,\bar{M}\right] - 
2\hat{\nabla}_1\phi \bar{M}\,, \qquad
\hat{\omega}_{\bar2} = \hat{\nabla}_{\bar2} \phi \left[M,\bar{M}\right] - 
2\hat{\nabla}_{\bar1}\phi M\,,
\end{eqnarray}
where $M,\bar{M},[M,\bar{M}]$ are the linearly independent matrices (with index 
ordering $(1,\bar1, 2, \bar2)$)
\begin{equation}
M=\left(\begin{array}{r r r r}
       0 & 0 & 1 & 0 \\
       0 & 0 & 0 & 0 \\
       0 & 0 & 0 & 0 \\
       0 & -1 & 0 & 0
      \end{array}\right)\,,
\qquad\bar{M}=
\left(\begin{array}{r r r r}
       0 &  0 & 0 & 0 \\
       0 & 0 & 0 &1 \\
       -1 & 0 & 0 & 0 \\
       0 & 0 & 0 & 0
      \end{array}\right)\,,
\qquad\left[M,\bar{M}\right]=\left(\begin{array}{r r r r}
       -1 & 0 & 0 & 0 \\
       0 & 1 & 0 & 0 \\
       0 & 0 & 1 & 0 \\
       0 & 0 & 0 & -1
      \end{array}\right)\,. 
\end{equation}
The nonzero components of the curvature two-form (with its coordinate indices 
flattened with the vielbein)
can then be written
\begin{eqnarray}
\hat{R}_{1\bar1}&=&-e^{-2\phi}\hat{\nabla}_{\bar1} \hat{\nabla}_{2}e^{2\phi}  M 
+ e^{-2\phi}\hat{\nabla}_{1}\hat{\nabla}_{\bar2} e^{2\phi} \bar{M} -\left( 
2\hat{\nabla}_{1}\hat{\nabla}_{\bar1}\phi - 
4\hat{\nabla}_{2}\phi\hat{\nabla}_{\bar2}\phi\right) [M,\bar{M}]\,,\nonumber\\
\hat{R}_{2\bar2}&=&-e^{-2\phi}\hat{\nabla}_{\bar1} \hat{\nabla}_{2}e^{2\phi}  M 
+ e^{-2\phi}\hat{\nabla}_{1}\hat{\nabla}_{\bar2} e^{2\phi} \bar{M} 
+\left(2\hat{\nabla}_{2}\hat{\nabla}_{\bar2}\phi - 
4\hat{\nabla}_{1}\phi\hat{\nabla}_{\bar1}\phi\right) [M,\bar{M}]\,,\nonumber\\
\hat{R}_{12}&=&-e^{-2\phi}\hat{\nabla}_2 \hat{\nabla}_{2}e^{2\phi} M - 
e^{-2\phi}\hat{\nabla}_1 \hat{\nabla}_{1}
e^{2\phi} \bar{M} - e^{-2\phi}\hat{\nabla}_{1}\hat{\nabla}_{2}e^{2\phi}  
[M,\bar{M}]\,,\nonumber\\
\hat{R}_{\bar1\bar2}&=& -e^{-2\phi}\hat{\nabla}_{\bar1} 
\hat{\nabla}_{\bar1}e^{2\phi} M - e^{-2\phi}\hat{\nabla}_{\bar2} 
\hat{\nabla}_{\bar2}e^{2\phi}\bar{M} + 
e^{2\phi}\hat{\nabla}_{\bar1}\hat{\nabla}_{\bar2}e^{2\phi}  
[M,\bar{M}]\,,\nonumber\\
\hat{R}_{1\bar2}&=&-\tfrac{1}{2}e^{2\phi}\hat{\nabla}^i\hat{\nabla}_ie^{-2\phi} 
M\,, \qquad
\hat{R}_{\bar1 2} = 
-\tfrac{1}{2}e^{2\phi}\hat{\nabla}^i\hat{\nabla}_ie^{-2\phi} \bar{M}\,.
\end{eqnarray}
Using the symmetries of the curvature tensor, in particular setting 
$\hat{R}_{ij}^{(-)}=0$ leads to
\begin{equation}
\hat{\nabla}^i \hat{\nabla}_{j}H^{-1}=0\,, \quad i\neq j\,, \qquad
\hat{\nabla}_1\hat{\nabla}^1 H^{-1}=\hat{\nabla}_2\hat{\nabla}^2 H^{-1}\,,
\end{equation}
and we find that the base space is locally flat, as we also have that $H$ is a 
positive harmonic function.
We can write $\hat\nabla^2H=0$ in terms of $H^{-1}$ as
\begin{equation}
\hat{\nabla}^i\hat{\nabla}_i H^{-1} + 2 H^{-1}\hat{\nabla}^i 
H^{-1}\hat{\nabla}_i H^{-1}=0\,,
\end{equation}
which allows us to rewrite the conditions on $H$ in the concise form that 
appears in \cite{Gauntlett:2002nw};
\begin{equation}
-\hat{\nabla}_i\hat{\nabla}_j H^{-1} + 
\frac{1}{2H}\delta_{ij}\delta^{pq}\hat{\nabla}_pH^{-1}
\hat{\nabla}_qH^{-1}=0\,.
\end{equation}
Solving this equation we have that $H=k$, or $H=\tfrac{2k}{r^2}$, where $k$ is 
a positive constant and $r^2=(x_1)^2 + \cdots + (x_4)^2$, and we have 
introduced coordinates such that the metric on the base is
$d\hat s^2=\delta_{ij}dx^idx^j$.

Let us first consider the case $dH=0$. We thus have $d\phi=0$, the connection 
and electric parts of $v$ and
$F^I$ vanish, as does the auxiliary scalar $D$, and we have two cases to 
consider, depending on whether
$G^{(-)}$ vanishes or not. In the case $G^{(-)}=0$, all of the gauge and 
auxiliary fields vanish, and we are left with five-dimensional Minkowski space.

Now let us take $G^{(-)}\neq 0$. Setting
$f^i=\{\textrm{Re}(\lambda),\textrm{Im}(\lambda),\textrm{Re}(\sigma),\textrm{Im}
(\sigma)\}$, we must have
$f^i\neq0 \,\forall i$ from (\ref{g-as-lambda}) and $\partial_0f^i=0$ from the 
first two eqns.~of (\ref{d0f}). 
Furthermore none of the $f^i$ may be proportional. One can see this by making a 
(rigid)
$\text{SU}(2)$ transformation in $\Sigma(\mathcal{P})$. In the case that any 
two of the $f^i$ are proportional, we may set one of them to zero and hence 
obtain $G^{(-)}=0$, without loss of generality.
$\hat{G}^{-}$ is now covariantly constant and can be written as a constant 
linear combination of the hyper-K\"ahler two-forms,
$\hat{G}^{(-)}=\sum_{(i)=(1)}^{(3)}c^{(i)} \hat{X}^{(i)}$. This implies
\begin{equation}
\hat{\nabla}\hat{\nabla}f^i=0\,.
\end{equation}
Hence a suitable solution for the parameters of the Killing spinors is 
$f^i=a^ix^i$ (no sum over $i$,
$a^i\neq 0 \;\forall i$) in Cartesian coordinates on the base, where $a^i$ are 
constants and
$(a^1)^2 + \cdots +(a^4)^2=\hat{G}^{(-)}{}^2=4\sum_{(i)=(1)}^{(3)}(c^{(i)})^2$. 
Following
\cite{Gauntlett:2002nw} we next introduce  $\text{SU}(2)$ right-invariant (or ``left'')
one-forms $\sigma^{(i)}_L$
on the base such that $X^{(i)}= \tfrac{1}{4} d(r^2\sigma^{(i)}_L)$, where 
from now on we
will leave the sum over $(i)$ implicit. Introducing Euler angles for SU(2) $0\leq \theta \leq \pi$, $0\leq \phi \leq 2\pi$, $0\leq\psi < 4\pi$,
which in terms of the cartesian coordinates are given by
\begin{eqnarray}
 x^1 + ix^2 = r\cos \tfrac{\theta}{2}e^{\tfrac{i}{2}(\psi+\phi)} \ , \nonumber\\
 x^3 + ix^4 = r\sin \tfrac{\theta}{2}e^{\tfrac{i}{2}(\psi-\phi)} \ ,
\end{eqnarray}
these 1-forms have the parametrization
\begin{eqnarray}
 \sigma_L^{(1)}&=& \sin \phi d\theta - \cos \phi \sin \theta d\psi \ ,\nonumber\\
 \sigma_L^{(2)}&=& \cos \phi d\theta + \sin \phi \sin \theta d\psi \ ,\nonumber\\
 \sigma_L^{(3)}&=&  d\phi + \cos \theta d\psi \ ,
\end{eqnarray}
and obey 
\begin{equation}
d\sigma_L^{(i)}=-\tfrac{1}{2}\epsilon^{(i)(j)(k)}\sigma_L^{(j)} \wedge \sigma_L^{(k)} \ .
\end{equation}

We can now solve for $\Omega$,
\begin{equation}
\Omega= \frac{k r^2}{4} c^{(i)}\sigma^{(i)}_L\,.
\end{equation}
Let us now turn to the case $H=\tfrac{2k}{r^2}$. In this case we have 
$\nabla(H G^{(-)})=0$. We 
introduce a new basis of anti-selfdual two-forms 
$Q^{(i)}=d(r^{-2}\sigma^{(i)}_R)$, where
$\sigma^{(i)}_R$ denote $\text{SU}(2)$ left-invariant (or ``right'') one-forms. In terms of the Euler angles these are parameterized by
\begin{eqnarray}
 \sigma_R^{(1)}&=& -\sin \psi d\theta + \cos \psi \sin \theta d\phi \ , \nonumber\\
 \sigma_R^{(2)}&=& \cos \psi d\theta + \sin \psi \sin \theta d\phi \ ,\nonumber\\
 \sigma_R^{(3)}&=&  d\psi + \cos \theta d\phi \ ,
\end{eqnarray}
which obey
\begin{equation}
d\sigma_R^{(i)}=\tfrac{1}{2}\epsilon^{(i)(j)(k)}\sigma_R^{(j)} \wedge \sigma_R^{(k)}.
\end{equation}
Then writing $\hat{G}^{(-)} = c^{(i)} r^2 \hat{Q}^{(i)}$,  we find
\begin{equation}
\Omega = \frac{2k}{ r^2} c^{(i)} \sigma^{(i)}_R\,.
\end{equation}
The five-dimensional spacetime geometry is given by
\begin{equation}
ds^2 = \frac{r^4}{4k^2} (dt + \frac{2k}{ r^2}
c^{(i)} \sigma^{(i)}_R)^2
-\frac{2k}{r^2} \left[ dr^2 +r^2 d \Omega_3{}^2 \right]\,.
\end{equation}
This is the near-horizon geometry of the rotating BMPV black hole 
\cite{Breckenridge:1996is}.
Setting $c^{(i)} = 0$ gives $\text{AdS}_2\times\text{S}^3$.

In summary, we have the following  cases:
\begin{itemize}
\item Five-dimensional Minkowski space. All coefficients of the Killing spinors 
are constants and all
auxiliary and gauge fields vanish.
\item The G\"odel-type solution \cite{Gauntlett:2002nw}. The scalars are 
constant, $dM^I=0$. The base
space is $\bR^4$, the electric parts of the fluxes vanish and $d\phi=0$. The 
metric can be written
\begin{equation}
ds^2 = k^{-2} (dt+\frac{kr^2}{4}c^{(i)}\sigma^{(i)}_L)^2-k\left[ dr^2 +r^2 d 
\Omega_3{}^2 \right]\,.
\end{equation}
Only the anti-selfdual parts of the magnetic components of $v,F^I$ are non zero 
and are given by
$\hat{F}^I=-\tfrac{4}{3}M^I \hat{v}^{(-)} = M^Ic^{(i)} \hat{X}^{(i)}$. 
 \item $\text{AdS}_2\times\text{S}^3$, 
\begin{equation}
ds^2 = \frac{r^4}{4k^2} dt^2
-\frac{2k}{ r^2} \left[ dr^2 +r^2 d \Omega_3{}^2 \right]\,. \label{ads2s3}
\end{equation}
The electric fluxes are non-zero and given by $F^I=\tfrac{1}{2k}M^Idt\wedge 
dr$.
\item Near-horizon geometry of the BMPV black hole,
\begin{equation}
ds^2 = \frac{r^4}{4k^2} (dt + \frac{2k}{ r^2}c^{(i)} \sigma^{(i)}_R)^2
-\frac{2k}{ r^2} \left[ dr^2 +r^2 d \Omega_3{}^2 \right]\,.
\end{equation}
We have electric and magnetic fluxes with $F^I=\tfrac{1}{2k}M^Idt\wedge dr + 
M^I\tfrac{c^{(i)}}{ r^2} \sigma^{(i)}_R \wedge dr$. 
\end{itemize}
We have derived these results off-shell in our consistent truncation, next we shall examine the 
equations of motion by
making use of the Killing spinor identities. The results for the system if the first Killing spinor is taken to be in the second orbit are similar, with self- and anti-self-dual forms interchanged.

\subsection{Killing spinor identities and equations of motion}
In addition to (and using) the conditions derived from the half-BPS time-like 
case in (\ref{halfbpstlksi}), we obtain
\begin{eqnarray}
\mathcal{E}(M_I)=0\,, \qquad \mathcal{E}(A_I)=0\,, \qquad
\left( \tfrac{1}{4}\mathcal{E}(v)+\mathcal{E}(D)v\right)^{\bar1 2}=0\,, 
\nonumber \\
\left( \tfrac{1}{4}\mathcal{E}(v)+\mathcal{E}(D)v\right)^{1\bar1} - \left( 
\tfrac{1}{4}\mathcal{E}(v)+\mathcal{E}(D)v\right)^{2\bar2} = 
\nabla^0\mathcal{E}(D)\,, \nonumber\\
\left( \tfrac{1}{4}\mathcal{E}(v)+\mathcal{E}(D)v\right)^{0i}= -\nabla^i 
\mathcal{E}(D)\,,
\end{eqnarray}
from which we immediately see that it is sufficient to impose the single equation of motion
\begin{equation}
\mathcal{E}(D)=0\,.
\end{equation}
This can be written as
\begin{eqnarray}
0 &=& \tfrac{1}{2}(\mathcal{N} - 1) + \tfrac{c_{2I}}{144}\left[ M^ID  
+2v^{0i}F^I_{0j} + v^{ij}F^I_{ij}\right]\,,
\nonumber\\
&=&(\mathcal{N} - 1) + \tfrac{c_{2I}}{72}M^I\left[ 
2e^{2\phi}\hat{\nabla}_i\phi\hat{\nabla}^i\phi + 
\tfrac{3}{2}e^{4\phi}\hat{G}^{(-)ij}\hat{G}^{(-)}_{ij}\right]\,.
\end{eqnarray}
Thus in the first case, Minkowski space, we obtain the usual very special 
geometry condition
\begin{equation}
\mathcal{N} = 1\,,
\end{equation}
while for the G\"odel-type solution and $\text{AdS}_2\times\text{S}^3$ we get 
respectively
\begin{equation}
\mathcal{N}= 1  - c_{2}\frac{c^{(i)}c_{(i)}}{12k^2}\,,
\end{equation}
\begin{equation}
\mathcal{N} = 1 - \frac{c_{2}}{144k}\,,
\end{equation}
where we defined $c_{2}=c_{2I}M^I$.
Finally for the near-horizon BMPV solution, we obtain
\begin{equation}
\mathcal{N} = 1 - \frac{c_{2}}{36}\left(\frac{1}{k} 
+\frac{3}{k^2}c^{(i)}c_{(i)}\right)\,.
\end{equation}

Note that these are all constant deformations of the very special geometry condition $\mathcal{N}=1$. One may wonder whether this is a coincidence for the invariants we have considered, or whether this will always be the case. Looking at the Killing spinor identities, tells us that
\begin{equation}
 \nabla\mathcal{E}(D)=0\ ,
\end{equation}
so that corrections to the equation of motion of $D$ and hence corrections to the very special geometry condition
\begin{equation}
 \mathcal{N}=1 +  \mathcal{O}(\alpha') + \cdots
\end{equation}
must be constant for the maximally supersymmetric time-like solutions. Again the results if we take the first Killing spinor to be in the second time-like representative are similiar, up to a reflection.

\section{Null supersymmetry and the Ricci scalar squared invariant}\label{nullriccisquared}
In this section we will show that the Ricci scalar squared invariant does not affect the equations of motion for the null class of supersymmetric solutions, without going into the details of the geometries. This shows the power of the Killing spinor identities in analysing higher derivative invariants.
As shown in detail in appendix \ref{spinors} a representative for the orbit of Spin$(1,4)$ in the space of spinors with stability subgroup $\mathbb{R}^3$ has first component
\begin{equation}
 \epsilon^{\mathbf{1}}=(1+e_1) \ .
\end{equation}
Using the adapted basis \eqref{nullbasis} we find the linear system presented in \ref{ksinull}. Taking $z_1=1$ all others vanishing in this system yields
\begin{eqnarray}
&&\mathcal{E}(M)_I = 0, \qquad
\mathcal{E}(A)_I^+ = 0, \qquad
\mathcal{E}(A)_I^i = 0, \qquad
\tfrac{1}{4} \mathcal{E}(v)^{+-} + \mathcal{E}(D) v^{+-} = 0, \nonumber\\
&&\nabla^+ \mathcal{E}(D) = 0, \qquad
\tfrac{1}{4} \mathcal{E}(v)^{+i} + \mathcal{E}(D) v^{+i} = 0, \qquad
\tfrac{1}{4} \mathcal{E}(v)^{ij} + \mathcal{E}(D) v^{ij} - \epsilon^{ijk} \nabla^k \mathcal{E}(D) = 0 ,\nonumber\\
&&a=+,-,i \quad  \left. \mathcal{E}(g)_{a-} \right|_{\text{other bosons on-shell}} = 0,  \qquad
\left. \mathcal{E}(g)_{aj} \right|_{\text{other bosons on-shell}} = 0, 
\end{eqnarray}
and we conclude that the equations that remain to be solved are
\begin{eqnarray}
\mathcal{E}(D) = 0 \ ,\qquad
\mathcal{E}(A)_I^- = 0 \ , \qquad &&
\mathcal{E}(v)^{-i} = 0 \ ,\qquad
\mathcal{E}(g)_{++} = 0 \; .
\end{eqnarray}
Notice however that the scalar equation is automatic, which imples that 
\begin{equation}
  R=\tfrac{4}{3}v^2-\tfrac{2}{3}D^2 \ ,
\end{equation}
just as in the time-like case. Note that since this must arise due to the supersymmetry conditions alone, and not any other equations of motion, that this is an identity for the null class whether we couple to the Ricci scalar squared invariant or not, i.e. whether $e_I$ vanishes or not.
This completes the proof that the Ricci scalar squared invariant does not contribute to the equations of motion of any supersymmetric solution in this consistent truncation, and thus to any supersymmetric solution at first order  in $\alpha'$.

\section{Maximal supersymmetry in the general case}\label{allordersmax}
In this section we will work with the untruncated theory in order to show that the maximally supersymmetric solutions of the two derivative supergravity theory are those of the minimal theory, i.e. the all order consistency of the maximally supersymmetric vacua.
This was discussed in \cite{Meessen:2007ef}, but there an on-shell hypermultiplet compensator was used. Due to the construction of supersymmetric higher derivative invariants using the compensator, it becomes important to have this multiplet off-shell.
Whilst we have shown the Ricci scalar invariant does not affect the solutions in the truncated case (and so to order $\alpha'$ in the presence of the invariants we have considered), other invariants involving the compensating multiplet may have some effect, as may the invariants we consider here when considering their contribution to higher order in $\alpha'$.
In fact it is well known that this occurs, since adding the cosmological constant density changes the theory in such a way that the only maximally supersymetric solution at two derivative level is AdS$_5$.
We also wish to generalize to the case in which the higher derivative supergravity need not be the usual two derivative one with perturbative corrections, but also allow the higher derivative terms to have large coefficients.
The equations we wish to solve are

\begin{align}
0 &= \nabla_\mu \epsilon^{\mathbf{i}} + \tfrac{1}{2} \gamma_{\mu ab} 
v^{ab} \epsilon^{\mathbf{i}} - \tfrac{1}{3} \gamma_\mu \gamma_{ab} v^{ab} \epsilon^{\mathbf{i}} + V^{\mathbf{ij}}_{\mu}\epsilon_{\mathbf{j}}+  
\tfrac{1}{6}\gamma_\mu(\slashed{P}+N)L^{\mathbf{ij}}\epsilon_{\mathbf{j}}- 
\tfrac{1}{3}\gamma_\mu\gamma^a{V'}_a^{\mathbf{ij}}\epsilon_{\mathbf{j}} \; ,\label{gengrav}\\
0 &= D \epsilon^{\mathbf{i}} - 2 \gamma^c \gamma^{ab} \nabla_{a}v_{bc} 
\epsilon^{\mathbf{i}} - 2  \epsilon_{abcde} v^{bc} v^{de} \gamma^a \epsilon^{\mathbf{i}} + 
\tfrac{4}{3} (v \cdot \gamma)^2 \epsilon^{\mathbf{i}} 
-\gamma^{ab}V_{ab}^{\mathbf{ij}}\epsilon_{\mathbf{j}}\nonumber\\
&-\tfrac{2}{3}\slashed{v}(\slashed{P}+N)L^{\mathbf{ij}}\epsilon_{\mathbf{j}}+\tfrac{4}{3}
\slashed{v}\gamma^a{V'}_a^{\mathbf{ij}}\epsilon_{\mathbf{j}} \ ,\label{genaux}\\
0&= - \tfrac{1}{4} F^I_{ab} \gamma^{ab} \epsilon^{\mathbf{i}} 
-\tfrac{1}{2} \gamma^\mu \partial_\mu M^I \epsilon^{\mathbf{i}} - Y^{I\mathbf{ij}}\epsilon_{\mathbf{j}}
-M^I\tfrac{1}{3} \slashed{v} \epsilon^{\mathbf{i}} +\tfrac{M^I}{6}(\slashed{P}+N)L^{\mathbf{ij}}\epsilon_{\mathbf{j}} 
-\tfrac{M^I}{3}\gamma^a{V'}_a^{\mathbf{ij}}\epsilon_{\mathbf{j}} \; .\label{gengaugino}
\end{align}
Following exactly the logic of \cite{Meessen:2007ef} we first consider the gaugino equation \eqref{gengaugino} and impose maximal supersymmetry.
Asumming that not all of the $M^I$ vanish we find
\begin{equation}
 F^{I}+\tfrac{4}{3}M^Iv=0 \ , \qquad Y^I=\tfrac{1}{6}M^IN \ ,\qquad {Y'}^{I\mathbf{ij}}={V'}_a^{\mathbf{ij}}=P_a=\partial_a M^I =0 \ , \label{maxgengaugino}
\end{equation}
whilst from the auxilary fermion equation we further obtain
\begin{equation}
 D=\tfrac{8}{3}v^2 \qquad dv=0 \qquad \nabla_bv^{ba}=\tfrac{1}{3}\epsilon^{abcde}v_{bc}v_{de} \quad \partial_{[a}V_{b]}=-\tfrac{1}{3}Nv_{ab} \ . \label{maxgenaux}
\end{equation}
The gravitino equation then resembles the Killing spinor equation of the ($U(1)$) \emph{gauged} theory. 

To proceed we consider the integrability condition of the gravitino Killing spinor equation, the scalar part of which yields $\partial_{[a}V_{b]}=0$ so $Nv_{ab}=0$ from \eqref{maxgenaux}. In the case $v=0$ the flux vanishes, and we obtain that $N$ is constant from the part of the integrability condition with one gamma matrix,
whilst from the part with two gamma matrices we obtain 
\begin{equation}
 R_{abcd}= -\tfrac{N^2}{9}(\eta_{a [c}\eta_{d]b}) \ ,
\end{equation}
so we have AdS$_5$ in the case of non-vanishing $N$ with radius $l=\tfrac{3\sqrt2}{N}$ and $Y^I=\tfrac{3\sqrt2}{l} M^I$ is constant. In the case that $N$ also vanishes the geometry is Minkowski space.
Substituting this information into the gravitino Killing spinor equation, we find that for both AdS$_5$ and Minkowski space that $V_\mu$ vanishes.

If, on the other hand, we assume $v_{ab}$ is non-zero, then $N$ vanishes. The integrability condition then reduces to that of the ungauged minimal theory, and in particular does not involve $V_\mu$.
This integrability condition was solved in \cite{Gauntlett:2002nw}, and leads to the maximally supersymmetric solutions of the ungauged theory. This then implies $V_\mu$ vanishes upon substitution into the gravitino equation.

If all of the $M^I$ vanish we find that $N=P_a=V^{\mathbf{ij}}_{\mu}=Y^{I\mathbf{ij}}=F^I_{ab}=0$. The solution of the Killing spinor equations yields exactly the maximally supersymmetric configurations of the minimal ungauged theory, with the two-form $v$, which is closed, playing the role of the gravi-photon field strength.

Turning to the Killing spinor identities we find from the gaugino KSI \eqref{gauginoksigen}
\begin{equation}
 \nabla \mathcal{E}(Y)_I^{\mathbf{ij}} = v\mathcal{E}(Y)_I^{\mathbf{ij}}=\mathcal{E}(A)^\mu_I=\mathcal{E}(M)_I=0 \ ,
\end{equation}
whilst from the auxiliary fermion KSI we obtain \eqref{auxfermksigen}
\begin{eqnarray}
\nabla\mathcal{E}(D)=\mathcal{E}(V)^{\mathbf{ij}}_\mu=0 \ ,&&\qquad M^I\mathcal{E}(Y')^{\mathbf{ij}}_I=0 \ , \nonumber\\
 \tfrac{1}{4}\mathcal{E}(v) +\mathcal{E}(D)v=0 \ ,&& \qquad \mathcal{E}(N)=\tfrac{1}{2}M^I\mathcal{E}(Y)_I \ ,  
\end{eqnarray}
and the gravitino Killing spinor identity tells us, at least, that the Einstein equation is automatic as long as we solve the other equations of motion.
Notice that we have not yet mentioned the equation of motion  for $P_\mu$.
This is because its variation does not involve the gaugino or the auxiliary fermion, and so information about its equation of motion may only come from the gravitino KSI.
In order to avoid working with the full gravitino KSI, we make the observation that in any case we need only solve the equations of motion of $D$, $P_\mu$ and $Y^{I\mathbf{ij}}$ as the others are then automatic from the proceeding discussion.
The vielbien equation of motion enters the gravitino KSI only with one gamma matrix so further information may be obtained from the scalar and two-form part of the gravitino KSI, ignoring the contributions from the other equations of motion.  
First note that the variation of $Y^{I\mathbf{ij}}$ does not contain the gravitino, so $\mathcal{E}(Y)^{I\mathbf{ij}}$ will not appear in the gravitino KSI. So we must solve this equation of motion iff $v$ vanishes, and this then implies the equation of motion of $N$ is satisfied.
In particular we must solve it in the cases of Minowski space or AdS$_5$.

Furthermore we shall choose to solve the $D$ equation of motion, and so may ignore this contribution to the KSIs, since we know from experience the $D$ equation is not automatic even in the two derivative theory, and this implies the equation of motion of $v$ is satisfied.
The relevant terms in the variation of $P_\mu$ are given by	
\begin{equation}
 \delta P^a =   2i\bar{\epsilon}_{\mathbf{i}}\gamma^{ab}( \tfrac{N}{2}\psi_b^{\mathbf{i}} 
+ 2(\gamma\cdot v)L^{\mathbf{ij}}\psi_{b\mathbf{j}} + 6L^{\mathbf{ij}}\phi_{b\mathbf{j}}) \ ,
\end{equation}
where
\begin{equation}
 \gamma^{ab}\phi^{\mathbf{i}}_{b}= \tfrac{1}{4}v^{ab}\psi^{\mathbf{i}}_b +\tfrac{1}{4}v_{cd}\gamma^{abcd}\psi^{\mathbf{i}}_b 
 -\tfrac{1}{6}v_{bc}\gamma^{bc}\psi^{a\mathbf{i}} - \tfrac{7}{6}{v^b}_c\gamma^{ac}\psi_b^{\mathbf{i}} 
 -\tfrac{1}{3}\gamma^{abc}\nabla_b\psi^{\mathbf{i}}_c  \ .
\end{equation}
We find
\begin{eqnarray}
 \tfrac{i}{2}\delta P^a &=&   \bar{\epsilon}^{\mathbf{i}}\left(\epsilon_{\mathbf{ji}} \tfrac{N}{2}\gamma^{ab} 
 +L_{\mathbf{ij}}\left( v_{cd}\gamma^{cd}\eta^{ab} - 4v^{ab} 
 +\tfrac{7}{2} v_{cd}\gamma^{abcd} +4{v^a}_c\gamma^{bc} 
 +3 {v^b}_c\gamma^{ac}\right)   \right)\psi^{\mathbf{j}}_b \nonumber\\
 &&-2\bar{\epsilon}^{\mathbf{i}}\gamma^{abc}\nabla_c\psi_b^{\mathbf{j}}L_{\mathbf{ij}} \ .
\end{eqnarray}
Integrating by parts, and using that we have
\begin{eqnarray}
  \gamma^{abc}\nabla_c \epsilon^{\mathbf{i}} &=&- \tfrac{1}{2} \gamma^{abc}\gamma_{c de} 
v^{de} \epsilon^{\mathbf{i}} + \tfrac{1}{3} \gamma^{abc}\gamma_c \gamma_{de} v^{de} \epsilon^{\mathbf{i}} -  
\tfrac{1}{6}\gamma^{abc}\gamma_c N L^{\mathbf{ij}}\epsilon_{\mathbf{j}} \nonumber\\
&=&(v^{ab} + \tfrac{1}{2} \gamma^{abcd}v_{cd})\epsilon^{\mathbf{i}} -\tfrac{N}{2}\gamma^{ab}L^{\mathbf{ij}}\epsilon_{\mathbf{j}} \ .
\end{eqnarray}
The part of the gravitino KSI without gamma matrices thus yields
\begin{equation}
 v_{ab}\mathcal{E}(P)^b=0 \ . \label{0ggravksi}
\end{equation}
From the part with one gamma matrix we obtain
\begin{equation}
 \mathcal{E}(P)\wedge v =0 \ .\label{EPwedgev}
\end{equation}
Note that this means that as long as we solve the non-trivial equation of motion of $D$, we do not have to solve the equation of motion for $P_{a}$
in order for the Einstein equation to be automatic for the maximally supersymmetric solutions, due to the appearance of $L^{\mathbf{ij}}$ in the relevant term of the Killing spinor identity.

Using this in the part with two gamma matrices we obtain
\begin{equation}
 N\mathcal{E}(P)_a = 0 \ , \qquad d\mathcal{E}(P)=0 \ , \qquad v_{cd}\mathcal{E}(P)_b=3 v_{b[c}\mathcal{E}(P)_{d]}\ . \label{2ggravksi}
\end{equation}
Clearly in Minowski space, where $N=v=0$ we must therefore solve the equation of motion for $P$, however we know that $d\mathcal{E}(P)=0$.
In AdS$_5$ the $P_a$ equation of motion is automatic, whilst in the case of the maximally supersymmetric solutions of the ungauged theory with flux comparing \eqref{EPwedgev} and the last equation of \eqref{2ggravksi}, we find that if $v_{ab}$ is non-vanishing then the equation of motion for $P_a$ is automatic.

In the case that all of the $M^I$ vanish, the Killing spinor identities imply that the equations of motion that remain to be solved are those of $D$, and also $Y^{I\mathbf{ij}}$ in the case that $v$ vanishes. Therefore the maximally supersymmetric configurations of the ungauged minimal supergravity are maximally supersymmetric configurations also in the case of $M^I$ all vanishing (with $F^I_{ab}=0$ but $v\neq0$),
whilst AdS$_5$ is not as in this case $N$ vanishes. Note that this may not occur in the two derivative case, as the equation of motion of $D$ is inconsistent at this level.

In summary, in the cases that $v$ vanishes we have Minkowski space or AdS$_5$. When $N$ vanishes we obtain Minkowski space and we must solve the equation of motion of $D$, $P_a$ and that of $Y^{I\mathbf{ij}}$, whilst for non-vanishing $N$ we obtain AdS$_5$ and only need solve the equation of motion for $D$ and $Y^{I\mathbf{ij}}$.
It is instructive to consider how this works in the two derivative case, with and without a cosomological constant. Consider the two derivative density of \eqref{2derivaction} in addition to the (bosonic part of) the cosmological constant density given by using the physical vector multiplets and the compensating linear multiplet directly in \eqref{actionprinciple},
\begin{equation}
\left. \mathcal{L}(\mathbf{L} \cdot \mathbf{V}) \right|_\text{bosonic} = g_I\left( Y^{I\mathbf{ij}} 
\cdot L_{\mathbf{ij}} - \tfrac{1}{2} A^I_a \cdot P^a + \tfrac{1}{2} M^I \cdot N\right) \; , \label{cosconstant}
\end{equation}
where we allow $g_I$ also to vanish, allowing us to consider the $U(1)$ gauged and ungauged cases together.
Now AdS$_5$ is a solution if and only if $N$ is non-zero, and $N$ must be constant and is inversely proportional to the AdS radius. In the two derivative case we have
$\mathcal{N}=1$ the very special geometry condition from the $D$ equation of motion and from the $Y^{I\mathbf{ij}}$ equation of motion we obtain $g_I=\mathcal{N}_{IJ}Y^{J}=\tfrac{6\sqrt{2}}{l}\mathcal{N}_I$ which contracting with $M^I$ implies $l=\tfrac{18\sqrt{2}}{g_IM^I}$ directly relating the coupling of the cosmological constant density to the AdS radius, and clearly in this case we must have $g_IM^I\neq0$.
In the general case of an arbitrary supersymmetric action, however, $g_I$ may be zero and we still have this solution, but the gauging will be higher derivative and the theory may contain ghosts.
In the case of Minkowski space in the two derivative case we have the very special geometry condition from the $D$ equation of motion, and $g_I=0$ from the $Y^I$ equation of motion and $g_IA^I_\mu =0$ from the $P_\mu$ equation of motion, so as expected we only have Minkowski space if we do not couple to the cosmological constant density at two derivative level.
In the general case however it is possible that there are Minkowski space solutions in theories which have non-zero coupling to the cosmological constant, if there is a suitable cancellation in the equations of motion.

In the case that the field $v$ and hence the flux does not vanish, it is clear that the only remaining equation to solve is that of $D$. However we immediately run into a contradiction. Examining the equations of motion for $P_a$ and $Y^I$
in the two derivative case we obtain $g_IA^I_\mu =0$ and $g_I=0$, but this contradicts the assumption that $v_{ab}$ is non-zero unless $g_I$ vanishes, so again these are only maximally supersymmetric solutions in the ungauged theory. In the general case however these may also be solutions whether or not the 
cosmological constant is included, but only if these contributions to the equations of motion are cancelled. This may be impossible given that the invariants that may be used to construct such a cancellation must be higher (than zero) derivative invariants. This leads us to question under what assumptions the Killing spinor identities are valid.
We should note that the Killing spinor identities for off-shell theories are a consequence of supersymmetry alone, and so they for hold for each supersymmetric density taken in isolation. However the equations of motion of $Y^I$ and $P_a$ for the cosmological constant density (with non-zero coupling) are singular in the sense that they imply $\det{e}=0$ when taken in isolation, and so the full equations need to be checked.
In particular if we include densities which have singular equations of motion individually, we must check each of these equations of motion, as the Killing spinor identities are no longer valid for them. The task is considerably simplified by noting that for any densities which do not have singular equations of motion taken in isolation, the Killing spinor identities hold, and the contributions from such invariants vanish.
In fact this also occurs with the equation of motion for $D$, which is why we have to introduce the compensator in the first place at two derivative level, but we have avoided this subtlety by choosing to always solve this equation. 
In all cases the corrections to the very special geometry condition will be constant, as will corrections to the effective cosmological constant. In the case of Minkowski space we also have that $d\mathcal{E}(P)=0$.
In particular we find that invariants with singular equations of motion, as defined above, play an important role in whether the maximally supersymmetric solutions of the theory are those of the gauged or ungauged two derivative theories.

\section{Conclusions}
In this paper we reexamined the supersymmetric solutions of higher derivative minimally supersymmetric five dimensional supergravity. In particular we have shown the power of the 
Killing spinor identities in analysing these solutions in the presence of higher derivative corrections,
particularly when combined with the spinorial geometry techniques. We have shown, as expected from string theory, that the Ricci scalar squared invariant does not affect the 
supersymmetric solutions of the ungauged theory at order $\alpha'$, as the corrections to the equations of motion for the supersymmetric solutions are trivial at this order.
This was quite easy to see from the form of the contributions to the equations of motion coming from this invariant, but was simplfied by using the Killing spinor identities. In fact, using the Killing spinor identities, we did not
even have to solve the Killing spinor equations to conclude this.

We reexamined the geometry of the time-like class of solutions, and were able to give compact expressions for the full equations of motion, without any simplifying assumptions,
complementing the analysis of \cite{Castro:2008ne}. We then examined the maximally supersymmetric solutions in the time-like class, streamlining the derivation
to avoid the additional solutions of \cite{Gauntlett:2002nw} which were later shown to be isometric to the near-horizon geometry of the BMPV black hole \cite{Fiol:2003yq}.
We then went on to show that the maximally supersymmetric solutions are unchanged apart from a constant deformation of the very special geometry condition and 
the cosmological constant, generalizing the work of Meessen \cite{Meessen:2007ef} to the case of an off-shell compensating multiplet. We found that the equation of motion of the auxilary field $P_\mu$ is automatic, with the exception of the Minkowski space solution.
However we also found that it was necessary to consider this equation of motion, as it leads, at two derivative level, to the fact that the solutions with flux of the ungauged two derivative theory, cannot be maximally supersymmetric solutions when we couple to the cosmological constant density.
In fact, as the Killing spinor identities are valid for any supersymmetric density with non-singular equations of motion (i.e. those which do not imply $det(e)=0$ for non-zero coupling when taken in isolation), we may quickly analyze the equations of motion of each invariant individually, to see if they present terms
which will exclude some of the solutions, if they are not cancelled by contributions from other densities. Note that this implies that there must be constraints on the couplings
of densities with singular equations of motion in order to achieve the desired cancellation for any particular maximally supersymmetric configuration to solve the equations of motion of the particular theory.
We note that the usual gauged or ungauged two derivative theories are given by a linear combination of such invariants, the zero derivative cosmological constant density, and the two derivative densities formed from the vector multiplets and the compensating multiplet. The former has singular equations of motion for $Y^{I\mathbf{ij}}$, whilst the latter two have singular equations of motion for $D$.
Indeed it is well known that it is necessary to take the latter two densities to both have non-zero couplings so that the $D$ equation is consistent. 

Whilst our analysis does not lead to new maximally supersymmetric solutions (apart from AdS$_5$, as off-shell there is no difference between the Abelian gauged and ungauged theories, and the possibility of the usual ungauged solutions, but with vanishing scalars, $M^I$ and $v$ playing the role of the gravi-photon field strength), 
the remaining equations of motion may lead to constraints, restricting the known geometries. Whilst this has no effect at leading order for the invariants we have considered one would expect this to become important 
at some finite order,
or for supergravities for which the higher derivative densities are not perturbative corrections to the two derivative action, at least in the case of invariants with singular equations of motion.
When considering higher derivative corrections from
string theory, the choice of effective Lagrangian, i.e. the choice of the couplings of the different invariant densities, may still have a dramatic effect on the supersymmetric spectrum,
the non-vanishing of ${V'}_\mu^{\mathbf{ij}}$ for example leading to solutions that only preserve one out of the eight supersymmetries. In the time-like case this leads to solutions for which the
complex structures on the base are not closed, but are instead parameterized by $V^{\mathbf{ij}}_\mu$ which vanishes to leading order in the ungauged case\footnote{In the U(1) gauged case, $V_\mu$ is non-zero at leading order whilst ${V'}_\mu^{\mathbf{ij}}=0$ at the two derivative level.}.

It would be particularly interesting to study the Ricci tensor squared invariant (or equivalently the Riemman tensor squared invariant),
that was constructed in superspace in \cite{Butter:2014xxa}, but has yet to appear in components, along with the $F^4$ and off-diagonal invariants constructed in \cite{Ozkan:2016csy}.
One wonders whether it is possible to choose the couplings of the invariants by field redefinitions allowed by string theory in higher dimensions, such that
the supersymmetric solutions are those of the truncated theory. 
In \cite{Coomans:2012cf} the off-shell version of the alternative supergravity of Nishino and Rajpoot \cite{Nishino:2000cz,Nishino:2001ji} with one vector multiplet was constructed,
and was extended to arbitrary number of Abelian vector multiplets in \cite{Sloane:2014xya}. Interestingly in these theories, which are constructed in the dilaton-Weyl multiplet, the 
Riemman tensor squared invariant is known in component form \cite{Bergshoeff:2011xn}\footnote{The Riemman tensor squared invariant is given in a particular gauge in \cite{Bergshoeff:2011xn,Ozkan:2013uk,Ozkan:2013nwa}, 
and it would be useful to have the explicit expression after the reversal of this gauge fixing.},
and can be added to the Weyl-squared invariant, resulting in the Gauss-Bonnet invariant \cite{Ozkan:2013uk}, which was generalized to an arbitrary number of Abelian vector multiplets in \cite{Ozkan:2013nwa}.
It turns out that for the particular case of Gauss-Bonnet the auxiliary fields $N$ and $P_a$  may be eliminated by their equations of motion in the absence of the cosmological constant invariant.
If this is again the case for the standard Weyl multiplet, and if the field $V_\mu^{\mathbf{ij}}$ can be treated in a similar way, then the off-shell supersymmetric spectrum will be the same as the truncated case discussed in \cite{Castro:2008ne} and in this work.
If this is not the case, the same effect would also occur if the coupling of the Ricci tensor squared invariant may be choosen to produce equations of motion for the auxiliary 
fields that only have $P_a=N={V}_\mu^{\mathbf{ij}}=Y^{I\mathbf{ij}}=0$ as solutions, in which case the Ricci scalar squared invariant would not affect the other equations of motion for the supersymmetric solutions, as we have discussed above.
In recent work \cite{Cano:2018qev} string theory corrections in the effective five dimensional theory coming from the Heterotic theory have been analysed, and it would be interesting to perform the same general analysis presented here, using the off-shell theory decribed in \cite{Sloane:2014xya} and references therein.

The gauged theory has been discussed before, in \cite{Cremonini:2008tw} black holes in the order $\alpha'$ U(1) gauged theory were discussed by integrating out the auxiliary fields after
the inclusion of the Weyl tensor squared invariant, whilst in \cite{Baggio:2014hua}, some supersymmetric solutions of the $U(1)$ gauged theory coupled to an abritary number of on-shell 
hypermultiplets were discussed in the presence of the Weyl squared and Ricci squared invariants. Clearly an off-shell classification of the supersymmetric solutions of the $U(1)$ gauged case would be desirable,
particularly in holographic applications, however a fuller understanding of the freedom to choose the couplings in the invariants in that case would also be useful, as the supersymmetric spectrum in the general
case is much more complicated, and in particular when $V_\mu^{\mathbf{ij}}$ does not vanish there may exist solutions that preserve only one of the eight supersymmetries, but this could be avoided by choosing a 
particular field redefinition allowing for an effective theory with supersymmetric solutions more similar to the two derivative case. 

\acknowledgments
The authors would like to thank Jan Gutowski for his collaboration in the early stages of this work. This work was partially supported by the INFN. The work of WS is supported in part by the National Science Foundation under grant number PHY-1620505. P.S. would like to acknowledge the support 
of the Consejo Nacional de Ciencia y Tecnologia (Conacyt), Mexico, the partial support of  
FONDECYT Postdoctorado Project number 3130541, CONICYT, Chile and  the 
Mesoamerican Centre for Theoretical Physics (MCTP), the Galileo Galilei Institute for Theoretical Physics (GGI) and the Mainz Institute for Theoretical Physics (MITP) for hospitality whilst this 
work was completed.

\appendix
\section{Action and equations of motion}\label{constructaction}

We shall briefly review the off-shell superconformal construction of two 
derivative, Weyl tensor squared and Ricci tensor squared supersymmetric action 
with arbitrarily many Abelian vector multiplets in the standard Weyl multiplet 
\cite{Fujita:2001kv, Hanaki:2006pj}. Our starting point is the rigid 
exceptional superalgebra $F(4)$, generated by
\begin{equation}
\mathbf{P}_a ,\; \mathbf{M}_{ab}, \; \mathbf{D}, \; \mathbf{K}_a, \; 
\mathbf{U}_{\mathbf{ij}}, \; \mathbf{Q}_{\mathbf{i}} , \; \mathbf{S}_{\mathbf{i}}
\end{equation}
where $a,b,...$ are flat Lorentz indices, $\mathbf{i},\mathbf{j},...$ are SU(2) indices, 
$\mathbf{Q}_{\mathbf{i}}$ and $\mathbf{S}_{\mathbf{i}}$ are symplectic-Majorana spinors in the 
fundamental of SU(2). We raise and lower the SU$(2)$ indices using the antsymnmetric tensor $\epsilon_{\mathbf{ij}}$ where $\epsilon_{\mathbf{12}}=\epsilon^\mathbf{12}=1$. 
We will also make use of the (NW)-(SE) convention so that for example $\bar{\chi}\chi=\bar{\chi}^{\mathbf{i}}\chi_{\mathbf{i}}=\bar{\chi}^{\mathbf{i}}\chi^{\mathbf{j}}\epsilon_{\mathbf{ji}}$. 
The geometrical interpretation of the generators is as 
follows:
\begin{itemize}
\item $\mathbf{P}_a$: spacetime translation
\item $\mathbf{M}_{ab}$: Lorentz transformation
\item $\mathbf{D}$: dilatation
\item $\mathbf{K}_a$: special conformal transformation
\item $\mathbf{U}_{\mathbf{ij}}$: internal SU(2) transformation
\item $\mathbf{Q}_{\mathbf{i}}$: Poincar\'e supersymmetry transformation
\item $\mathbf{S}_{\mathbf{i}}$: conformal supersymmetry transformation.
\end{itemize}

In order to upgrade to the local theory, a gauge field is introduced for each of the 
generators; we have respectively
\begin{equation}
e^a_\mu, \; \omega_{\mu}^{\phantom{\mu}ab}, \; b_\mu, \; f^a_\mu, \; 
V^{\mathbf{ij}}_\mu, \; \psi^{\mathbf{i}}_\mu, \; \phi^{\mathbf{i}}_\mu \; .
\end{equation}
Conventional constraints in this case are taken to be
\begin{equation}
 \hat{R}(P)^a_{\phantom{a}\mu\nu} = 0 \ ,\qquad
 \gamma^\mu \hat{R}(Q)^{\mathbf{i}}_{\mu\nu} = 0 \ ,\qquad
 e_b^\nu\hat{R}(M)^{ab}_{\phantom{ab}\mu\nu} = 0 \ ,
\end{equation}
which make $\omega_{\mu}^{\phantom{\mu}ab}, f^a_\mu$ and, $\phi^{\mathbf{i}}_\mu$ into 
composite fields. As discussed in \cite{Fujita:2001kv} these constraints are 
avoidable, however in the following we will use them to simplify the derivation.
Covariant derivatives $\hat{\mathcal{D}}$ and $\mathcal{D}$ are defined as
\begin{align}
\hat{\mathcal{D}}_\mu &:= \partial_\mu - \sum_{\mathbf{X}_A = \mathbf{M}_{ab}, 
\mathbf{D}, \mathbf{U}_{\mathbf{ij}},\mathbf{K}_a, \mathbf{Q}_{\mathbf{i}}, \mathbf{S}_{\mathbf{i}}} h^A_\mu 
\mathbf{X}_A \ ,\nonumber\\
\mathcal{D}_\mu &:= \partial_\mu - \sum_{\mathbf{X}_A = \mathbf{M}_{ab}, 
\mathbf{D}, \mathbf{U}_{ij}} h^A_\mu \mathbf{X}_A \; .
\end{align}

Auxiliary fields have to be introduced as we can see counting bosonic and 
fermionic degrees of freedom. The total number of components of the bosonic 
gauge fields (not including the composite $\omega_{\mu}^{\phantom{\mu}ab}, 
f^a_\mu$) is
$
25 + 5 + 15 = 45
$,
 which must be reduced by the total number of bosonic generators (including 
$\mathbf{M}_{ab}, \mathbf{K}_a$)
$
5 + 10 + 1 + 5 + 3 = 24
$,
giving 21 degrees of freedom. On the fermionic side we have
$
40
$
components from the gravitino, and 
$
8 + 8 = 16 
$
real supercharges, hence 24 fermionic degrees of freedom. We can bring the 
number of both bosonic and fermionic degrees of freedom to 32 by adding a 
two-form, a scalar and an SU(2)-Majorana spinor
\begin{equation}
v_{ab}, \; D, \; \chi^{\mathbf{i}} \; .
\end{equation}
We thus obtain the standard-Weyl superconformal multiplet
\begin{equation}
e^a_\mu,\; b_\mu, \; V^{\mathbf{ij}}_\mu,\; v_{ab},\; D, \; \psi^{\mathbf{i}}_\mu,\; \chi^{\mathbf{i}} \; ;
\end{equation}
for which we record only transformation rules which will be useful for our 
discussion:
\begin{align}
\delta e^a_\mu &= -2 i \bar{\epsilon} \gamma^a \psi_\mu \ ,\nonumber \\
\delta v_{ab} &= -\tfrac{1}{8}i \bar{\epsilon} \gamma_{ab} \chi - \tfrac{3}{2} 
i \bar{\epsilon} \hat{R}(Q)_{ab}\ ,\nonumber\\
\delta D &= -i \bar{\epsilon} \gamma^a \hat{\mathcal{D}}_a \chi - 8 i v^{ab} 
\bar{\epsilon} \hat{R}(Q)_{ab} + i \bar{\eta} \chi \ ,\nonumber\\
\delta \psi^{\mathbf{i}}_\mu &= \mathcal{D}_\mu \epsilon^{\mathbf{i}} + \tfrac{1}{2} \gamma_{\mu 
ab}v^{ab} \epsilon^{\mathbf{i}} - \gamma_\mu \eta^{\mathbf{i}} \ ,\nonumber\\
\delta \chi^{\mathbf{i}} &= D \epsilon^{\mathbf{i}} -2 \gamma^c \gamma^{ab} \epsilon^{\mathbf{i}} 
\hat{\mathcal{D}}_a v_{bc} + \gamma^{\mu\nu} 
\hat{R}(U)^{\mathbf{ij}}_{\phantom{\mathbf{ij}}\mu\nu} \epsilon^{\mathbf{k}} \epsilon_{\mathbf{jk}} -2 \gamma^a 
\epsilon^{\mathbf{i}} \epsilon_{abcde} v^{bc} v^{de} + 4v^{ab} \gamma_{ab} \eta^{\mathbf{i}}\ , 
\nonumber\\
\delta V^{\mathbf{ij}}_\mu &= - 
6i\bar{\epsilon}^{\mathbf{(i}}\phi^{\mathbf{j)}}_\mu + 
4i\bar{\epsilon}^{\mathbf{(i}}\gamma\cdot v \psi^{\mathbf{j)}}_\mu 
-\tfrac{i}{4}\bar{\epsilon}^{\mathbf{(i}}\gamma_\mu\chi^{\mathbf{j)}} 
+6i\bar{\eta}^{\mathbf{(i}}\psi_\mu^{\mathbf{j)}}\ ,
\end{align}
where $\epsilon^{\mathbf{i}}, \eta^{\mathbf{i}}$ are infinitesimal parameters of $\mathbf{Q}_{\mathbf{i}}, 
\mathbf{S}_{\mathbf{i}}$ transformations respectively.

The explicit expressions
\begin{align}
\phi^{\mathbf{i}}_\mu &= \left( -\frac{1}{3}e^a_\mu \gamma^b + \frac{1}{24} \gamma_\mu 
\gamma^{ab} \right) \left.\hat{R}(Q)^{\mathbf{i}}_{ab} \right|_{\phi^{\mathbf{i}}_\mu=0} \ ,\nonumber\\
\hat{R}(Q)^{\mathbf{i}}_{\mu\nu} &= 2 \nabla_{[\mu} \psi^{\mathbf{i}}_{\nu]} + b_{[\mu} \psi^{\mathbf{i}}_{\nu]} 
-2 V^{\mathbf{ij}}_{[\mu} \psi^{\mathbf{k}}_{\nu]} \epsilon_{\mathbf{jk}} + v^{ab} \gamma_{ab[\mu} 
\psi_{\nu]}^{\mathbf{i}} - 2 \gamma_{[\mu} \phi^{\mathbf{i}}_{\nu]} \ ,\nonumber\\
\hat{R}(U)^{\mathbf{ij}}_{\phantom{\mathbf{ij}}\mu\nu} &= 2 \nabla_{[\mu} V^{\mathbf{ij}}_{\nu]} -2 
V_{[\mu| \mathbf{k}}^{\mathbf{i}} V^{\mathbf{kj}}_{\nu]} +12i \bar{\psi}^{(\mathbf{i}}_{[\mu} \phi^{\mathbf{j})}_{\nu]} - 4i 
v^{ab} \bar{\psi}^{(\mathbf{i}}_{[\mu} \gamma_{ab} \psi^{\mathbf{j})}_{\nu]} +\tfrac{1}{2}i 
\bar{\psi}^{(\mathbf{i}}_{[\mu} \gamma_{\nu]} \chi^{\mathbf{j})} \ ,\nonumber\\
\hat{\mathcal{D}}_\mu v_{ab} &= \nabla_\mu v_{ab} - b_\mu v_{ab} + \tfrac 
{1}{8} i \bar{\psi}_\mu \gamma_{ab} \chi + \tfrac{3}{2} i \bar{\psi}_\mu 
\hat{R}(Q)_{ab} \ ,\nonumber\\
\hat{\mathcal{D}}_\mu \chi^{\mathbf{i}} &= \mathcal{D}_\mu \chi^{\mathbf{i}} - D \psi^{\mathbf{i}}_\mu + 2 
\gamma^c \gamma^{ab} \psi^{\mathbf{i}}_\mu \hat{\mathcal{D}}_a v_{bc} - \gamma^{\nu\rho} 
\hat{R}(U)^{\mathbf{ij}}_{\phantom{\mathbf{ij}}\nu\rho} \psi^{\mathbf{k}}_\mu \epsilon_{\mathbf{jk}} \nonumber\\
&+ 2 \gamma^a \psi^{\mathbf{i}}_\mu \epsilon_{abcde} v^{bc} v^{de} - 4v^{ab} \gamma_{ab} 
\phi^{\mathbf{i}}_\mu \ ,
\end{align}
will also be needed during Poincar\'e gauge-fixing. $\nabla$ will always refer 
to the spin covariant derivative.

Abelian vector fields will be introduced by means of 
superconformal vector multiplets
\begin{equation}
A^I_\mu, \; M^I, \; \Omega^{I\mathbf{i}}, \; Y^I_{\mathbf{ij}}\ ,
\end{equation}
consisting of a 1-form, a scalar, an SU(2)-Majorana spinor and an auxiliary 
symmetric SU(2)-triplet of Lorentz scalars. These transform as
\begin{align}
\delta A^I_\mu & = -2i \bar{\epsilon} \gamma_\mu \Omega^I + 2i 
M^I\bar{\epsilon} \psi_\mu \ ,\nonumber\\
\delta M^I &= 2i \bar{\epsilon} \Omega^I \ , \nonumber\\
\delta Y^{I\mathbf{ij}} 
&=2i\bar{\epsilon}^{\mathbf{(i}}\gamma^a\hat{\mathcal{D}}_a\Omega^{\mathbf{j})I}
 - i\bar{\epsilon}^{\mathbf{(i}}\gamma\cdot 
v\Omega^{\mathbf{j})I}-\tfrac{i}{4}\bar{\epsilon}^{\mathbf{(i}}\chi^{\mathbf{j})}M^I
-2i\bar{\eta}^{\mathbf{(i}}\Omega^{\mathbf{j})I} \ ,\nonumber\\
\delta \Omega^{I\mathbf{i}} &= - \tfrac{1}{4} F^I_{ab} \gamma^{ab} \epsilon^{\mathbf{i}} - 
\tfrac{1}{2} \gamma^a \hat{\mathcal{D}}_a M^I \epsilon^{\mathbf{i}} + 
{Y^{I\mathbf{i}}}_{\mathbf{j}} \epsilon^{\mathbf{j}} - M^I \eta^{\mathbf{i}} 
\; .
\end{align}

We shall also introduce an off-shell linear multiplet as our compensator as was 
done in \cite{Coomans:2012cf, Ozkan:2013nwa}.\footnote{In \cite{Castro:2008ne, 
Meessen:2007ef} a hyper-multiplet was taken as compensator however to avoid 
subtleties arising from central charge and constraints for the closure of the 
superconformal algebra off-shell we shall instead use a linear multiplet. One 
can easily map to a hypermultiplet compensator and due to the gauge fixing this 
seems to change very little. In the component formalism that we adopt it is 
only known how to take a single hypermultiplet off-shell without resorting to 
an infinite number of auxiliary fields. To our knowledge this was first done in 
the superconformal formalism in \cite{deWit:1980gt}. In superspace however an 
off-shell formalism for general hypermultiplets is known \cite{Kuzenko:2007hu, 
Kuzenko:2008wr, Kuzenko:2014eqa}, and is discussed at length in the 
interesting papers \cite{Butter:2014xua, Butter:2014xxa}.}
The linear multiplet is also a key ingredient for finding supersymmetric actions
\begin{equation}
L^{\mathbf{ij}}, \; \varphi^{\mathbf{i}}, \; E_a,\; N \; 
\end{equation}
and consists of a $\mathrm{SU}(2)$-symmetric real scalar, an SU(2)-Majorana 
spinor, a vector, and a scalar. The importance of linear multiplets can be 
understood by looking at the supersymmetry transformation of $L^{\mathbf{ij}}$, which 
reads
\begin{equation}
\delta L^{\mathbf{ij}} = 2i \bar{\epsilon}^{(\mathbf{i}} \varphi^{\mathbf{j})} \; .
\end{equation}
Note the invariance under $\mathbf{S}_{\mathbf{i}}$ supersymmetry. Suppose we have a 
composite real symmetric bosonic field which is $\mathbf{S}_{\mathbf{i}}$-invariant, and 
let us denote it $L^{\mathbf{ij}}$: its supersymmetry transformation must be of the form 
$2i \bar{\epsilon}^{(\mathbf{i}} \phi^{\mathbf{j})}$ for some suitable fermion $\phi^{\mathbf{i}}$. We therefore 
have found the first two elements of a linear multiplet. In order to close the 
multiplet one has to look at $\phi^{\mathbf{i}}$ supersymmetry transformation, on the 
right hand side of which one can read off $E_a, N$. This procedure can be used to 
embed Weyl and vector multiplets into a linear multiplet. The remaining 
tranformation rules under supersymmetry and special supersymmetry read
\begin{align}
\delta \varphi^{\mathbf{i}} &= 
-\gamma^a\hat{\mathcal{D}}_aL^{\mathbf{ij}}\epsilon_{\mathbf{j}} + 
\tfrac{1}{2}\gamma^aE_a\epsilon^{\mathbf{i}}
+\tfrac{N}{2}\epsilon^{\mathbf{i}} + 2(\gamma \cdot 
v)L^{\mathbf{ij}}\epsilon_{\mathbf{j}} -6L^{\mathbf{ij}}\eta_{\mathbf{j}}\ , 
\nonumber\\
\delta E^a &= 2i\bar{\epsilon}\gamma^{ab}\hat{\mathcal{D}}_b\varphi 
-2i\bar{\epsilon}\gamma^{abc}v_{bc}\varphi + 
6i\bar{\epsilon}\gamma_bv^{ab}\varphi -8i\bar{\eta}\gamma^a \varphi \ ,\nonumber\\
\delta N &= -2i\bar{\epsilon}\gamma^a\hat{\mathcal{D}}_a\varphi 
-3i\bar{\epsilon}(\gamma\cdot v)\varphi 
+\tfrac{i}{2}\bar{\epsilon}^{\mathbf{i}}\chi^{\mathbf{j}}L_{\mathbf{ij}} - 
6i\bar{\eta}\varphi\ .
\end{align}
where
\begin{align}
 \hat{\mathcal{D}}_\mu L^{\mathbf{ij}}&= \partial_\mu L^{\mathbf{ij}} - 3b_\mu L^{\mathbf{ij}}- 2V^{\mathbf{(i}}_{\mu\phantom{\mathbf{i}}\mathbf{k}}L^{\mathbf{j)k}}- 2i\bar{\psi}_{\mu}^{(\mathbf{i}}\varphi^{\mathbf{j})}\nonumber\\
\hat{\mathcal{D}}_\mu \varphi^{\mathbf{i}}&= \mathcal{D}_\mu \varphi^{\mathbf{i}} - i\slashed{\mathcal{D}}L^{\mathbf{ij}}\psi_{\mu\mathbf{j}} - \tfrac{1}{2}(\slashed{E}+N)\psi_\mu^{\mathbf{i}} 
- 2(\gamma\cdot v)L^{\mathbf{ij}}\psi_{\mu\mathbf{j}} - 6L^{\mathbf{ij}}\phi_{\mu\mathbf{j}} \nonumber\\
\mathcal{D}_\mu \varphi^{\mathbf{i}}&= \nabla_\mu \varphi^{\mathbf{i}} -\tfrac{7}{2}b_\mu\varphi^{\mathbf{i}} +V^{\mathbf{ij}}_{\mu}\varphi_{\mathbf{j}}
\end{align}

\subsection{Superconformal action}\label{superconformaltheory}

The starting point of determination of supersymmetric actions is the 
construction of a supersymmetric Lagrangian (up to surface terms) out of a given 
linear and vector multiplet:
\begin{align}
\mathcal{L}(\mathbf{L} \cdot \mathbf{V}) &= Y^{\mathbf{ij}} \cdot L_{\mathbf{ij}} + 2i 
\bar{\Omega} \cdot \phi + 2i \overline{\psi}_{\mathbf{i}}^a \gamma_a \Omega_{\mathbf{j}} \cdot L^{\mathbf{ij}} 
\nonumber\\
&- \tfrac{1}{2} A_a \cdot \left( E^a - 2i \bar{\psi}_b \gamma^{ba} \phi +2i 
\bar{\psi}^{(\mathbf{i}|}_b \gamma^{abc} \psi^{|\mathbf{j})}_c L_{\mathbf{ij}} \right) \nonumber \\
&+ \tfrac{1}{2} M \cdot \left( N -2i \bar{\psi}_a \gamma^a \phi - 2i 
\bar{\psi}^{(\mathbf{i}|}_a \gamma^{ab} \psi^{|\mathbf{j})}_b L_{\mathbf{ij}} \right) \; . \label{sca}
\end{align} 
In this equation we adopt the notation
\begin{equation}
Z \cdot \left( \dots \right):= Z^I T_I \left( \dots \right)\ ,
\end{equation}
where $Z$ stands for a member of vector multiplet and $T_I$ are $U(1)^{n_V+1}$ 
generators.
Truncating fermions we have 
\begin{equation}
\left. \mathcal{L}(\mathbf{L} \cdot \mathbf{V}) \right|_\text{bosonic} = Y^{\mathbf{ij}} 
\cdot L_{\mathbf{ij}} - \tfrac{1}{2} A_a \cdot E^a + \tfrac{1}{2} M \cdot N \; . \label{actionprinciple}
\end{equation}

All terms in the supersymmetric action we are going to study are of this form. 
They differ because of the different composition of the linear multiplet and 
vector multiplet. In particular, in addition to a vector-linear coupling, we 
will consider the following compositions
\begin{itemize}
\item Linear multiplet composed of two vector multiplets, 
$\mathbf{L}[\mathbf{V}, \mathbf{V}]$. This composition is well known and is 
given in \cite{Fujita:2001kv, Kugo:2000hn}.
The resulting Lagrangian turns out to be totally symmetric in the three vector 
multiplets and is given by 
\begin{align}
\mathcal{L}_V &= -Y^{\mathbf{ij}} \cdot L_{\mathbf{ij}}[\mathbf{V}, \mathbf{V}] + \tfrac{1}{2} 
A_a \cdot E^a[\mathbf{V}, \mathbf{V}] - \tfrac{1}{2} M \cdot N[\mathbf{V}, 
\mathbf{V}] \nonumber\\
&= \mathcal{N} \left( \tfrac{1}{2}D - \tfrac{1}{4} R + 3 v^2 \right) + 2 
\mathcal{N}_I v^{ab} F^I_{ab} + \tfrac{1}{4} \mathcal{N}_{IJ} F^{I}_{ab}F^{Jab} 
\nonumber\\
&- \mathcal{N}_{IJ} \left( \tfrac{1}{2} \mathcal{D}_a M^I \mathcal{D}^aM^J + 
Y^I_{\mathbf{ij}} Y^{J\mathbf{ij}} \right) + \tfrac{1}{24} e^{-1} \epsilon^{abcde} c_{IJK} A_a^I 
F_{bc}^J F_{de}^K \; .
\end{align}
where $v^2:=v_{ab}v^{ab}$ and
$\mathcal{N} = \tfrac{1}{6} c_{IJK} M^I M^J M^K$ is an arbitrary cubic function 
of the scalars, and subscripts $I,J,...$ denote partial derivatives with 
respect to $M^I$:
\begin{align}
\mathcal{N}_I &:= \frac{\partial}{\partial M^I} \mathcal{N} = 
\frac{1}{2}c_{IJK} M^J M^K \; ,&
\mathcal{N}_{IJ} &:= \frac{\partial}{\partial M^I} \frac{\partial}{\partial 
M^J} \mathcal{N} = c_{IJK}M^K \; .
\end{align}
\item Vector multiplet composed of a linear multiplet, which leads to a 
linear-linear action. Only the leading component of this composition was given 
in \cite{Fujita:2001kv}, but was given completely in 
\cite{Coomans:2012cf} in different conventions.\footnote{One can check this by 
using appendix B of \cite{Ozkan:2013nwa}, where we take an additional minus sign for all fields in the vector multiplet i.e. 
take $A_\mu=-{A'}_\mu$, $\Omega^{\mathbf{i}}=\tfrac{1}{2}\lambda^{\mathbf{i}}$, 
$Y^{\mathbf{ij}}={Y'}^{\mathbf{ij}}$ and $M=\rho$, since with this choice we arrive at the same first component 
of the embedding as in \cite{Fujita:2001kv}.} Defining 
$L=\sqrt{L_{\mathbf{ij}}L^{\mathbf{ij}}}$ in the current 
conventions\footnote{It is useful to note the SU(2) index identity 
$L_{\mathbf{ik}}{L^{\mathbf{k}}}_{\mathbf{j}}=\tfrac{1}{2}\epsilon_{\mathbf{ij}} 
L_{\mathbf{kl}}L^{\mathbf{kl}}$.} this reads
\begin{align}
 M &= L^{-1}N 
+iL^{-3}\bar{\varphi}^{\mathbf{i}}\varphi^{\mathbf{j}}L_{\mathbf{ij}} \ 
,\nonumber\\
 \Omega^{\mathbf{i}}&= -L^{-1}(\hat{\slashed{\mathcal{D}}} \varphi^{\mathbf{i}} 
+\tfrac{1}{2}(v\cdot\gamma)\varphi^{\mathbf{i}} 
 +\tfrac{1}{4}L^{\mathbf{ij}}\chi_{\mathbf{j}}) 
+L^{-3}((\hat{\slashed{\mathcal{D}}} 
L^{\mathbf{ij}})L_{\mathbf{jk}}\varphi^{\mathbf{k}} 
+\tfrac{1}{2}(N-\slashed{E}) L^{\mathbf{ij}}\varphi_{\mathbf{j}})\nonumber\\
&+iL^{-3}\varphi^{\mathbf{j}}\bar{\varphi}^{\mathbf{i}}\varphi_{\mathbf{j}} 
+3iL^{-5}L^{\mathbf{ij}}L^{\mathbf{kl}}\varphi_{\mathbf{j}}\bar{\varphi}_{
\mathbf{k}}\varphi_{\mathbf{l}} \ , \nonumber\\
 \hat{F}_{\mu\nu} &= 2\mathcal{D}_{[\mu}(L^{-1}E_{\nu]}) - 
2L^{-1}\hat{R}_{\mu\nu}^{\mathbf{ij}}(U)L_{\mathbf{ij}} + 
2L^{-3}L^{\mathbf{l}}_{\mathbf{k}}\mathcal{D}_{[\mu}L^{\mathbf{kp}}\mathcal{D}_{
\nu]}L_{\mathbf{lp}}\nonumber\\ 
&+2i\mathcal{D}_{[\mu}(L^{-3}\bar{\varphi}^{\mathbf{i}}\gamma_{\nu]}{\varphi}^{
\mathbf{j}}L_{\mathbf{ij}}) +iL^{-1}\bar{\varphi}\hat{R}_{\mu\nu}(Q) \ 
,\nonumber\\
 Y_{\mathbf{ij}}&= -L^{-1}(\Box^C L_{\mathbf{ij}} 
+\tfrac{1}{2}v^2L_{\mathbf{ij}} - 
\tfrac{D}{4}L_{\mathbf{ij}})+L^{-3}\mathcal{D}_aL_{\mathbf{k(i}}\mathcal{D}^aL_{
\mathbf{j)m}}L^{\mathbf{km}} 
  \nonumber\\
 &+ \tfrac{1}{4}L^{-3}(E^2-N^2)L_{\mathbf{ij}} 
 +L^{-3}E_a L_{\mathbf{k(i}}\mathcal{D}^a{L_{\mathbf{j)}}}^{\mathbf{k}} \ 
+\cdots \ ,
  \label{linearinvector}
\end{align}
where the first three expressions are given in their entirety, but we have not 
given fermion bilinear terms in the last expression\footnote{The first three 
expressions can be used along with the supersymmetry variations to reproduce 
these terms, and as we will gauge fix $\varphi^{\mathbf{i}}=0$, which appears 
at least once in all such terms, they will not contribute to our analysis.}. 
In order to use this embedding it is essential to note that for the closure of 
the algebra, the contraint $\mathcal{D}^aE_a$ is necessary. This constraint can 
of course be solved in terms of a three form
\begin{equation}
E^\mu = \tfrac{1}{12}\epsilon^{\mu\nu\rho\sigma\tau}\mathcal{D}_\nu 
E_{\rho\sigma\tau}\ ,
\end{equation}
which exhibits the gauge symmetry
\begin{equation}
\delta_{\Lambda^{(2)}} E_{\mu\nu\rho} = \partial_{[\mu}\Lambda^{(2)}_{\nu\rho ]}\ .
\end{equation}
Defining a two form $E_{\mu\nu}$ by
\begin{equation}
E^\mu=\mathcal{D}_\nu E^{\mu\nu} \ ,\qquad 
E_{\mu\nu\rho}=\epsilon_{\mu\nu\rho\sigma\tau}E^{\sigma\tau}\ ,
\end{equation}
we can rewrite the action formula \eqref{sca} by partial integration as
\begin{align}
\mathcal{L}_{VL} &= -Y^{\mathbf{ij}} \cdot L_{\mathbf{ij}} + \tfrac{1}{24} 
\epsilon^{\mu\nu\rho\sigma\tau} A_\mu  \partial_\nu E_{\rho\sigma\tau} - 
\tfrac{1}{2} M \cdot N  \ ,\nonumber\\
&=-Y^{\mathbf{ij}} \cdot L_{\mathbf{ij}} + \tfrac{1}{4} F_{\mu\nu} E^{\mu\nu} - \tfrac{1}{2} M 
\cdot N \ ,
\end{align}
which allows us to use the embedding \eqref{linearinvector} directly to 
obtain the linear-linear action, for which we record the bosonic part
 \begin{align}
  e^{-1}\mathcal{L}_L &= L^{-1}L_{\mathbf{ij}}\Box L^{\mathbf{ij}} 
  - L^{\mathbf{ij}}\mathcal{D}_{\mu}L_{\mathbf{k(i}}\mathcal{D}^\mu 
L_{\mathbf{j)m}}L^{\mathbf{km}} L^{-3} -N^2 L^{-1} \nonumber\\
  &- \tfrac{1}{4}P_\mu P^\mu L^{-1} +\tfrac{1}{2}L {v}^2 -\tfrac{1}{4}DL  
+\tfrac{1}{4}L^{-3}P^{\mu\nu}L^{\mathbf{l}}_{\mathbf{k}}\partial_\mu 
L^{\mathbf{kp}}\partial_\nu L_{\mathbf{pl}} \nonumber\\
  &+ \tfrac{1}{2}P^{\mu\nu}\partial_\mu(L^{-1} P_\nu 
+2V_\nu^{\mathbf{ij}}L_{\mathbf{ij}}L^{-1}) \ ,  \label{LAGLstandard}
 \end{align}
 where $L^2=L_{\mathbf{ij}}L^{\mathbf{ij}}$, $P^\mu$, $P^{\mu\nu}$ are the 
bosonic parts 
of 
$E^\mu$,$E^{\mu\nu}$ and the bosonic part of $L_{\mathbf{ij}}\Box 
L^{\mathbf{ij}}$ is given by
 \begin{align}
  L_{\mathbf{ij}}\Box L^{\mathbf{ij}} &= L_{\mathbf{ij}}(\partial^m +4b^m 
+{\omega_n}^{nm})\mathcal{D}_mL^{\mathbf{ij}} 
-2L_{\mathbf{ij}}V_{n\phantom{\mathbf{i}}\mathbf{k}}^{\mathbf{i}}\mathcal{D}^nL^
{\mathbf{jk}} 
  -\tfrac{3}{8}L^{2}R
  \ ,
 \end{align}
 and where the superconformal deriviative of $L^{\mathbf{ij}}$ is given by 
 \begin{equation}
  \hat{\mathcal{D}}_\mu L^{\mathbf{ij}}=(\partial_\mu-3b_\mu)L^{\mathbf{ij}} 
-2V^{\mathbf{(i}}_{\mu \; \mathbf{k}}L_{\phantom{\mu 
\mathbf{k}}}^{\mathbf{j)k}} - 
2i\bar{\psi}_\mu^{\mathbf{(i}}\varphi^{\mathbf{j)}}\ .
 \end{equation}

 We can also use the emdedding \eqref{linearinvector} in the vector multiplet 
action to produce the Ricci scalar squared invariant coupled to vector 
multiplets. Labelling the 
 composite vector multiplet $V_\sharp$ 
and considering the coupling $C_{I\sharp\sharp}$ we may obtain this invariant, 
however it is easier to construct using gauge fixed quatities, so we shall give 
its gauge fixed form in the next section.
\item Linear multiplet constructed from Weyl multiplet squared, 
$\mathbf{L}[\mathbf{W}^2]$. In order to get a mixed Chern-Simons gravitational 
term the embedding of the square of the Weyl multiplet into the linear 
multiplet is realized schematically as
\begin{align}
L^{\mathbf{ij}} \sim i \hat{\bar{R}}(Q)^{(\mathbf{i}}_{\phantom{(\mathbf{i}}\mu\nu} \hat{R}(Q)^{\mathbf{j})\mu\nu} 
&\Rightarrow 
\phi^{\mathbf{i}} \sim \hat{R}(M)^{ab\mu\nu} \gamma_{ab}\hat{R}(Q)^{\mathbf{i}}_{\phantom{\mathbf{i}}\mu\nu} 
\nonumber\\
&\Rightarrow E_a \sim \epsilon_{abcde}\hat{R}(M)^{bc\mu\nu} 
\hat{R}(M)^{de}_{\phantom{de}\mu\nu} \; .
\end{align}
This embedding is given in its entirety in \cite{Hanaki:2006pj}.
Here arbitrary constants $c_{2I}$ are used in order to contract $I,J,...$ 
indices of the vector multiplet. One obtains
\begin{align}
\mathcal{L}_{C^2} &= \frac{c_{2I}}{24} \left( -Y^{I\mathbf{ij}} L_{\mathbf{ij}}[\mathbf{W}^2] - 
\tfrac{1}{2} A^I_a E^a[\mathbf{W}^2] + \tfrac{1}{2} M^I N[\mathbf{W}^2] \right) 
 \nonumber\\
&= \frac{c_{2I}}{24} \left\{ \frac{1}{16} \epsilon^{abcde} A^I_a 
\hat{R}(M)_{bcfg} \hat{R}(M)_{de}^{\phantom{de}fg} - \frac{1}{12} 
\epsilon^{abcde} A^I_a \hat{R}(U)^{\mathbf{ij}}_{\phantom{\mathbf{ij}}bc} \hat{R}(U)_{\mathbf{ij}de} 
\right. \nonumber\\
&+ \frac{1}{8} M^I \hat{R}(M)^{abcd} \hat{R}(M)_{abcd} - \frac{1}{3} M^I 
\hat{R}(U)^{ijab} \hat{R}(U)_{ijab} + \frac{1}{12} M^I D^2 \nonumber\\
&+ \frac{1}{6} D v^{ab} F^I_{ab} - \frac{1}{3} M^I \hat{R}(M)_{abcd} v^{ab} 
v^{cd} -\frac{1}{2} \hat{R}(M)_{abcd}F^{Iab} v^{cd} \nonumber \\
&+ \frac{8}{3} M^I v_{ab} \hat{\mathcal{D}}^b \hat{\mathcal{D}}_c v^{ac} + 
\frac{4}{3} M^I \hat{\mathcal{D}}_a v_{bc} \hat{\mathcal{D}}^a v^{bc} + 
\frac{4}{3} M^I \hat{\mathcal{D}}_a v_{bc} \hat{\mathcal{D}}^b v^{ca} 
\nonumber\\
& - \frac{2}{3} M^I \epsilon^{abcde}v_{ab} v_{cd} \hat{\mathcal{D}}^f v_{ef} + 
\frac{2}{3} \epsilon^{abcde} F^I_{ab} v_{cf} \hat{\mathcal{D}}^f v_{de} 
\nonumber\\
&+ \epsilon^{abcde} F^I_{ab} v_{cf} \hat{\mathcal{D}}_d v_{e}^{\phantom{e}f} 
-\frac{4}{3} F^I_{ab} v^{ac} v_{cd} v^{db} - \frac{1}{3} F^I_{ab}v^{ab} v_{cd} 
v^{cd} \nonumber\\
&\left. + 4 M^I v_{ab} v^{bc} v_{cd} v^{da} - M^I v_{ab} v^{ab} v_{cd} v^{cd} 
-\tfrac{4}{3}Y^I_{\mathbf{ij}}v^{ab}\hat{R}(U)^{\mathbf{ij}}_{\phantom{\mathbf{ij}}ab}\phantom{\frac{1}{1}}
\right\} \; .
\end{align}
\end{itemize}
\subsection{Poincar\'e gauge-fixing} \label{gaugefix}

We are now in a position to break superconformal invariance down to 
super-Poincar\'e invariance. First of all, we set the gauge field of 
dilatations to zero,
$b_\mu =0$, 
which can be done consistently since it appears in our Lagrangian only in 
covariant derivatives of matter fields, not in curvatures. Note that under a 
special conformal transformation of parameter $\xi^a$ we have
\begin{equation}
\delta b_\mu = -2 \xi_\mu \; , 
\end{equation}
so our gauge fixing choice breaks invariance under conformal boosts. Next, we 
set
\begin{equation}
\partial_\mu L_{\mathbf{ij}}=0 \ , \qquad L^2=1 \ ,
\end{equation}
which breaks local SU(2) down to global SU(2)\footnote{Choosing a particular value for $L^{\mathbf{ij}}$, for example $L_{\mathbf{ij}}=\tfrac{1}{\sqrt2}\delta_{\mathbf{ij}}$ would further break this down to U(1), but doesn't simplify the expressions.} and breaks dilatational 
invariance respectively.
As far as the fermion is concerned, we set $\varphi^i=0$. Since its Q-, 
S-supersymmetry transformation before gauge-fixing is
\begin{equation}
\delta \varphi^{\mathbf{i}} = 
-\gamma^a\hat{\mathcal{D}}_aL^{\mathbf{ij}}\epsilon_{\mathbf{j}} + 
\tfrac{1}{2}\gamma^aE_a\epsilon^{\mathbf{i}}
+\tfrac{N}{2}\epsilon^{\mathbf{i}} + 2(\gamma \cdot 
v)L^{\mathbf{ij}}\epsilon_{\mathbf{j}} -6L^{\mathbf{ij}}\eta_{\mathbf{j}}
\end{equation}
consistency requires $\eta$ to be fixed in terms of $\epsilon$ in order to make 
this variation vanish. Multiplying this expression with $L_{\mathbf{ij}}$ our 
gauge choices imply
\begin{equation}
\eta^{\mathbf{i}} = \tfrac{1}{3} v^{ab} \gamma_{ab} \epsilon^{\mathbf{i}} 
-\tfrac{1}{6}(\slashed{E}+N)L^{\mathbf{ij}}\epsilon_{\mathbf{j}}+\tfrac{1}{6}\gamma^a({V'
}_a^{\mathbf{ij}}-2L^{\mathbf{ik}}L^{\mathbf{jl}}{V'}_{a\mathbf{kl}})\epsilon_{
\mathbf{j}}\; ,\label{fixedeta}
\end{equation}
where we found it useful to define a splitting of the SU(2) field 
$V_\mu^{\mathbf{ij}} = {V'}_\mu^{\mathbf{ij}}+L^{\mathbf{ij}}V_\mu$, where 
$V_\mu=V_\mu^{\mathbf{ij}}L_{\mathbf{ij}}$  so that 
${V'}^{\mathbf{ij}}_\mu L_{\mathbf{ij}}=0$.
Examining the last term we find that 
$L^{\mathbf{ik}}L^{\mathbf{jl}}{V'}_{a\mathbf{kl}}=-\tfrac{1}{2}{V'}_a^{\mathbf{
ij}}$ so we obtain
\begin{equation}
\eta^{\mathbf{i}} = \tfrac{1}{3} v^{ab} \gamma_{ab} \epsilon^{\mathbf{i}} 
-\tfrac{1}{6}(\slashed{E}+N)L^{\mathbf{ij}}\epsilon_{\mathbf{j}}+\tfrac{1}{3}\gamma^a{V'}
_a^{\mathbf{ij}}\epsilon_{\mathbf{j}}\; .\label{fixedetasimplified}
\end{equation}
We can immediately write down the supersymmetry transformations of the funfbein 
and of the gravitino as
\begin{align}
\delta e^a_\mu &= -2i \bar{\epsilon} \gamma^a \psi_\mu \ , \nonumber\\
\delta \psi^{\mathbf{i}}_\mu &= \nabla_\mu \epsilon^{\mathbf{i}} + \tfrac{1}{2} \gamma_{\mu ab} 
v^{ab} \epsilon^{\mathbf{i}} - \tfrac{1}{3} \gamma_\mu \gamma_{ab} v^{ab}\epsilon^{\mathbf{i}} \nonumber\\
&+ V^{\mathbf{ij}}_{\mu}\epsilon_{\mathbf{j}}+  
\tfrac{1}{6}\gamma_\mu(\slashed{E}+N)L^{\mathbf{ij}}\epsilon_{{\mathbf{j}}}- 
\tfrac{1}{3}\gamma_\mu\gamma^a{V'}_a^{\mathbf{ij}}\epsilon_{\mathbf{j}} \; .
\end{align}
Next we consider the auxiliary fermion: since we will be concerned with the 
bosonic sector of the theory we can write
\begin{align}
\delta \chi^{\mathbf{i}} &= D \epsilon^{\mathbf{i}} - 2 \gamma^c \gamma^{ab} \nabla_{a}v_{bc} 
\epsilon^{\mathbf{i}} - 2  \epsilon_{abcde} v^{bc} v^{de} \gamma^a \epsilon^{\mathbf{i}} + 
\tfrac{4}{3} (v \cdot \gamma)^2 \epsilon^{\mathbf{i}} 
-\gamma^{ab}V_{ab}^{\mathbf{ij}}\epsilon_{\mathbf{j}}\nonumber\\
&-\tfrac{2}{3}\slashed{v}(\slashed{E}+N)L^{\mathbf{ij}}\epsilon_{\mathbf{j}}+\tfrac{4}{3}
\slashed{v}\gamma^a{V'}_a^{\mathbf{ij}}\epsilon_{\mathbf{j}}+ \text{fermion 
bilinears}
\end{align}
and discard such bilinears, where we defined 
$V^{\mathbf{ij}}_{\mu\nu}=2\partial^{\phantom{ij}}_{[\mu}V_{\nu]}^{\mathbf{ij}} 
+ 2V^{\mathbf{ik}}_{[\mu}V^{\mathbf{j}}_{\nu]\mathbf{k}}$ and at this point we 
do not expand this quantity in terms of the $V_\mu$ and 
${V'}_\mu^{\mathbf{ij}}$ fields. Let us now examine the auxiliary 2-form: its 
supersymmetry transformation is determined by the equations
\begin{align}
& \delta v_{ab} = -\tfrac{1}{8} i \bar{\epsilon} \gamma_{ab} \chi - 
\tfrac{3}{2} i \bar{\epsilon} \hat{R}(Q)_{ab} \ ,\nonumber\\
& \hat{R}(Q)^i_{\phantom{i}\mu\nu} = 2 \nabla_{[\mu} \psi^i_{\nu]} 
+2V^{\mathbf{ij}}_{[\mu}\psi^{\mathbf{k}\phantom{\mathbf{j}}}_{\nu]}\epsilon_{
\mathbf{jk}}+ v^{ab} \gamma_{ab[\mu} \psi_{\nu]}^i  - 2 \gamma_{[\mu} 
\phi^i_{\nu]} \ ,\nonumber\\
& \phi^i_\mu = \left( -\frac{1}{3}e^a_\mu \gamma^b + \frac{1}{24} \gamma_\mu 
\gamma^{ab} \right) \left.\hat{R}(Q)^i_{ab} \right|_{\phi^i_\mu=0}\ .
\end{align}
A straightforward calculation gives
\begin{align}
\delta v_{ab} &= \tfrac{1}{2}i v_{ab} \bar{\epsilon} \gamma^\mu \psi_\mu +i 
v_{[a|\mu} \bar{\epsilon} \gamma^\mu \psi_{b]} - \tfrac{1}{2}i v_{[a|\mu} 
\bar{\epsilon} \gamma_{b]} \psi^\mu - \tfrac{1}{8}i \bar{\epsilon} \gamma_{ab} 
\chi \nonumber\\
&- \tfrac{3}{2}i \bar{\epsilon} \nabla_{[a} \psi_{b]} 
- \tfrac{3}{4} i\bar{\epsilon} \gamma_{[a|\mu} \nabla_{b]} \psi^\mu + 
\tfrac{3}{4}i \bar{\epsilon} \gamma_{[a|\mu} \nabla^\mu \psi_{b]} \nonumber\\
&- \tfrac{3}{2}i \bar{\epsilon}^{\mathbf{i}} V_{\mathbf{ij}[a} 
\psi^{\mathbf{j}}_{b]} 
- \tfrac{3}{4} i\bar{\epsilon}^{\mathbf{i}} \gamma_{[a|\mu} V_{b]\mathbf{ij}} 
\psi^{\mu\mathbf{j}} + \tfrac{3}{4}i \bar{\epsilon}^{\mathbf{i}} 
\gamma_{[a|\mu} V^{\mu}_{\mathbf{ij}} \psi^{\mathbf{j}}_{|b]}\; .
\end{align}
Next we turn to the auxiliary scalar $D$. We should compute 
$\hat{\mathcal{D}}_\mu \chi$ and then gauge fix. To this end note that in 
\begin{align}
\hat{\mathcal{D}}_\mu \chi^{\mathbf{i}} &= \mathcal{D}_\mu \chi^{\mathbf{i}} - D \psi^{\mathbf{i}}_\mu + 2 
\gamma^c \gamma^{ab} \psi^{\mathbf{i}}_\mu \hat{\mathcal{D}}_a v_{bc} - \gamma^{ab} 
\hat{R}(U)^{\mathbf{ij}}_{\phantom{ij}ab} \psi^{\mathbf{k}}_\mu 
\epsilon_{\mathbf{jk}}\nonumber\\
& + 2 \gamma^a \psi^{\mathbf{i}}_\mu \epsilon_{abcde} v^{bc} v^{de} - 4v^{ab} \gamma_{ab} 
\phi^{\mathbf{i}}_\mu
\end{align}
one has $\hat{\mathcal{D}}_\mu v_{ab} = \nabla_\mu v_{ab}$ up to fermion bilinears, 
so that
\begin{align}
\hat{\mathcal{D}}_\mu \chi^{\mathbf{i}} &= \nabla_\mu \chi^{\mathbf{i}} 
+V_\mu^{\mathbf{ij}}\chi_{\mathbf{j}} - D \psi^i_\mu + 2 \gamma^c \gamma^{ab} 
\psi^{\mathbf{i}}_\mu \nabla_a v_{bc} \nonumber\\
&+ 2 \gamma^a \psi^{\mathbf{i}}_\mu \epsilon_{abcde} v^{bc} v^{de} - 4v^{ab} \gamma_{ab} 
\phi^{\mathbf{i}}_\mu - \gamma^{ab} V^{\mathbf{ij}}_{\phantom{ij}ab} \psi^{\mathbf{k}}_\mu 
\epsilon_{\mathbf{jk}}+ \text{fermion trilinears}\ .
\end{align}
One can thus write
\begin{align}
\delta D &= -i \bar{\epsilon}^{\mathbf{i}} \gamma^f e_f^\mu \left( \nabla_\mu 
\chi_{\mathbf{i}} -V_{\mu\mathbf{ij}}\chi^{\mathbf{j}} - \gamma^{ab} 
V_{\mathbf{ij}ab} \psi^{\mathbf{j}}_\mu - D \psi_{\mathbf{i}\mu} + 2 \gamma^c 
\gamma^{ab} \psi_{\mathbf{i}\mu} \nabla_a v_{bc} \right.\nonumber\\ 
&+\left. 2 \gamma^a \psi_{\mathbf{i}\mu} \epsilon_{abcde} v^{bc} v^{de} - 
4v^{ab} \gamma_{ab} \phi_{\mathbf{i}\mu}\right)  - 8 i v^{ab} \bar{\epsilon} 
\hat{R}(Q)_{ab} - \tfrac{1}{3}i \bar{\epsilon}v^{ab} \gamma_{ab}  \chi 
\nonumber\\
&-\tfrac{i}{6}\bar{\epsilon}^{\mathbf{i}}(\slashed{E}+N)L_{\mathbf{ij}}\chi^{
\mathbf{j}}+\tfrac{i}{3}\bar{\epsilon}^{\mathbf{i}}\gamma^a{V'}_{a\mathbf{ij}}
\chi^{\mathbf{j}} + \bar{\epsilon}(\text{fermion trilinears}) \; .
\end{align}
Once again straightforward computation gives
\begin{align}
\delta D &=4i\bar{\epsilon} \psi_\mu \nabla_\nu v^{\nu\mu} - 2i 
\epsilon^{\mu\nu\rho\sigma\tau} \bar{\epsilon}\psi_\mu v_{\nu\rho} 
v_{\sigma\tau}  + i\left( D - \tfrac{2}{3} v^2 \right) \bar{\epsilon} \gamma^\mu \psi_\mu + 
\tfrac{22}{3}i v_{\mu\rho}v_{\nu}^{\phantom{\nu}\rho} \bar{\epsilon} \gamma^\mu 
\psi^\nu \nonumber\\
&- 2i \epsilon^{\nu\lambda\rho\sigma\tau} v_{\lambda\rho} 
v_{\sigma\tau} \bar{\epsilon} \gamma_{\mu\nu} \psi^\mu -2 i \bar{\epsilon} \gamma^{\rho\sigma} \psi^\mu \nabla_\mu v_{\rho\sigma} +4 
i \bar{\epsilon} \gamma^{\mu\nu} \psi_\mu \nabla^\rho v_{\nu\rho} -4i 
\bar{\epsilon}\gamma^{\nu\rho} \psi^\mu \nabla_\rho v_{\mu\nu}\nonumber\\
&- 12i v_{\mu\nu} \bar{\epsilon} \nabla^\mu \psi^\nu + 4i v^{\mu\rho} 
\bar{\epsilon} \gamma_{\nu\rho} \nabla^\nu \psi_\mu - 4 i v^{\mu\rho} 
\bar{\epsilon} \gamma_{\nu\rho} \nabla_\mu \psi^\nu - 12i v_{\mu\nu} \bar{\epsilon}^{\mathbf{i}} V_{\mathbf{ij}}^\mu 
\psi^{\mathbf{j}\nu} \nonumber\\
&+ 4i v^{\mu\rho} \bar{\epsilon}^{\mathbf{i}} 
\gamma_{\nu\rho} V_{\mathbf{ij}}^\nu \psi^{\mathbf{j}}_\mu - 4 i v^{\mu\rho} 
\bar{\epsilon}^{\mathbf{i}} \gamma_{\nu\rho} V_{\mathbf{ij}\mu} 
\psi^{\mathbf{j}\nu} -\tfrac{1}{3}i \bar{\epsilon}\gamma^{\mu\nu} \chi v_{\mu\nu} -i 
\bar{\epsilon}\gamma^\mu \nabla_\mu \chi  +i\bar{\epsilon}^{\mathbf{i}}\gamma^\mu 
V_{\mathbf{ij}\mu}\chi^{\mathbf{j}}\nonumber\\
&-\tfrac{i}{6}\bar{\epsilon}^{\mathbf{i}}(\slashed{E}+N)L_{\mathbf{ij}}\chi^{
\mathbf{j}} 
+\tfrac{i}{3}\bar{\epsilon}^{\mathbf{i}}\gamma^a{V'}_{a\mathbf{ij}}\chi^{\mathbf
{j}}-i\bar{\epsilon}^{\mathbf{i}}\gamma^c\gamma^{ab}V_{\mathbf{ij}ab}\psi_c^{
\mathbf{j}}+ \bar{\epsilon}(\text{fermion trilinears}) \; .
\end{align}
Finally for the Weyl multiplet we compute
\begin{align}
 \delta 
V_\mu^{\mathbf{ij}}=-\tfrac{i}{4}\bar{\epsilon}^{\mathbf{(i}}\gamma_\mu\chi^{\mathbf{j)}} 
+\textrm{terms involving the gravitino} \ ,
\end{align}
where we will not need the gravitino terms in our analysis.

Now consider the vector multiplet. In this case we just have to replace $\eta$ 
and note that
$
\hat{\mathcal{D}}_a M^I = \nabla_a M^I = e_a^\mu \partial_\mu M^I \; .
$
We obtain
\begin{align}
\delta A^I_\mu & = -2i \bar{\epsilon} \gamma_\mu \Omega^I + 2i 
M^I\bar{\epsilon} \psi_\mu \ , \nonumber\\
\delta M^I &= 2i \bar{\epsilon} \Omega^I \ ,\nonumber\\ 
\delta \Omega^{I\mathbf{i}} &= - \tfrac{1}{4} F^I_{ab} \gamma^{ab} \epsilon^{\mathbf{i}} 
-\tfrac{1}{2} \gamma^\mu \partial_\mu M^I \epsilon^{\mathbf{i}} - Y^{I\mathbf{ij}}\epsilon_{\mathbf{j}} \nonumber\\
&-M^I\tfrac{1}{3} v^{ab}\gamma_{ab} \epsilon^{\mathbf{i}} +\tfrac{M^I}{6}(\slashed{E}+N)L^{\mathbf{ij}}\epsilon_{\mathbf{j}} 
-\tfrac{M^I}{3}\gamma^a{V'}_a^{\mathbf{ij}}\epsilon_{\mathbf{j}} \; ,\nonumber\\
\delta Y^{I\mathbf{ij}}&=2i\bar{\epsilon}^{(\mathbf{i}}\gamma^a \nabla_a 
\Omega^{\mathbf{j})I}
-2i\bar{\epsilon}^{(\mathbf{i}}\gamma^a 
V_{a\phantom{\mathbf{j})}\mathbf{k}}^{\phantom{a}\mathbf{j})} 
\Omega^{\mathbf{k}I}
-\tfrac{2i}{3}V_a^{\mathbf{k}(\mathbf{i}}\bar{\epsilon}_{\mathbf{k}}
\gamma_a\Omega^{\mathbf{j})}
-\tfrac{i}{3}\bar{\epsilon}^{(\mathbf{i}}\gamma_{ab}v^{ab}\Omega^{\mathbf{j})I} 
\nonumber\\ 
&-\tfrac{i}{4}\bar{\epsilon}^{(\mathbf{i}}\chi^{\mathbf{j})}M^I  \ .
\end{align}
Finally we need the transformation rules for the unfixed fields in the compensating linear 
multiplet. The non-trival transformations are
\begin{align}
 \delta N &= \tfrac{i}{2}L_{\mathbf{ij}}\bar{\epsilon}^{\mathbf{i}}\chi^{\mathbf{j}} + \textrm{gravitino terms} \nonumber \ ,\\ 
 \delta P_a&= \textrm{gravitino terms} \ .
\end{align}
We will only consider the gravitino terms, which arise from the non-vanishing of $\mathcal{D}\varphi$ even after setting $\varphi=0$, in the special case of maximal supersymmetry, and so we will not give the full expressions here, but to derive them it is useful to note that
\begin{eqnarray}
\phi^{\mathbf{j}}_\mu &=&\tfrac{1}{4}{v_\mu}^a\psi^{\mathbf{j}}_a - \tfrac{1}{2}\gamma^a(\nabla_{[\mu}\psi^{\mathbf{j}}_{a]} 
+ V^{\mathbf{ij}}_{[\mu}\psi_{a]\mathbf{i}}) 
- \tfrac{1}{6}v_{bc}{\gamma_{\mu}}^{abc}\psi^{\mathbf{j}}_a 
+ \tfrac{5}{12}v^{ab}\gamma_{b\mu}\psi^{\mathbf{j}}_a \nonumber\\
&&+\tfrac{1}{4}\slashed{v}\psi^{\mathbf{j}}_\mu 
+ \tfrac{1}{6}v_{\mu a}\gamma^{ab}\psi^{\mathbf{j}}_b + \tfrac{1}{12}{\gamma_{\mu}}^{ab}(\nabla_a\psi^{\mathbf{j}}_b +V^{\mathbf{ij}}_a\psi_{b\mathbf{i}})\ . \label{explicitphi}
\end{eqnarray}

We now summarize the effect of gauge-fixing on the superconformal Lagrangians 
constructed above. The Lagrangian $\mathcal{L}_V$ is virtually unchanged, the 
only difference being the removal of the gauge field $b_\mu$ from the 
supercovariant derivatives.
The compensating linear-linear action now becomes
 \begin{align}
  e^{-1}\mathcal{L}_L &=-(\tfrac{3}{8}R +\tfrac{1}{4}D -\tfrac{1}{2} {v}^2)
  -\tfrac{3}{2}{V'}_\mu^{\mathbf{ij}}{V'}_{\mathbf{ij}}^\mu 
  -N^2 
  + \tfrac{1}{4}P_\mu P^\mu   + P^{\mu}V_\mu \ ,  \label{LLfixed}
 \end{align}
As far as Weyl-squared Lagrangian is considered one finds (modulo fermions)
\begin{align}
\mathcal{L}_{C^2} = \tfrac{c_{2I}}{24}& \left\{ \tfrac{1}{16} \epsilon^{abcde} 
A^I_a C_{bcfg} C_{de}^{\phantom{de}fg} + \tfrac{1}{8} M^I C^{abcd} C_{abcd} 
 + \tfrac{1}{12} M^I D^2 + \tfrac{1}{6} D v^{ab} F^I_{ab} \right. \nonumber\\
&  + \tfrac{1}{3} M^I 
C_{abcd} v^{ab} v^{cd} + \tfrac{1}{2} C_{abcd}F^{Iab} v^{cd} + \tfrac{8}{3} M^I v_{ab} \nabla^b \nabla_c v^{ac} 
-\tfrac{16}{9}M^I v^{ab}v_{bc}R_a^{\phantom{a}c}  \nonumber\\
& - \tfrac{2}{9} M^I v^2 R  + \tfrac{4}{3} M^I \nabla_a v_{bc} \nabla^a v^{bc} +\tfrac{4}{3} M^I \nabla_a 
v_{bc} \nabla^b v^{ca} - \tfrac{2}{3} M^I \epsilon^{abcde}v_{ab} v_{cd} \nabla^f v_{ef}\nonumber\\
&  + 
\tfrac{2}{3} \epsilon^{abcde} F^I_{ab} v_{cf} \nabla^f v_{de} + \epsilon^{abcde} F^I_{ab} v_{cf} \nabla_d v_{e}^{\phantom{e}f} -\tfrac{4}{3} 
F^I_{ab} v^{ac} v_{cd} v^{db} - \tfrac{1}{3} F^I_{ab}v^{ab} v_{cd} v^{cd} \nonumber\\
& + 4 M^I v_{ab} v^{bc} v_{cd} v^{da} - M^I v_{ab} v^{ab} v_{cd} v^{cd} - \tfrac{1}{12} \epsilon^{abcde} A^I_a V^{\mathbf{ij}}_{\phantom{ij}bc} 
V_{\mathbf{ij}de}
- \tfrac{1}{3} M^I V^{\mathbf{ij}ab} V_{\mathbf{ij}ab} \nonumber\\
&\left.
-\tfrac{4}{3}Y^I_{\mathbf{ij}}v^{ab}V^{\mathbf{ij}}_{\phantom{ij}ab}\phantom{
\tfrac{1}{1}}\right\}\; .
\end{align}
$C$ denotes the Weyl tensor: it appears because the conventional constraints imply 
$\hat{R}(M)$ is traceless. Note also that in the first term the Weyl and 
Riemman tensors may be used interchangeably.  The new terms with the Ricci 
tensor and Ricci scalar arise by virtue of the identity
\begin{equation}
v_{ab} \hat{\mathcal{D}}^b \hat{\mathcal{D}}_c v^{ac} = v_{ab} \nabla^b 
\nabla_c v^{ac} - \frac{2}{3} v^{ac} v_{cb} R_a^{\phantom{s}b} - \frac{1}{12} 
v^2 R \; ,
\end{equation}
which arises because whilst we have set $b_\mu=0$ its full superconformally covariant 
derivative does not vanish.
Finally, note the change of sign in terms containing one Weyl tensor, which is due 
to our conventions for the Riemann and Weyl tensors, which are those of 
\cite{Wald:1984rg} and are different from those of \cite{Hanaki:2006pj}.

We have yet to construct the Ricci squared invariant. By gauge fixing using the 
compensating linear multiplet the bosonic parts of the embedding into the 
vector multiplet become
\begin{align}
 M^\sharp &= N  \ ,\nonumber\\
F^\sharp_{\mu\nu} &= 2\partial_{[\mu}P_{\nu]} - 4\partial_{[\mu}V_{\nu]} \ 
,\nonumber\\
 Y^\sharp_{\mathbf{ij}}&= 
2\nabla^\mu{V'}_{\mu\mathbf{k}}^{\mathbf{(i}}L^{\mathbf{j)k}}
  + \tfrac{1}{4}(P^2 +4V\cdot P -N^2 -2v^2 +D + 
6{V'}_a^{\mathbf{kl}}{V'}^a_{\mathbf{kl}}+ \tfrac{3}{2}R)L_{\mathbf{ij}} 
  \ .
  \label{linearinvectorfixed}
\end{align}
Using this composite vector multiplet, which we denote $\mathbf{V}_\sharp$, in 
the vector multiplet action with the coupling $C_{I\sharp\sharp}=e_I$ we obtain 
the density 
\begin{align}
 e^{-1}\mathcal{L} =& \mathcal{E} \left[ N^2 (\tfrac{1}{4}D - \tfrac{1}{8}R 
+\tfrac{3}{2}v^2) +2N v\cdot(dP-2dV) +\tfrac{1}{4}(dP-2dV)^2 - 
\tfrac{1}{2}(dN)^2 \right. \nonumber\\
 &\left. -\tfrac{1}{16}(P^2 +4V\cdot P -N^2 -2v^2 +D + 
6{V'}_a^{\mathbf{ij}}{V'}^a_{\mathbf{ij}}+ \tfrac{3}{2}R)^2 +2\nabla^a {V'}_a^{\mathbf{ij}}\nabla_b {V'}^b_{\mathbf{ij}} \right]  \nonumber\\
 &+e_I\left[ N^2 F^I\cdot v + \tfrac{N}{2} F^I\cdot(dP -2dV) -N dN \cdot d M^I 
\right. \nonumber\\
 &\left. -\tfrac{1}{2}NY^I (P^2 +4V\cdot P -N^2 -2v^2 +D + 
6{V'}_a^{\mathbf{kl}}{V'}^a_{\mathbf{kl}}+ \tfrac{3}{2}R) \right. \nonumber\\
&\left.- 4N{Y'}^{I}_{\mathbf{ij}}\nabla^\mu{V'}_{\mu\mathbf{k}}^{\mathbf{(i}}L^{\mathbf{j)k}}
+\tfrac{1}{8}e^{-1}\epsilon^{abcde}A_a^I(dP-2dV)_{bc}(dP-2dV)_{de} \right] \ . \label{ricci_inv}
\end{align}

If one considers the two-derivative theory with Lagrangian
\begin{align}
\mathcal{L}_2 &= \mathcal{L}_V +2 \mathcal{L}_L = \nonumber\\
&= \tfrac{1}{2}D(\mathcal{N} -1 ) - \tfrac{1}{4} (\mathcal{N} +3) R + 
(3\mathcal{N} +1)v^2 + 2 \mathcal{N}_I v \cdot F^I \nonumber\\
&+ \mathcal{N}_{IJ} \left( \tfrac{1}{4} F^I \cdot F^J - \tfrac{1}{2} \partial 
M^I \cdot \partial M^J -Y^{I\mathbf{ij}}Y^J_{\mathbf{ij}}\right) + 
\tfrac{1}{24}\tfrac{1}{\sqrt{|g|}} C_{IJK} \epsilon^{\mu\nu\rho\sigma\tau} A_\mu^I 
F_{\nu\rho}^J F^K_{\sigma \tau}\nonumber\\
&-3{V'}^{\mathbf{ij}}{V'}_{\mathbf{ij}} - 2N^2 + \tfrac{1}{2}P_\mu P^\mu + 2 
P^\mu V_\mu \ , \label{2derivaction}
\end{align}
one finds non-propagating equations of motion for auxiliary fields. In 
particular note that $D$ acts as a Lagrange multiplier in order to implement 
the constraint
\begin{equation}
\mathcal{N} = 1 \, ,
\end{equation}
and that thanks to this constraint the Ricci scalar acquires the canonical 
normalization.
Similarly to what was shown in \cite{Cremonini:2008tw} for a hypermultiplet 
compensator, the auxiliary fields $N, P, V, V', Y^I$ can be completely 
eliminated from the Lagrangian, and we arrive at the on-shell ungauged 
Poincar\'e supergravity coupled to Abelian vector multiplets.
\subsection{Equations of motion}\label{eoms}
Here we record the equations of motion for the Lagrangian \eqref{2plus4Lagrangian} which is a consistent truncation of the sum of two derivative theory with the four derivate Lagrangians derived above.
Luckily we will not have to solve all of these equations as the 
Killing spinor identities imply
 that some of their components are automatic for supersymmetric solutions.
Denoting the two derivative action $S_2$ and the four derivative pieces of the 
action $S_{C^2}$ and $S_{R_s^2}$ so that the action for this theory is $S=S_2+S_{C^2}+S_{R_s^2}$ and taking as 
the independent fields\footnote{As we are concerned with the Einstein equation only in the case where all other bosons are on-shell we can interpret $\mathcal{E}(v), \mathcal{E}(D), \mathcal{E}(A), 
\mathcal{E}(M)$ as variational derivatives with respect to either $(e^a_\mu, 
v_{ab}, D, M^I, A^I_\mu)$ or $(g_{\mu\nu}, v_{\mu\nu}, D, M^I, A^I_\mu)$ 
indifferently.
} $D, M^I, v_{\mu\nu}, A^I_\mu, g_{\mu\nu}$
 the equations of motion for the two derivative theory are given by
\begin{eqnarray}
\tfrac{1}{\sqrt{|g|}}\frac{\delta S_2}{\delta D} &=& \tfrac{1}{2} \left( 
\mathcal{N} -1 \right), \qquad
\tfrac{1}{\sqrt{|g|}}\frac{\delta S_2}{\delta v_{\mu\nu}}  =2 ( \mathcal{N}_I 
F^{I\mu\nu} + (3\mathcal{N} +1) v^{\mu\nu}), \nonumber\\
\tfrac{1}{\sqrt{|g|}}\frac{\delta S_2}{\delta M^I}  &=& (\tfrac{1}{2} D - 
\tfrac{1}{4} R  + 3 v^2) \mathcal{N}_I +  c_{IJK}(\tfrac{1}{4} F^J\cdot F^{K}
 + \tfrac{1}{2}\nabla M^J\cdot \nabla M^K ) \nonumber\\&&+  \mathcal{N}_{IJ}(2 
F^J_{a b} v^{a b} + \nabla^2 M^J)  \ ,\nonumber\\
\tfrac{1}{\sqrt{|g|}}\frac{\delta S_2}{\delta A^I_{\mu}} &=& c_{IJK}( 
\tfrac{1}{8} \epsilon^{\mu abcd} F^J_{ab} F^K_{cd} +  F^{J\mu a} \nabla_a M^K ) 
+ 4 \mathcal{N}_I \nabla_a v^{\mu a} \\
&&+ \mathcal{N}_{IJ}(4 v^{\mu a} \nabla_a M^J + \nabla_a F^{J\mu a} )\ ,\nonumber\\
\tfrac{1}{\sqrt{|g|}}\frac{\delta S_2}{\delta g^{\mu\nu}} &=& 
-\tfrac{1}{4}(\mathcal{N}+3) (R_{\mu\nu}-\tfrac{1}{2}g_{\mu\nu}R) - \tfrac{1}{4} D (\mathcal{N} - 
1)g_{\mu\nu}  + 2(1+3\mathcal{N}) ( v_{a \mu} v^{a}_{\phantom{a}\nu} - 
\tfrac{1}{4} v^2 g_{\mu\nu} )\nonumber\\
&&+  \mathcal{N}_{IJ} (\tfrac{1}{2} F^I_{a \mu } F^{J a}_{\phantom{J a}\nu}  + 
4F^I_{a(\mu} v^{a}_{\phantom{a}\nu)}    - \tfrac{1}{2} \nabla_\mu M^I 
\nabla_\nu M^J  ) \nonumber\\
&&-\mathcal{N}_{IJ} (\tfrac{1}{8} F^I \cdot F^J +  F^I \cdot v - \tfrac{1}{4} 
\nabla M^I\cdot \nabla M^J) g_{\mu\nu}  + \tfrac{1}{4} ( \nabla_\mu \nabla_\nu 
\mathcal{N} - \nabla^2 \mathcal{N} g_{\mu\nu} ) \; \nonumber.
\end{eqnarray}
where lower case latin indices refer to the vielbein, and greek indices refer 
to the coordinates and we have found it convenient to express all contracted 
indices in terms of the veilbein.
For the contraction of two p-forms 
$\alpha,\beta$ we use the notation $\alpha\cdot\beta:=\alpha_{a_1\cdots 
a_p}\beta^{a_1\cdots a_p}$ and $\alpha^2:=\alpha\cdot\alpha$.

The additional contributions from the Weyl-squared Langrangian are given by

\begin{align}
\lefteqn{\frac{1}{\sqrt{|g|}}\frac{\delta S_{C^2}}{\delta g^{\mu\nu}} = 
\frac{c_{2I}}{24}
 \left\{ \phantom{\frac{1}{8}} \right.
-\tfrac{1}{8}\left[			\epsilon^{abcd}_{\phantom{abcd}(\mu|} 
\nabla_e F_{ab}^I R_{cd\phantom{e}|\nu)}^{\phantom{bc}e}\right]  }\nonumber\\ 
&+\tfrac{1}{4}\left[ 			
M^I\left(-C_{abc(\mu|}R^{abc}_{\phantom{abc}|\nu)} + \tfrac{4}{3}R_{ab} 
C_{\mu\phantom{a}\nu}^{\phantom{\mu}a\phantom{\nu}b} 
					+ 2C_{\mu}^{\phantom{\mu}bcd}C_{\nu 
bcd} -\tfrac{1}{4}g_{\mu\nu}C^{abcd}C_{abcd}\right) 
					\right. \nonumber\\
&\left.\phantom{+\tfrac{1}{4}}+ 2\nabla_a\nabla_b M^I 
C^{a\phantom{\mu}b}_{\phantom{a}\mu\phantom{b}\nu}\right] 
-\tfrac{1}{24}\left[g_{\mu\nu}M^ID^2 \right]
 +\tfrac{1}{3}\left[  			D {v_{(\mu}}^{a}F^I_{\nu)a}- 
\tfrac{1}{4}g_{\mu\nu}Dv^{ab}F^I_{ab} \right]\nonumber\\
&+\tfrac{1}{3}\left[ 			
M^I\left(\left(R_{abc(\mu}-4C_{abc(\mu}\right)v^{ab}{v_{\nu)}}^c + 
\tfrac{4}{3}R_{ab}{v_\mu}^a{v_\nu}^b 
					 -\tfrac{1}{3}R{v_\mu}^av_{\nu a} 
+\tfrac{1}{6}R_{\mu\nu}v^2 \right.
\right.\nonumber\\
&\left.\phantom{+\tfrac{1}{3}\;}	
\left.-\tfrac{1}{2}g_{\mu\nu}C_{abcd}v^{ab}v^{cd}\right) +2\nabla_a\nabla_b{v_{\mu}}^{a}{v_{\nu}}^{b}M^I 
+\tfrac{4}{3}\nabla_a\nabla_{(\mu} v_{\nu)b}v^{ab}M^I
					-\tfrac{2}{3}\nabla^2{v_\mu}^av_{\nu 
a}M^I  \right. \nonumber\\
&\left.\phantom{+\tfrac{1}{3}\;}	+ \tfrac{2}{3}g_{\mu\nu}\nabla_a\nabla_bv^{ac}{v_c}^bM^I 
+\tfrac{1}{6}\left(g_{\mu\nu}\nabla^2 
-\nabla_\mu\nabla_\nu \right) v^{ab}v_{ab}M^I		    \right]\nonumber\\
&+\left[				\tfrac{1}{2}R_{abc(\mu} {v_{\nu)}}^c 
F^{Iab} + \nabla_a\nabla_b {v_{(\mu}}^a {F^I_{\nu)}}^b
					+\tfrac{1}{3}\nabla_a\nabla_{(\mu|} 
v_{|\nu) b}F^{Iab} +\tfrac{1}{3}\nabla_a\nabla_{(\mu}{F^{Ib}}_{\nu)}{v_b}^a		
	 \right.\nonumber\\
&\left.\phantom{+\;} +\tfrac{1}{3}\nabla^2{F^{Ia}}_{(\mu}v_{\nu)a} - \tfrac{1}{3}g_{\mu\nu}\nabla_a 
\nabla_b {v^a}_{c}F^{Ibc}					
+\tfrac{2}{3}R_{ab}{F^{Ia}}_{(\mu}{v^{b}}_{\nu)} \right.  \nonumber\\
&\left.\phantom{+\;} +\tfrac{1}{12}\left(R_{\mu\nu} 
-\nabla_\mu \nabla_\nu
+g_{\mu\nu}\nabla^2\right)v_{ab}F^{Iab} 
+\tfrac{1}{6}R{F^{Ia}}_{(\mu} v_{\nu)a} \right.\nonumber\\ 
&\left.\phantom{+\;}- \left({F^{Ia}}_{(\mu}v^{bc}+v^a_{(\mu}F^{Ibc} \right)C_{|\nu)abc} 
-\tfrac{1}{4}g_{\mu\nu}F^{Iab}v^{cd}C_{abcd}      \right]\nonumber\\
&+\tfrac{8}{3}\left[			M^I\left( 
v_{a(\mu}\nabla_{\nu)}\nabla_bv^{ab} + v_{ab}\nabla^b\nabla_{(\mu}{v^a}_{\nu)} 
					+{v_{(\mu|}}^a\nabla_a\nabla_b 
{v_{|\nu)}}^b -\tfrac{1}{2}g_{\mu\nu}v_{ab}\nabla^b\nabla_cv^{ac}  \right) 
					\right.\nonumber\\
&\left.\phantom{+\tfrac{8}{3}\;}+\nabla_a{v_{(\mu|}}^a \nabla_b M^I 
{v_{|\nu)}}^b	-\nabla_{(\mu}v_{\nu)a}\nabla_b M^I 
v^{ab}
					+\tfrac{1}{2}g_{\mu\nu}\nabla_a {v^a}_b 
\nabla_c M^I v^{bc} -\nabla_a M^Iv^{ab}\nabla_{(\mu}v_{\nu)b} \right]\nonumber\\
&-\tfrac{16}{9}\left[ 			M^I \left( {v^a}_\mu {v_\nu}^b R_{ab} - 
2v^{ab}v_{a(\mu}R_{\nu)b} -\tfrac{1}{2}g_{\mu\nu}v^{ab}{v_{b}}^c R_{ac} \right) 
					+ \tfrac{1}{2}\nabla^2 M^I 
{v_{(\mu|}}^{a}v_{a|\nu)} \right.\nonumber\\
&\left. \phantom{-\tfrac{16}{9}}\;	+ 
\tfrac{1}{2}g_{\mu\nu}\nabla_a\nabla_bM^Iv^{ac}{v_c}^b - 
\nabla_a\nabla_{(\mu|}M^Iv^{ab}v_{b|\nu)}  \right]\nonumber\\
&-\tfrac{2}{9}\left[ 			M^I\left( 2{v_\mu}^a v_{\nu a}R + 
v_{ab}v^{ab}R_{\mu\nu} - \tfrac{1}{2}g_{\mu\nu}Rv_{ab}v^{ab} \right) 
					-\left(\nabla_\mu\nabla_\nu - 
g_{\mu\nu}\nabla^2\right)M^I v_{ab}v^{ab} \right]\nonumber\\
&+\tfrac{4}{3}\left[			M^I\left( (\nabla_\mu 
v_{ab})(\nabla_\nu v^{ab}) + 2(\nabla_av_{b\mu})(\nabla^a{v^b}_\nu ) 
-\tfrac{1}{2}g_{\mu\nu}(\nabla_a v_{bc})(\nabla^a v^{bc})\right) \right. \nonumber\\
&\left. \phantom{+\tfrac{4}{3}\;}	+ 2 \nabla_a M^I(\nabla^a 
{v_{(\mu|}}^b)v_{ b|\nu)} + 2\nabla_aM^I(\nabla_{(\mu|}v^{ab})v_{b|\nu)} - 
2\nabla_aM^I(\nabla_{(\mu|}v_{b|\nu)}))v^{ab} \right]\nonumber\\
&+\tfrac{4}{3}\left[			
M^I\left(2(\nabla_{(\mu|}v^{ab})(\nabla_a v_{b|\nu)}) + (\nabla_av_{b(\mu|}) 
(\nabla^b{v_{|\nu)}}^{a}) - \tfrac{1}{2	}g_{\mu\nu}(\nabla_a 
v_{bc})(\nabla^bv^{ca})\right)\right.\nonumber\\
&\left.	\phantom{+\tfrac{4}{3}\;}	+\nabla_a\left( 
M^Iv_{b(\mu}\nabla_{\nu)}v^{ba} + M^Iv_{b(\mu}\nabla^a {v^b}_{\nu)} - M^I 
v^{ba} \nabla_{(\mu|}v_{b|\nu)} \right)\right]\nonumber\\
&-\tfrac{2}{3}\left[ 			
M^I\epsilon^{abcde}v_{ab}v_{cd}\nabla_{(\mu |}v_{e|\nu )} - 
\epsilon^{abcde}\nabla_{(\mu|}M^Iv_{ab}v_{cd}v_{e|\nu)} \right.\nonumber\\
&\left.\phantom{-\tfrac{2}{3}\;}	-{\epsilon^{abcd}}_{(\mu |}\nabla_e M^I 
v_{ab}v_{cd}{v_{|\nu)}}^e + \tfrac{1}{2}g_{\mu\nu}\epsilon^{abcde}\nabla^f 
M^Iv_{ab}v_{cd}v_{ef}  \right]\nonumber\\
&+\tfrac{2}{3}\left[ 			
\epsilon^{abcde}F^{I}_{ab}v_{c(\mu}\nabla_{\nu)}v_{de} -2 
{\epsilon^{abcd}}_{(\mu |}\nabla_eF^I_{ab}{v_c}^ev_{d|\nu)}  \right]\nonumber\\
&+\left[ 				\epsilon^{abcde} F^I_{ab} v_{c(\mu|} 
\nabla_d v_{e|\nu)} + {\epsilon^{abcd}}_{(\mu |} \nabla_e F^I_{ab} {v_c}^e 
v_{d|\nu)}   \right]\nonumber\\
&-\tfrac{4}{3}\left[ 			2F^I_{a(\mu}{v_{\nu)}}^bv_{bc}v^{ac} - 
2F^I_{ab}{v^a}_{(\mu} v_{\nu) c}v^{bc} 
-\tfrac{1}{2}g_{\mu\nu}F^I_{ab}v^{ac}v_{cd}v^{db} \right]\nonumber\\
&-\tfrac{1}{3}\left[ 			2F^{I}_{a(\mu}{v^a}_{\nu)}v_{bc}v^{bc} 
+ 2F^{Iab}v_{ab}v_{c\mu}{v^c}_\nu - 
\tfrac{1}{2}g_{\mu\nu}F^{Iab}v_{ab}v^{cd}v_{cd} \right]\nonumber\\
&+\left[				16M^Iv_{ab}{v^{b}}_{(\mu} v_{\nu) 
c}v^{ca} -2g_{\mu\nu}M^Iv_{ab}v^{bc}v_{cd}v^{da} \right]\nonumber\\
&\left.+\left[				4M^Iv_{ab}v^{ab}v_{c\mu}{v_{\nu}}^c 
+\tfrac{1}{2}g_{\mu\nu}M^Iv_{ab}v^{ab}v_{cd}v^{cd} \right] \right\} \ ,
\end{align}
\begin{align}
\tfrac{1}{\sqrt{|g|}}\frac{\delta S_{C^2}}{\delta D} &= \tfrac{c_{2I}}{144}\left\{ 
D M^I +v \cdot F^I \right\} \ , \\
\tfrac{1}{\sqrt{|g|}}\frac{\delta S_{C^2}}{\delta M^I} &= \tfrac{c_{2I}}{24} 
\left\{  \tfrac{1}{8}  C^{abcd} C_{abcd} 
 + \tfrac{1}{12}  D^2  + \tfrac{1}{3}  C_{abcd} v^{ab} v^{cd} 
+ \tfrac{8}{3}  v_{ab} \nabla^b \nabla_c v^{ac} -\tfrac{16}{9} 
v^{ab}v_{bc}R_a^{\phantom{a}c}   \right.\nonumber\\
&\phantom{\tfrac{c_{2I}}{24}}\; - \tfrac{2}{9}  v^2 R + \tfrac{4}{3} ( \nabla_a v_{bc} )(\nabla^a 
v^{bc}) +\tfrac{4}{3}  (\nabla_a v_{bc}) (\nabla^b v^{ca})  - \tfrac{2}{3} 
e^{-1}\epsilon^{abcde}v_{ab} v_{cd} \nabla^f v_{ef} 	 \nonumber\\
&\phantom{\tfrac{c_{2I}}{24}}\;\left.  +  4v_{ab} 
v^{bc} v_{cd} v^{da} -  (v^2)^2  \phantom{\tfrac{1}{1}}\right\} \ ,\\
\tfrac{1}{\sqrt{|g|}}\frac{\delta S_{C^2}}{\delta v_{\mu\nu}} & = 
\tfrac{c_{2I}}{24} \left\{ \tfrac{1}{6}DF^{I\mu\nu} + 
\tfrac{2}{3}M^IC^{\mu\nu}_{\phantom{\mu\nu}ab}v^{ab} + 
\tfrac{1}{2}C^{\mu\nu}_{\phantom{\mu\nu}ab}F^{Iab} + 
\tfrac{8}{3}M^I\nabla^{[\mu|}\nabla_av^{|\nu]a} \right. \nonumber\\
&\phantom{\tfrac{c_{2I}}{24}}\; - 
\tfrac{8}{3}\nabla^{[\mu|}\nabla_aM^Iv^{|\nu]a} +\tfrac{32}{9}M^I{v^{[\mu}}_aR^{\nu]a} - 
\tfrac{4}{9}M^IRv^{\mu\nu} -\tfrac{8}{3}\nabla_aM^I\nabla^av^{\mu\nu} 
\nonumber\\
&\phantom{\tfrac{c_{2I}}{24}}\; -\tfrac{8}{3}\nabla_aM^I\nabla^{[\mu} v^{\nu] a} 
-\tfrac{4}{3}M^I\epsilon^{\mu\nu abc}v_{ab}\nabla^dv_{cd} + 
\tfrac{2}{3}\epsilon^{abcd[\mu}\nabla^{\nu]}M^Iv_{ab}v_{cd} \nonumber\\
&\phantom{\tfrac{c_{2I}}{24}}\; + 
\tfrac{2}{3}\epsilon^{abcd[\mu}F^{I}_{ab}\nabla^{\nu]}v_{cd} 
-\tfrac{2}{3}\epsilon^{abc\mu\nu}\nabla^{d}F^{I}_{ab}v_{cd}
+ \epsilon^{abcd[\mu}F^I_{ab}\nabla_c{v_d}^{\nu]} 
+\epsilon^{abcd[\mu}\nabla_cF^I_{ab}{v_d}^{\nu]}  \nonumber\\
&\phantom{\tfrac{c_{2I}}{24}}\; \left. + 
\tfrac{8}{3}F^{I[\mu}_{\phantom{I[\mu}a}{v^{\nu]}}_bv^{ab}
-\tfrac{4}{3}F^I_{ab}v^{a\mu}v^{\nu b} -\tfrac{1}{3}v^2F^{I\mu\nu}- \tfrac{2}{3}\left( F^I\cdot v 
\right)v^{\mu\nu} -16M^Iv_{ab}v^{a\mu}v^{\nu b}  \right. \nonumber\\
&\phantom{\tfrac{c_{2I}}{24}}\; \left.- 4M^I v^2 v^{\mu\nu} \phantom{\tfrac{1}{1}}\right\} \ ,\\
\tfrac{1}{\sqrt{|g|}}\frac{\delta S_{C^2}}{\delta A^I_{\mu}} & = \tfrac{c_{2I}}{24} 
 \left\{ \tfrac{1}{16} \epsilon^{\mu abcd}  C_{abef} C_{cd}^{\phantom{cd}ef} 
-\tfrac{1}{3}\nabla_aDv^{a\mu} -\nabla_a C^{a\mu}_{\phantom{a\mu} bc}v^{bc} + 
\tfrac{4}{3} \epsilon^{\mu a bcd}\nabla_a  v_{be} \nabla^e v_{cd} \right.\nonumber\\
&\phantom{\tfrac{c_{2I}}{24}}\; \left. + 2\epsilon^{\mu a bcd }\nabla_a v_{be} 
\nabla_c v_{d}^{\phantom{d}e} +\tfrac{8}{3} \nabla_a v^{ab} v_{bc} v^{c\mu} + 
\tfrac{2}{3} \nabla_a v^{a\mu} v^2\right\} \ ,
\end{align}
where we have used the convention in the higher devivative corrections that the covariant derivative acts on all quantities to its right, unless the brackets indicate otherwise.
From the Ricci scalar squared density we obtain
\begin{align}
\tfrac{1}{\sqrt{|g|}}\frac{\delta S_{Rs^2}}{\delta D} &=   \tfrac{4}{3}\mathcal{E}D(\tfrac{2}{3}D - \tfrac{4}{3}v^2 + R) \ , \nonumber\\
\tfrac{1}{\sqrt{|g|}}\frac{\delta S_{Rs^2}}{\delta M^I} &= e_I (\tfrac{2}{3}D - \tfrac{4}{3}v^2 + R)^2 \ , \nonumber\\
\tfrac{1}{\sqrt{|g|}}\frac{\delta S_{Rs^2}}{\delta v_{\mu\nu}} & = -\tfrac{16}{3}\mathcal{E}(\tfrac{2}{3}D - \tfrac{4}{3}v^2 + R)v^{\mu\nu} \ , \nonumber\\
\tfrac{1}{\sqrt{|g|}}\frac{\delta S_{Rs^2}}{\delta A^I_{\mu}} & = 0 \ ,\nonumber\\
\tfrac{1}{\sqrt{|g|}}\frac{\delta S_{Rs^2}}{\delta g^{\mu\nu}} &= \mathcal{E} \left\{ 2(\tfrac{2}{3}D - \tfrac{4}{3}v^2 + R)(R_{\mu\nu}-\tfrac{8}{3}v_{\mu a}{v_\nu}^{a}) - \tfrac{1}{2}g_{\mu\nu}(\tfrac{2}{3}D - \tfrac{4}{3}v^2 + R)^2 \right\} \nonumber\\
&+2(\nabla_\mu\nabla_\nu - g_{\mu\nu}\nabla^2)\mathcal{E}(\tfrac{2}{3}D - \tfrac{4}{3}v^2 + R)^2 \ .
\end{align}

\section{Spinors and forms}
\label{spinors}

In this appendix, we summarize the essential information needed to realize
spinors of Spin(1,4) in terms of forms
and we review some facts about the orbits of the action of Spin(1,4)
on spinors.

\subsection{Conventions}
Let $V = \bR^4$ be a real vector space with orthonormal basis
$e^1, e^2, e^3, e^4$, and consider the subspace $U$
spanned by the first two basis vectors $e^1, e^2$. The space of Dirac spinors
is $\Delta_c = \Lambda^{\ast}(U\otimes \bC)$, with basis
$1, e^1, e^2, e^{12} = e^1 \wedge e^2$.
The gamma matrices are represented on $\Delta_c$ as
\begin{equation}
\gamma_i\eta = i(e^i\wedge\eta + e^i\lrcorner\eta)\ , \qquad
\gamma_{i+2}\eta = -e^i\wedge\eta + e^i\lrcorner\eta\ ,
\end{equation}
where $i=1,2$. $\gamma_0$ is defined by
\begin{equation}
\gamma_0 = \gamma_{1234}\ .
\end{equation}
Here,
\begin{equation}
\eta = \frac 1{k!}\eta_{j_1\ldots j_k} e^{j_1}\wedge\ldots\wedge e^{j_k}
\end{equation}
is a $k$-form and
\begin{equation}
e_i\lrcorner\eta = \frac 1{(k-1)!}\eta_{ij_1\ldots j_{k-1}} e^{j_1}\wedge\ldots
                  \wedge e^{j_{k-1}}\,.
\end{equation}
One easily checks that this representation of the gamma matrices satisfies
the Clifford algebra relations $\{\gamma_a, \gamma_b\} = 2\eta_{ab}$,
where $\eta_{ab}=\text{diag}(1,-1,-1,-1,-1)$. Note that $\gamma_0$ is Hermitian,
while $\gamma_1,\ldots,\gamma_4$ are anti-Hermitian. Moreover,
\begin{equation}
\gamma_0^T = \gamma_0\ , \qquad \gamma_i^T = \gamma_i\ , \qquad \gamma_{i+2}^T
= -\gamma_{i+2}\ . \label{gamma^T} 
\end{equation}
The Dirac, complex and charge conjugation matrices satisfy
\begin{equation}
D_{\pm}\gamma_a D_{\pm}^{-1} = \pm\gamma_a^{\dagger}\ , \qquad
B_{\pm}\gamma_a B_{\pm}^{-1} = \pm\gamma_a^{\ast}\ , \qquad
C_{\pm}\gamma_a C_{\pm}^{-1} = \pm\gamma_a^T\ . \label{conj-matr}
\end{equation}
A natural choice for the Dirac conjugation matrix is
\begin{equation}
D = i\gamma_0\ ,
\end{equation}
which corresponds to $D=D_+$ and leads to the desired (anti-)Hermiticity
properties
mentioned above. The other conjugation matrices are related to $D$ by
\begin{equation}
C_{\pm} = B^T_{\pm}D\ ,
\end{equation}
but it can be shown that in this case only $C=C_+$ and $B=B_+$ exist and are
both
antisymmetric. We take them to be
\begin{equation}
C = -\gamma_{34}\ , \qquad B = i\gamma_{12}\ ,
\end{equation}
which is compatible with \eqref{gamma^T}. The action of $B$ and $C$ on the basis
forms is
\begin{equation}
B1 = -ie^{12}\ , \qquad Be^j = i\epsilon_{jk}e^k\ , \qquad Be^{12} = i1\ ,
\end{equation}
\begin{equation}
C1 = -e^{12}\ , \qquad Ce^j = -\epsilon_{jk}e^k\ , \qquad Ce^{12} = 1\ ,
\end{equation}
where $\epsilon_{ij}=\epsilon^{ij}$ is antisymmetric with $\epsilon_{12}=1$.
Due to $B^{\ast}B=-1$, the Majorana condition $i\psi^{\dagger}\gamma^0=\psi^TC$
is inconsistent. One introduces therefore an SU$(2)$ doublet
$\psi^{\mathbf{i}}$ of spinors, and imposes the symplectic Majorana
condition $i\psi^{\mathbf{i}\dagger}\gamma^0=\epsilon_{\mathbf{ij}}\psi^{\mathbf{j}T}C$, or equivalently
\begin{equation}
\psi^{\mathbf{i}\ast} = B\epsilon_{\mathbf{ij}}\psi^{\mathbf{j}}\ . \label{majo}
\end{equation}
For an arbitrary spinor $\psi$ with first component
\begin{equation}
\psi^{\mathbf{1}} = \lambda 1 + \mu_1e^1 + \mu_2 e^2 + \sigma e^{12}\ ,
\end{equation}
where $\lambda,\mu_i$ and $\sigma$ are complex-valued functions, \eqref{majo}
implies
\begin{equation}
\psi^{\mathbf{2}} = i\sigma^{\ast}1 - i\mu_2^{\ast}e^1 + i\mu_1^{\ast}e^2 -
i\lambda^{\ast}e^{12}\ .
\end{equation}
Let us define the auxiliary inner product
\begin{equation}
\langle\alpha_i e^i,\beta_j e^j\rangle = \sum_{i=1}^2
\alpha_i^{\ast}\beta_i
\end{equation}
on $U\otimes \bC$, and then extend it to $\Delta_c$. A Spin$(1,4)$ invariant
inner product on $\Delta_c$ is then given by
\begin{equation}
{\cal B}(\zeta, \eta) = \langle C\zeta^{\ast}, \eta\rangle\,. \label{Majorana}
\end{equation}
Notice that Spin$(1,4)$ invariance of \eqref{Majorana} is equivalent to
\begin{equation}
{\cal B}(\zeta,\gamma_{ab}\eta) + {\cal B}(\gamma_{ab}\zeta,\eta) = 0\ ,
\end{equation}
which can be easily shown using \eqref{conj-matr}. 
Let us also point out that, since the pairing $\langle \cdot , \cdot \rangle$
is antilinear in its first argument, ${\cal B}(\zeta, \eta)$ is a bilinear pairing
which only depends on the spinors $\zeta$, $\eta$ 
and not their complex conjugates $\zeta^*$, $\eta^*$,
and is therefore a Majorana bilinear.
Let us use the symbol $\tilde {\mathcal B}$
to denote the paring of symplectic Majorana
spinors constructed with $\mathcal B$
by contraction of SU(2) indices,
\begin{equation} \label{SymplMajPairing}
\tilde {\cal B}(\zeta, \eta) =
\tfrac 12 \epsilon_{\mathbf {ij}}\,   \mathcal B(\zeta^{\mathbf i} , \eta^{\mathbf j})
=
 \tfrac{1}{2}\epsilon_{\mathbf{ij}}\langle C\zeta^{\mathbf{i}\ast}, \eta^{\mathbf{j}}\rangle\ .
\end{equation}
Let us record the symmetry and reality properties of this pairing,
\begin{equation}
\tilde {\cal B}(\zeta, \gamma_{a_1 \dots a_p}\eta) = 
s_G \, \tilde {\cal B}(\eta , \gamma_{a_p \dots a_1}   \zeta)
\ , \qquad
\tilde {\cal B}(\zeta, \gamma_{a_1 \dots a_p}\eta)^* =
 - \tilde {\cal B}(\eta , \gamma_{a_p \dots a_1}   \zeta) \ ,
\end{equation}
where $s_G = +1$ if the spinors are Grassmann-even,
$s_G = -1$ if they are Grassmann-odd.
We have assumed $(ab)^* = b^* a^*$
to derive the second identity.

\subsection{Review of the orbits of Spin\texorpdfstring{$(1,4)$}{(1,4)}}
We wish to simplify the task of solving the Killing spinor equations by using 
the gauge freedom Spin$(1,4)$.
There are four orbits of Spin$(1,4)$ in $\De_c$, the zero spinor which we 
disregard, two with isotropy group SU(2) and one with isotropy group $\mathbb{R}^3$.

To see this first we shall investigate the stability subgroup of the spinor 
$1$, i.e. the subgroup of $\mathrm{Spin}(1,4)$ which leaves $1,e_{12}$ 
invariant. Let
\begin{equation}
S(\lambda) := \exp \left( \tfrac{1}{2} \lambda^{ab} \Sigma_{ab}\right)
\end{equation}
be a $\mathrm{Spin}(1,4)$ transformation; it leaves $1$ invariant if and only if
\begin{equation}
\tfrac{1}{2} \lambda^{ab} \Sigma_{ab} 1 = 0 \; .
\end{equation}
Thus an element of the stability subgroup of $1$ can be written as
\begin{equation}
S(\lambda) = \exp \left( i \tfrac{\theta}{2}\vec{n} \cdot \vec{\Sigma}^{(-)} 
\right) \ ,
\end{equation}
where $\theta \in [0,4\pi]$, $\vec{n}$ is an Euclidean unit three-vector and
\begin{align}
\Sigma^{(-)}_1 &:= -\tfrac{i}{2} (\gamma_{14} + \gamma_{23}) \ , \nonumber\\
\Sigma^{(-)}_2 &:= \tfrac{i}{2} (\gamma_{12} + \gamma_{34}) \ , \nonumber\\
\Sigma^{(-)}_3 &:= -\tfrac{i}{2} (\gamma_{13} - \gamma_{24}) \; .
\end{align}
The label $(-)$ refers to the fact that these operators act non-trivially only 
on the subspace
\begin{equation}
\Delta^{(-)} := \left\{ \psi \in \Delta : \gamma_0 \psi = - \psi \right\} = 
\mathrm{span}(e_1,e_2) \ , 
\end{equation}
while they annihilate 
\begin{equation}
\Delta^{(+)} := \left\{ \psi \in \Delta : \gamma_0 \psi = \psi \right\} = 
\mathrm{span}(1,e_{12}) \; .
\end{equation}
We can represent the $\Delta = \Delta^{(+)} + \Delta^{(-)}$ decomposition by 
means of a matrix block-diagonal representation of gamma matrices and 
generators in the ordered basis $\{1, e_{12},e_1, e_2 \}$. The matrix 
representations of the Hermitian generators $\vec{\Sigma}^{(-)}$ and of the 
stability transformations turn out to be
\begin{align}
\vec{\Sigma}^{(-)} &= \left( \begin{array}{cc}
0 & 0\\
0 & \vec{\sigma} 
\end{array} \right) \ ,\nonumber\\
\exp \left( i \tfrac{\theta}{2} \vec{n} \cdot \vec{\Sigma}^{(-)} \right) &= 
\left( \begin{array}{cc}
\mathbb{I} & 0\\
0 & \cos \tfrac{\theta}{2} + i \sin \tfrac{\theta}{2} \vec{n} \cdot \vec{\sigma}
\end{array} \right) \ .
\end{align}
Thus the stability subgroup of 1 is isomorphic to SU(2). One can 
verify that this SU(2) is also the stability subgroup of $e_{12}$. 

Similarly acting on $e^1$ we find
\begin{align}
\Sigma^{(+)}_1 &:= -\tfrac{i}{2} (\gamma_{23} - \gamma_{14}) \ , \nonumber\\
\Sigma^{(+)}_2 &:= -\tfrac{i}{2} (\gamma_{12} - \gamma_{34}) \ , \nonumber\\
\Sigma^{(+)}_3 &:= \tfrac{i}{2} (\gamma_{13} + \gamma_{24}) \ ,
\end{align}
and we obtain another SU(2);
\begin{align}
\vec{\Sigma}^{(+)} &= \left( \begin{array}{cc}
\vec{\sigma} & 0\\
0 & 0 
\end{array} \right) \ ,\nonumber\\
\exp \left( i \tfrac{\theta}{2} \vec{n} \cdot \vec{\Sigma}^{(+)} \right) &= 
\left( \begin{array}{cc}
\cos \tfrac{\theta}{2} + i \sin \tfrac{\theta}{2} \vec{n} \cdot \vec{\sigma} & 
0\\
0 & \mathbb{I}
\end{array} \right) \; .
\end{align}
This SU(2) is also the stability subgroup of $e_2$.

It is evident from their block-diagonal form that these SU(2)-isomorphic 
subgroups of $\mathrm{Spin}(1,4)$ commute, thus we have an explicit 
representation of the well known isomorphism
\begin{equation}
\mathrm{Spin}(4) \cong \mathrm{SU}(2) \times \mathrm{SU}(2) \; .
\end{equation}

Now
let $\mathrm{SU}(2)$ act on $\mathbb{C}^2$ in the fundamental representation 
and let us write $z \sim z' $ if $z,z' \in \mathbb{C}$ lie in the same orbit. We 
then have 
\begin{equation}
\left( \begin{array}{c}
z_1 \\
z_2
\end{array} \right) \sim \left( \begin{array}{c}
\sqrt{|z_1|^2 + |z_2|^2} \\
0
\end{array} \right) \; \qquad \forall z_1, z_2 \in \mathbb{C}.
\end{equation}
To see this note that
the following identity holds for $\beta , \theta, \alpha \in \mathbb{R}$ and 
$\lambda \ge 0$:
\begin{equation}
e^{i\beta \sigma_3} e^{i \theta \sigma_1} e^{i\alpha \sigma_3} \left( 
\begin{array}{c}
\lambda\\
0
\end{array} \right) = \left( \begin{array}{c}
\lambda \cos\theta e^{i(\alpha+\beta)} \\
\lambda \sin\theta e^{i(\alpha - \beta +\tfrac{\pi}{2})}
\end{array} \right) \; .
\end{equation}
On the right hand side we can recognize the general element of $\mathbb{C}^2$ 
satisfying $|z_1|^2 + |z_2|^2 = \lambda^2$. 

Thus we can conclude that given 
\begin{equation}
\psi = z1 + we_{12} + z^1 e_1 + z^2 e_2 \in \Delta \ ,
\end{equation}
we are always able to perform a $\mathrm{Spin}(1,4)$ transformation which 
carries $\psi$ to
\begin{equation}
\psi' = \lambda_\psi 1 + \mu_\psi e_1 \ ,
\end{equation}
where
\begin{equation}
\lambda_\psi := \sqrt{|z|^2 + |w|^2} \;,\quad \mu_\psi := \sqrt{|s|^2+|t|^2} \;  
.
\end{equation} 
Hence there will be no loss in generality restricting to $\psi = \lambda 1 + 
\mu e_1$ with $\lambda,\mu \ge 0$ in the following.

Let us now act on $\psi$ with a Lorentz boost generated by $\gamma_{03}$:
$$
\exp\left(x \gamma_{03} \right) \psi= (\lambda \cosh x + \mu \sinh x)1 + 
(\lambda \sinh x + \mu \cosh x) e_1 =: \lambda'(x) 1 + \mu'(x) e_1 \; .
$$
Four cases are possible:
\begin{itemize}
\item $\lambda = \mu = 0: $\\
$\psi$ is the zero spinor and constitutes an orbit of its own;
\item $\lambda = \mu >0: $ \\
we have $\lambda'(x) = \mu'(x) = \lambda e^x$ and hence we can always set 
$\lambda'(x) = \mu'(x) = 1$ by choosing $x =- \log \lambda$;
\item $\lambda > \mu:$ \\
under this assumption equation $\mu'(x)=0$ has exactly one root given by
$$
x_0 = - \mathrm{arctanh} \frac{\mu}{\lambda} \; ;
$$
one has $\lambda'(x_0) = \sqrt{\lambda^2 - \mu^2}$; 
\item $\lambda < \mu:$ \\
under this assumption equation $\lambda'(x)=0$ has exactly one root given by
$$
x_0 = - \mathrm{arctanh} \frac{\lambda}{\mu} \; ;
$$
one has $\mu'(x_0) = \sqrt{\mu^2 - \lambda^2}$ .
\end{itemize}

To summarize we have the following:

Let $\mathrm{Spin}(1,4)$ act on $\Delta$ and let us write $\psi \sim \psi' $ if 
$\psi,\psi' \in \Delta$ lie in the same orbit. Given $\psi = z1 + we_{12} + z^1 
e_1 + z^2 e_2$,
\begin{align*}
&\text{if} & & |z|^2+|w|^2=|z^1|^2+|z^2|^2 = 0 & &\text{then} & &\psi=0 \; ,\\
&\text{if} & & |z|^2+|w|^2=|z^1|^2+|z^2|^2 > 0 & &\text{then} & &\psi \sim 1+ 
e_1 \; ,\\ 
&\text{if} & & |z|^2+|w|^2>|z^1|^2+|z^2|^2 & &\text{then} & &\psi \sim 
1\sqrt{|z|^2+|w|^2-|z^1|^2-|z^2|^2} \; , \\
&\text{if} & & |z|^2+|w|^2<|z^1|^2+|z^2|^2 & &\text{then} & &\psi \sim 
e_1\sqrt{|z^1|^2+|z^2|^2-|z|^2-|w|^2} \; .
\end{align*}

As a consequence, in order to study Killing spinor equations we will be able to 
set the Killing spinor equal to $e^{\phi(x)} 1$, $e^{\phi(x)} e_1$ and $1+e_1$ 
in turn exhausting all inequivalent possibilities under local Lorentz 
transformations.

It remains to find the stability subgroup of $1+e_1$. Examining 
\begin{equation}
\tfrac{1}{2} \lambda^{ab} \Sigma_{ab}(1+e_1) = 0 \ ,
\end{equation}
we see that the stability subgroup of $1+e_1$ is generated by
\begin{eqnarray}
X &:=& \gamma_{34} - \gamma_{04} \ ,\nonumber\\
Y &:=& \gamma_{13} + \gamma_{01} \ ,\nonumber\\
Z &:=& \gamma_{23} + \gamma_{02} \ .
\end{eqnarray}
which satisfy
\begin{equation}
X^2=Y^2=Z^2=XY=YX=YZ=ZY=XZ=ZX=0 \; .
\end{equation}
We see that for $\mu,\nu,\rho \in \mathbb{R}$,
\begin{equation}
\exp(\mu X + \nu Y + \rho Z) = 1+ \mu X + \nu Y + \rho Z \ ,
\end{equation}
and so the stability subgroup of $1+e_1$ is isomorphic to the Abelian additive 
group $\mathbb{R}^3$. Note that this is also the stability subgroup of 
$(e^2-e^{12})$.

We may therefore always choose, up to a Spin$(1,4)$ transformation, the first component of the first
Killing spinor to be
\begin{equation}
\epsilon = (e^{\phi} 1,-ie^{\phi} e^{12}) \label{SU(2)-rep1} \ ,
\end{equation}
or
\begin{equation}
\epsilon=(e^{\phi} e^1,ie^{\phi} e^2) \ ,
\end{equation}
which have stability subgroup SU(2),
or
\begin{equation}
\ep=((1+e^1),i(e^2-e^{12})) \ ,
\end{equation}
with stability subgroup $\mathbb{R}^3$.

Consider the two different SU(2) orbits. They are not related by a 
Spin$^0(1,4)$ transformation,
 the connected to the identity component of Spin$(1,4)$.
 Instead they are related by a Pin$(4)$ transformation followed by an SU(2)$\subset$Spin$(1,4)$ transformation
\begin{equation}
\frac{1}{2}\left(\ga_{13} + \ga_{24}\right) \gamma_1 (e^{\phi}e^1,ie^{\phi}e^2) 
= (e^{\phi}1,-ie^{\phi}e^{12}) .
\end{equation} 

Spin$^0(1,4)$ transformations are those that project onto proper orthochronous 
Lorentz rotations of the frame,
 SO$(1,4)^+$.
Note that Pin$(4)$ is generated by $\ga_i$, where $i=1,\cdots,4$, 
and is associated with a spatial reflection. Indeed the Pin$(4)$ transformation
\begin{eqnarray}
\ep &\rightarrow& \ga_1\ep \ ,\nonumber\\
\ga_\mu &\rightarrow&  \ga_1 \ga_\mu (\ga_1)^{-1} \ ,
\end{eqnarray}
acts on the gamma matrices as 
\begin{equation}
 \ga_0 \rightarrow -\ga_0, \; \ga_1 \rightarrow \ga_1, \; \ga_2 \rightarrow  
-\ga_2, \; \ga_3 \rightarrow -\ga_3, \; \ga_4 \rightarrow -\ga_4 \ .
\end{equation}
Note that this preserves $C$ but changes the sign of $B$ and $D$. 
Hence we will consider the two representatives 
$\ep=(e^{\phi}1,-ie^{\phi}e^{12})$ and $\ep=(e^{\phi}e^1,ie^{\phi}e^2)$ to be 
equivalent, up to local orthogonal transformations. Given this, we will focus
 on the representative $e^{\phi}1$, however for completeness we will give the 
conditions arising from choosing a Killing spinor in the second 
orbit.

\subsection{Useful bases for \texorpdfstring{SU$(2)$}{SU(2)} and \texorpdfstring{$\mathbb R^3$}{R3} orbits}
In the case of the SU$(2)$ orbits, it will prove useful to work in an oscillator
basis of gamma matrices, defined by
\begin{equation}
\Gamma_{\alpha} = \frac1{\sqrt2}(\gamma_{\alpha+2} + i\gamma_{\alpha})\ , \qquad
\Gamma_{\bar\alpha} = \frac1{\sqrt2}(-\gamma_{\alpha+2} + i\gamma_{\alpha})\ ,
\qquad \alpha = 1,2\ .\label{tlbasis}
\end{equation}
Furthermore, let us define $\Gamma_0=\gamma_0$. Note that
$\Gamma_{\alpha}^{\dagger}=\Gamma_{\bar\alpha}$.
The Clifford algebra relations in this  basis are
$\{\Gamma_{\alpha},\Gamma_{\bar\beta}\} = 2g_{\alpha\bar\beta}$ and
$\{\Gamma_{\alpha},\Gamma_{\beta}\}=\{\Gamma_{\bar\alpha},\Gamma_{\bar\beta}\}
=0$,
where the nonvanishing components of the hermitian metric $g_{\alpha\bar\beta}$
read
$g_{1\bar1} = g_{\bar11} = g_{2\bar2} = g_{\bar22} = 1$.
The spinor 1 is a Clifford vacuum, $\Gamma_{\bar1}1 = \Gamma_{\bar2}1 = 0$, 
and the representation $\Delta_c$ can be constructed by acting on 1 with the
creation operators $\Gamma_1,\Gamma_2$.
The action of the new
gamma matrices and the Spin$(1,4)$ generators on the basis spinors is summarized
in table \ref{tab:gamma}.

\begin{table}[ht]
\begin{center}
\begin{tabular}{|c||c|c|c|c|}
\hline
& 1 & $e^1$ & $e^2$ & $e^{12}$\\
\hline\hline
$\Gamma_0$ &1 & $-e^1$ & $-e^2$ & $e^{12}$\\
\hline
$\Gamma_1$ & $-\sqrt2 e^1$ & 0 & $-\sqrt2 e^{12}$ & 0\\
\hline
$\Gamma_{\bar 1}$ & 0 & $-\sqrt2$ & 0 & $-\sqrt2 e^2$\\
\hline
$\Gamma_2$ & $-\sqrt2 e^2$ & $\sqrt2 e^{12}$ & 0 & 0\\
\hline
$\Gamma_{\bar 2}$ & 0 & 0 & $-\sqrt2$ & $\sqrt2 e^1$\\
\hline\hline
$\Gamma_{01}$ & $\sqrt2 e^1$ & 0 & $-\sqrt2 e^{12}$ & 0\\
\hline
$\Gamma_{0\bar 1}$ & 0 & $-\sqrt2$ & 0 & $\sqrt2 e^2$\\
\hline
$\Gamma_{02}$ & $\sqrt2 e^2$ & $\sqrt2 e^{12}$ & 0 & 0\\
\hline
$\Gamma_{0\bar 2}$ & 0 & 0 & $-\sqrt2$ & $-\sqrt2 e^1$\\
\hline
$\Gamma_{1\bar 1}$ & $-1$ & $e^1$ & $-e^2$ & $e^{12}$\\
\hline
$\Gamma_{12}$ & $2e^{12}$ & 0 & 0 & 0\\
\hline
$\Gamma_{1\bar 2}$ & 0 & 0 & $2e^1$ & 0\\
\hline
$\Gamma_{\bar 12}$ & 0 & $-2e^2$ & 0 & 0\\
\hline
$\Gamma_{\bar 1\bar 2}$ & 0 & 0 & 0 & $-2$\\
\hline
$\Gamma_{2\bar 2}$ & $-1$ & $-e^1$ & $e^2$ & $e^{12}$\\
\hline
\end{tabular}
\end{center}
\caption{The action of the gamma matrices and the Spin$(1,4)$ generators
on the different basis elements. \label{tab:gamma}}
\end{table}
The bilinears of section \ref{halftl} 
are built with the pairings $\mathcal B$, $\tilde {\mathcal B}$ 
introduced in \eqref{Majorana}, \eqref{SymplMajPairing} starting from the spinor
$\epsilon^{\mathbf i}$ specified in \eqref{SU(2)-rep1}.
More explicitly, treating $\epsilon^{\mathbf i}$ as Grassmann even,
one finds
\begin{align}
e^{2\phi} &= -i \, \tilde {\mathcal B}(\epsilon, \epsilon) \ , \quad
V = e^{2\phi} e^0 = - i \, \tilde {\mathcal B}(\epsilon , \Gamma^0 \epsilon)
\nonumber \\
X^{(1)} &= - e^{2\phi } (e^1 \wedge e^2 + e^{\bar 1 } \wedge e^{\bar 2})
= \tfrac 14 \mathcal B( \epsilon^{\mathbf 1} , \gamma_{\mu\nu} \epsilon^{\mathbf 1}) e^\mu \wedge e^\nu - \tfrac 14 
 \mathcal B( \epsilon^{\mathbf 2} , \gamma_{\mu\nu} \epsilon^{\mathbf 2})
 e^\mu \wedge e^\nu
\nonumber \\
X^{(2)} &= - i e^{2\phi} (e^1 \wedge e^2 - e^{\overline{1}} \wedge e^{\overline{2}}) =
 \tfrac i4 \mathcal B( \epsilon^{\mathbf 1} , \gamma_{\mu\nu} \epsilon^{\mathbf 1}) e^\mu \wedge e^\nu + \tfrac i4 
 \mathcal B( \epsilon^{\mathbf 2} , \gamma_{\mu\nu} \epsilon^{\mathbf 2})
 e^\mu \wedge e^\nu
\nonumber \\
X^{(3)} & =  -ie^{2\phi} (e^1 \wedge e^{\overline{1}} + e^{2} \wedge 
e^{\overline{2}})
= - \mathcal B( \epsilon^{\mathbf 1} , \gamma_{\mu\nu} \epsilon^{\mathbf 2}) e^\mu \wedge e^\nu
 \ ,
\end{align}
where $\mu$, $\nu$ are five-dimensional spacetime indices,
and $\{ e^0, e^1, e^2, e^{\bar 1}, e^{\bar 2}\}$ is a f\"unfbein
adapted to the oscillator basis of gamma matrices $\{ \Gamma_0, \Gamma_1, \Gamma_2,
\Gamma_{\bar 1}, \Gamma_{\bar 2} \}$
constructed above.

For the orbit with stabilizer $\mathbb{R}^3$ we will use the basis
\begin{eqnarray}
\Gamma_{\pm} &:=& \tfrac{1}{\sqrt{2}} (\gamma_0 \pm \gamma_3) \ ,\nonumber\\
\Gamma_1 &:=& - \gamma_4 \ ,\nonumber\\ 
\Gamma_2 &:=& - \gamma_2 \ ,\nonumber\\
\Gamma_3 &:=& - \gamma_1 \; .\label{nullbasis}
\end{eqnarray}
where we have
$
\epsilon_{-+123} = +1 \; .
$

The associated (real) f\"{u}nfbein turns out to be
\begin{align}
E^{\pm} &= \tfrac{1}{\sqrt{2}} (e^0 \pm e^3)\ , &
E^1 &= - e^4 \ ,& 
E^2 &= - e^2 \ ,&
E^3 &= - e^1\ .
\end{align}
 The new form of the flat metric is
\begin{equation}
\eta_{AB} = \left( \begin{array}{ccccc}
0 & 1 & & & \\
1 & 0 & & & \\
& & -1 & & \\
& & & -1 & \\
& & & & -1 
\end{array} \right)= \eta^{AB} \ , \qquad A,B=-,+,1,2,3 \; .
\end{equation}

It will be convenient to write the spinors in the basis
\begin{equation}
\{ 1 + e^1, e^{12} - e^2, 1-e^1, e^{12} + e^2 \} \; ,
\end{equation}
 with the first component of a generic spinor written as
\begin{equation}
\epsilon^{\mathbf{1}}= z_1 (1 + e^1) + z_2(e^{12} - e^2) + z_3(1-e^1) + z_4(e^{12} + e^2) 
\; , \label{gennullspinor}
\end{equation}
where the $z_i$ are complex spacetime functions. The symplectic-Majorana 
conjugate of this spinor is
\begin{equation}
\epsilon^{\mathbf{2}} = i z_2^* (1 + e^1) - i z_1^*(e^{12} - e^2) +  i z_4^* (1 - e^1) - 
iz_3^*(e^{12} + e^2) \ .\label{nullmajcong}
\end{equation}
The action of the new
gamma matrices and the Spin$(1,4)$ generators on these basis spinors is 
summarized
in table \ref{tab:gammaN}.
\begin{table}[ht]
\begin{center}
\begin{tabular}{|c||c|c|c|c|}
\hline
& $(1+e^1)$ & $(e^{12}-e^2)$ & $(1-e^1)$& $(e^{12}+e^2)$ \\
\hline\hline
$\Gamma_-$ &0 & 0 & $\sqrt{2}(1+e^1)$ & $\sqrt{2}(e^{12}-e^2)$\\
\hline
$\Gamma_+$ &$\sqrt{2}(1-e^1)$ & $\sqrt{2}(e^{12}+e^2)$ & 0 & 0\\
\hline
$\Gamma_1$ & $-(e^{12}-e^2)$ & $(1+e^1)$ & $(e^{12}+e^2)$ & $-(1-e^1)$ \\
\hline
$\Gamma_2$ & $i(e^{12}-e^2)$ & $i(1+e^1)$ & $-i(e^{12}+e^2)$ &$-i(1-e^1)$\\
\hline
$\Gamma_3$ & $-i(1+e^1)$ & $i(e^{12}-e^2)$ & $i(1-e^1)$& $-i(e^{12}+e^2)$ \\
\hline\hline
$\Gamma_{-+}$ & $(1+e^1)$ & $(e^{12}-e^2)$ & $-(1-e^1)$& $-(e^{12}+e^2)$ \\
\hline
$\Gamma_{-1}$ & 0 & 0 &$\sqrt{2}(e^{12}-e^2)$  & $-\sqrt{2}(1+e^1)$\\
\hline
$\Gamma_{-2}$ & 0 & 0 & $-i\sqrt{2}(e^{12}-e^2)$ & $-i\sqrt{2}(1+e^1)$\\
\hline
$\Gamma_{-3}$ & 0 & 0 & $i\sqrt{2}(1+e^1)$ &$-i\sqrt{2}(e^{12}-e^2)$ \\
\hline
$\Gamma_{+1}$ & $-\sqrt{2}(e^{12}+e^2)$ & $\sqrt{2}(1-e^1)$ & 0 & 0\\
\hline
$\Gamma_{+2}$ & $i\sqrt{2}(e^{12}+e^2)$ & $i\sqrt{2}(1-e^1)$ & 0 & 0\\
\hline
$\Gamma_{+3}$ & $-i\sqrt{2}(1-e^1)$ & $i\sqrt{2}(e^{12}+e^2)$ & 0 & 0\\
\hline
$\Gamma_{12}$ & $i(1+e^1)$ & $-i(e^{12}-e^2)$ & $i(1-e^1)$ & $-i(e^{12}+e^2)$\\
\hline
$\Gamma_{13}$ & $i(e^{12}-e^2)$ & $i(1+e^1)$& $i(e^{12}+e^2)$ & $i(1-e^1)$\\
\hline
$\Gamma_{23}$ & $(e^{12}-e^2)$ & $-(1+e^1)$  &  $(e^{12}+e^2)$ & $-(1-e^1)$\\
\hline
\end{tabular}
\end{center}
\caption{The action of the gamma matrices and the Spin$(1,4)$ generators
on the different basis elements. \label{tab:gammaN}}
\end{table}

\section{Killing spinor equations in a time-like basis}\label{tlsystem}

\subsection*{Gravitino equation}

Demanding the vanishing of the gravitino variation for a bosonic background 
implies
\begin{equation}
\delta \psi_\mu^{\mathbf{i}} = \left[\nabla_\mu + \frac{1}{2} v^{ab} \gamma_{\mu ab} - 
\frac{1}{3} v^{ab} \gamma_\mu \gamma_{ab} \right]\epsilon^{\mathbf{i}} =0\ .
\end{equation}
Focusing on the first symplectic Majorana component and making use of the 
identities
\begin{equation}
\gamma_a \gamma_{bc} = \eta_{ab} \gamma_c - \eta_{ac} \gamma_b + \gamma_{abc}\ 
, \qquad
\gamma_{abc} = -\frac{1}{2} \epsilon_{abcde} \gamma^{de}\ , \label{gamma-id}
\end{equation}
one gets
\begin{eqnarray}
\left[\partial_0 -\tfrac{2}{3}{v_0}^i\ga_i 
-\tfrac{1}{2}\om_{0,0}^{\phantom{0,0}i}\ga_i\ga_0 + 
(\tfrac{1}{4}\om_{0,}^{\phantom{0,}ij} - \tfrac{1}{6}v_{(+)}^{ij} + 
\tfrac{1}{6}v_{(-)}^{ij})\ga_{ij}\right]\epsilon&=&0\ , \nonumber\\
\left[\partial_i + \tfrac{2}{3}v_{0i}\ga_0  -\tfrac{2}{3}{v_i}^j\ga_j - 
(\tfrac{1}{2}\om_{i,0}^{\phantom{i,0}j} +\tfrac{1}{3}v^{(+)j}_i 
-\tfrac{1}{3}v^{(-)j}_i)\ga_j\ga_0 + 
\tfrac{1}{4}\om_{i,}^{\phantom{i,}jk}\ga_{jk} + 
\tfrac{1}{6}{v_0}^j\epsilon_{ijkl}\ga^{kl}\right]\epsilon&=&0\ , \nonumber\\
\end{eqnarray}
where we defined $\omega_{a,bc}=e^{\mu}_a\omega_{\mu, bc}$.
Decomposing this in the time-like oscillator basis for a generic spinor,
\begin{equation}
\ep = \la 1 + \mu_1 e^1 +\mu_2 e^2 + \sigma e^{12}\ ,
\end{equation}
we obtain the linear system
\begin{eqnarray}
 \partial_0 \lambda - \lambda\left( \tfrac{1}{2}{{\omega_0,}^\gamma}_\gamma +
\tfrac{1}{3} {v^\gamma}_\gamma \right)
 -\tfrac{\mu_1}{\sqrt{2}} \left( \omega_{0,01} - \tfrac{4}{3}v_{01}
\right) &&\nonumber\\
 -\tfrac{\mu_2}{\sqrt{2}} \left( \omega_{0,02} -\tfrac{4}{3}v_{02}
\right)
 -\sigma \left( \omega_{0,12} + \tfrac{2}{3}v_{12} \right) &=& 0\ , \nonumber \\
 -\lambda\left( \tfrac{1}{2} \omega_{0,0\bar{1}} + \tfrac{2}{3} v_{0\bar{1}} 
\right)
 -\tfrac{\partial_0 \mu_1}{\sqrt{2}}
 +\tfrac{\mu_1}{\sqrt{2}} \left( \tfrac{1}{2} \left( \omega_{0,1\bar{1}}
-\omega_{0,2\bar{2}}  \right) - \tfrac{1}{3} \left( v_{1\bar{1}} - v_{2\bar{2}}
\right)  \right) &&\nonumber\\
 -\tfrac{\mu_2}{\sqrt{2}}\left( \omega_{0,\bar{1}2} - \tfrac{2}{3}v_{\bar{1}2}
\right)
 +\sigma \left( \tfrac{1}{2}\omega_{0,02} + \tfrac{2}{3} v_{02} \right) &=&0\ , 
\nonumber \\
 -\lambda\left( \tfrac{1}{2}\omega_{0,0\bar{2}} + \tfrac{2}{3} v_{0\bar{2}} 
\right)
 +\tfrac{\mu_1}{\sqrt{2}} \left( \omega_{0,1\bar{2}} -\tfrac{2}{3}v_{1\bar{2}}
\right)
 -\tfrac{\partial_0\mu_2}{\sqrt{2}} &&\nonumber\\
 -\tfrac{\mu_2}{\sqrt{2}} \left( \tfrac{1}{2} \left( \omega_{0,1\bar{1}} -
\omega_{0,2\bar{2}} \right) - \tfrac{1}{3} \left( v_{1\bar{1}} -v_{2\bar{2}}
\right)    \right)
 -\sigma \left( \tfrac{1}{2}\omega_{0,01} + \tfrac{2}{3} v_{01} \right) &=& 0\ 
, \nonumber \\
 \lambda \left( \tfrac{1}{2} \omega_{0,\bar{1}\bar{2}} +
\tfrac{1}{3}v_{\bar{1}\bar{2}} \right)
 +\tfrac{\mu_1}{\sqrt{2}} \left( \tfrac{1}{2}\omega_{0,0\bar{2}} - 
\tfrac{2}{3}v_{0\bar{2}}
\right) &&\nonumber\\
 +\tfrac{\mu_2}{\sqrt{2}} \left( -\tfrac{1}{2}\omega_{0,0\bar{1}} + 
\tfrac{2}{3}v_{0\bar{1}}
\right)
 +\tfrac{\partial_0 \sigma}{2} + \sigma \left(
\tfrac{1}{4}{{\omega_{0,}}^\gamma}_\gamma + \tfrac{1}{6}{v^\gamma}_\gamma
\right)&=&0\ , 
 \end{eqnarray}
 \begin{eqnarray}
 \partial_\alpha \lambda 
 - \lambda \left( \tfrac{1}{2} {{\omega_{\alpha,}}^\gamma}_\gamma -  
v_{0\alpha} 
\right)
 - \tfrac{\mu_1}{\sqrt{2}} \left( \omega_{\alpha,01} +2 \delta_{\alpha2}v_{12}
\right) &&\nonumber\\
 -\tfrac{\mu_2}{\sqrt{2}} \left( \omega_{\alpha,02} - 2 \delta_{1\alpha} v_{12}
\right)
 -\sigma{\omega_{\alpha,12}}&=&0\ , \nonumber \\
 -\lambda\left( \tfrac{1}{2}\omega_{\alpha,0\bar{1}} +
\tfrac{1}{3}\delta_{1\alpha}\left( 2 v_{1\bar{1}} - v_{2\bar{2}} \right)
-\delta_{2\alpha}v_{\bar{1}2} \right) \nonumber\\
 -\tfrac{\partial_\alpha\mu_1}{\sqrt{2}} +\tfrac{\mu_1}{\sqrt{2}} \left(
\tfrac{1}{2}\left( \omega_{\alpha,1\bar{1}} - \omega_{\alpha,2\bar{2}} \right) +
\tfrac{1}{3}\delta_{1\alpha}v_{01} + \delta_{\alpha2}v_{02} \right) 
&&\nonumber\\
 -\tfrac{\mu_2}{\sqrt{2}} \left( \omega_{\alpha,\bar{1}2} +
\tfrac{2}{3}\delta_{\alpha1}v_{02} \right)
 +\sigma \left( \tfrac{1}{2}\omega_{\alpha,02} +\tfrac{1}{3}v_{\alpha 2} \right)
&=&0\ , \nonumber \\
  -\lambda\left(\tfrac{1}{2}\omega_{\alpha,0\bar2}+\delta_{\alpha 1}v_{1\bar{2}}
-\tfrac{1}{3}\delta_{\alpha 2}\left(v_{1\bar{1}} -2v_{2\bar{2}}  \right) \right)
+\tfrac{\mu_1}{\sqrt{2}} \left( \omega_{\alpha,1\bar{2}} -
\tfrac{2}{3}\delta_{\alpha 2}v_{01}  \right) &&\nonumber\\
-\tfrac{\partial_\alpha \mu_2}{\sqrt{2}}
+\tfrac{\mu_2}{\sqrt{2}} \left(-\tfrac{1}{2}\left(\omega_{\alpha,1\bar{1}} -
\omega_{\alpha,2\bar{2}}  \right) +\delta_{\alpha 1}v_{01} +
\tfrac{1}{3}\delta_{\alpha 2}v_{02} \right)
-\sigma\left( \tfrac{1}{2}\omega_{\alpha, 01} - \tfrac{1}{3}v_{1\alpha} \right) 
&=& 
0\ , \nonumber \\
 \lambda\left( \tfrac{1}{2}\omega_{\alpha, \bar{1}\bar{2}} -
\tfrac{1}{3}\epsilon_{\alpha\beta} {v_{0}}^\beta \right)
+ \tfrac{\mu_1}{\sqrt{2}} \left( \tfrac{1}{2}\omega_{\alpha, 0 \bar{2}} -
\tfrac{1}{3}\delta_{\alpha1}v_{1\bar{2}} - \tfrac{1}{3}\delta_{\alpha 2} \left(
v_{1\bar{1}} +2v_{2\bar{2}}  \right) \right)
&&\nonumber\\
- \tfrac{\mu_2}{\sqrt{2}} \left( \tfrac{1}{2}\omega_{\alpha, 0 \bar{1}} -
\tfrac{1}{3}\delta_{\alpha1} \left( 2v_{1\bar{1}} +v_{2\bar{2}}  \right)
+\tfrac{1}{3}\delta_{\alpha 2}v_{\bar{1}2}\right)
+\tfrac{\partial_\alpha \sigma}{2} + \sigma\left(
\tfrac{1}{4}{{\omega_{\alpha,}}^{\gamma}}_\gamma  + 
\tfrac{1}{2}v_{0\alpha}\right)
&=&0\ , 
\end{eqnarray}
\begin{eqnarray}
 \partial_{\bar{\alpha}}\lambda 
+ \lambda\left( - \tfrac{1}{2}{{\omega_{\bar{\alpha},}}^\gamma}_\gamma +
\tfrac{1}{3}v_{0\bar{\alpha}} \right)
+ \tfrac{\mu_1}{\sqrt{2}}\left( -\omega_{\bar{\alpha},01} -
\tfrac{2}{3}\delta_{\bal \bar1}\left( 2v_{1\bar1} + v_{2\bar2} \right) -
\tfrac{2}{3}\de_{\bal \bar 2}v_{1\bar2} \right) &&\nonumber\\
+ \tfrac{\mu_2}{\sqrt{2}}\left( -\omega_{\bar{\alpha},02} +
\tfrac{2}{3}\delta_{\bal \bar1}v_{\bar1 2} - \tfrac{2}{3}\de_{\bal \bar2} \left(
v_{1\bar1} + 2v_{2\bar2}\right)\right)  
+\si \left( -\omega_{\bal,12} + \tfrac{2}{3}\ep_{\bal \bga} {v_{0}}^{\bga} 
\right)
&=&0\ , \nonumber \\
\la \left( -\tfrac{1}{2}\om_{\bal, 0 \bar1}+\tfrac{1}{3} v_{\bar1 \bal} \right) 
-\tfrac{\partial_{\bal} \mu_1}{\sqrt2} &&\nonumber\\
+\tfrac{\mu_1}{\sqrt2}\left( \tfrac{1}{2}\left(\om_{\bal, 1\bar1}-\om_{\bal,
2\bar2} \right) + \de_{\bal\bar1}v_{0\bar1}
+\tfrac{1}{3}\de_{\bal\bar2}v_{0\bar2}\right)
+\tfrac{\mu_2}{\sqrt2}\left( -\om_{\bal, \bar1 2} +\tfrac{2}{3}
\de_{\bal\bar2}v_{0\bar1}\right) &&\nonumber\\
+\si\left( \tfrac{1}{2}\om_{\bal_,02} + \de_{\bal \bar1}v_{\bar1 2} +
\tfrac{1}{3}\de_{\bal\bar2}\left( v_{1\bar1}-2v_{2\bar2} \right) \right)&=&0\ , 
\nonumber \\
-\la\left( \tfrac{1}{2}\om_{\bal,0\bar2} + \tfrac{1}{3} v_{\bal \bar2} \right)
+\tfrac{\mu_1}{\sqrt2}\left( \om_{\bal, 1\bar2} +
\tfrac{2}{3}\de_{\bal\bar1}v_{0\bar2} \right) &&\nonumber\\
-\tfrac{\partial_{\bal} \mu_2}{\sqrt2} + \tfrac{\mu_2}{\sqrt2}\left(
-\tfrac{1}{2}\left( \om_{\bal, 1 \bar1}- \om_{\bal, 2\bar2}\right) +
\tfrac{1}{3}\de_{\bal \bar1}v_{0\bar1}+ \de_{\bal\bar2}v_{0\bar2}  \right)
&&\nonumber\\
+\si \left( -\tfrac{1}{2}\om_{\bal,01} + \tfrac{1}{3}\de_{\bal\bar1}\left(
2v_{1\bar1} - v_{2\bar2} \right) + \de_{\bal \bar2}v_{1\bar2} \right)&=&0\ , 
\nonumber \\
\tfrac{\la}{2}\om_{\bal, \bar1 \bar2} + \tfrac{\mu_1}{\sqrt2}\left( 
\tfrac{1}{2}\om_{\bal, 0\bar2} - \de_{\bal \bar1}v_{\bar1 \bar2} \right)
-\tfrac{\mu_2}{\sqrt2}\left(\tfrac{1}{2}\om_{\bal, 
0\bar1}+\de_{\bal\bar2}v_{\bar1 \bar2} \right) &&\nonumber\\
+\tfrac{1}{2}\partial_{\bal}\si + \si\left( \tfrac{1}{4}{{\om_{\bal}}^\ga}_\ga 
+ \tfrac{1}{2}v_{0\bal}\right)&=&0 \ .
\label{lin-sys-time-grav}
\end{eqnarray}
Notice that taking the dual of the complex conjugate of this system, we obtain 
the system for the symplectic Majorana conjugate of $\epsilon$.
This implies that if a spinor $\epsilon$ solves the gravitino equation, then so 
does its symplectic Majorana conjugate.

\subsection*{Gaugino equation}

From the vanishing of the gaugino variation for a bosonic background one has
\begin{equation}
\delta \Omega^{I{\mathbf{i}}} = \left[-\frac{1}{4} F^I_{ab} \gamma^{ab} - \frac{1}{2} 
\gamma^\mu \partial_\mu M^I - \frac{1}{3}M^I v^{ab} \gamma_{ab} 
\right]\epsilon^{\mathbf{i}} = 0\ .
\end{equation}
Defining
\begin{equation}
\cF^{I ab} = \tfrac{1}{4}F^{I\;ab} + \tfrac{1}{3}M^I v^{ab}\ , \label{curlyF}
\end{equation}
and expanding in the oscillator basis we obtain
\begin{eqnarray}
&&\la \left( \tfrac{1}{2}\partial_0 M^I - 2{\cF^{I\al}}_\al \right)
-\tfrac{\mu_1}{\sqrt2}\left( \partial_1 M^I + 4\cF^{I}_{01} \right)
-\tfrac{\mu_2}{\sqrt2}\left( \partial_2 M^I + 4\cF^{I}_{02} \right)
-4\si \cF^I_{12} = 0\ , \nonumber \\
&&\la\left( \tfrac{1}{2}\partial_{\bar1}M^I - 2\cF^I_{0\bar1} \right)
+\tfrac{\mu_1}{\sqrt2}\left(\tfrac{1}{2}\partial_0 M^I + 2\left(\cF^I_{1\bar1} 
- \cF^I_{2\bar2}\right)\right)
-\tfrac{4\mu_2}{\sqrt2}\cF^I_{\bar1 2} + \si 
\left(2\cF^I_{02}-\tfrac{1}{2}\partial_2 M^I\right) = 0\ , \nonumber \\
&&\la\left( \tfrac{1}{2}\partial_{\bar2}M^I - 2 \cF^I_{0\bar2} \right)
+\tfrac{4\mu_1}{\sqrt2}\cF^I_{1\bar2}
+\tfrac{\mu_2}{\sqrt2}\left( \tfrac{1}{2}\partial_0 M^I - 2 \left( 
\cF^I_{1\bar1} -\cF^I_{2\bar2} \right) \right)
+\si \left( \tfrac{1}{2}\partial_1 M^I - 2\cF^I_{01} \right) = 0\ , \nonumber \\
&&2\la\cF^I_{\bar1 \bar2} + 
\tfrac{\mu_1}{\sqrt2}\left(\tfrac{1}{2}\partial_{\bar2}M^I + 2 \cF^I_{0\bar2} 
\right)
+\tfrac{\mu_2}{\sqrt2}\left( -\tfrac{1}{2}\partial_{\bar1}M^I - 2\cF^I_{0\bar1} 
\right)
+\si\left( \tfrac{1}{4}\partial_0 M^I + {\cF^{I\al}}_{\al} \right) = 0\ . 
\label{lin-sys-time-gaug}
\end{eqnarray}

\subsection*{Auxiliary fermion equation}

From the vanishing of the auxilary fermion variation for a bosonic background 
we get
\begin{equation}
\delta \chi^{\mathbf{i}} = \left[D - 2 \gamma^c \gamma^{ab} \nabla_a v_{bc} -2 \gamma^a 
\epsilon_{abcde} v^{bc} v^{de} + \frac{4}{3} (v \cdot \gamma)^2 \right] 
\epsilon^{\mathbf{i}} = 0\ .
\end{equation}
By making use of identities \eqref{gamma-id} together with
\begin{eqnarray}
\gamma_{ab} \gamma_{cd} &=& \eta_{ad}\eta_{bc} - \eta_{ac} \eta_{bd} - 
\eta_{ac} \gamma_{bd} + \eta_{ad} \gamma_{bc} + \eta_{bc} \gamma_{ad} - 
\eta_{bd} \gamma_{ac} + \gamma_{abcd}\ , \nonumber \\
\gamma_{abcd} &=& \epsilon_{abcde}\gamma^e\ ,
\end{eqnarray}
this can be cast into the form
\begin{equation}
\delta \chi^{\mathbf{i}} = \left[D - \frac{8}{3} v^2 + \left(2\nabla_b v^{ba} - \frac{2}{3} 
\epsilon^{abcde} v_{bc} v_{de}\right)
\gamma_a + \epsilon^{abcde}\gamma_{ab}\nabla_c v_{de}\right] \epsilon^{\mathbf{i}} = 0\ . 
\label{lin-sys-time-aux}
\end{equation}
Acting on a generic spinor \eqref{lin-sys-time-aux} becomes
\begin{eqnarray}
\A(\la1 + \sigma e^{12}) + (\B + \B^i\ga_i)(\mu_1 e^1 + \mu_2 e^2) + 
\A^i\ga_i(\la 1+\sigma e^{12})&&
\nonumber \\
+ \A^{ij}\ga_{ij}(\la 1 + \mu_1 e^1 + \mu_2 e^2 + \sigma e^{12}) &=& 0\ , 
\label{lin-sys-time-aux'}
\end{eqnarray}
where we defined
\begin{eqnarray}
\A &=& D - \tfrac{16}{3}v_{(0)}^2 - 4v^2_{(+)} - \tfrac{4}{3}v^2_{(-)} - 
2\nabla_iv^{0i}\ , \nonumber \\
\A^i &=& 2\nabla_0v^{0i}  + 2\nabla_j v^{ji} + 
\tfrac{8}{3}\epsilon^{ijkl}v_{0j}v_{kl} -2\epsilon^{ijkl}
\nabla_j v_{kl}\ , \nonumber \\
\A^{ij}&=& \epsilon^{ijkl}\left(\nabla_0 v_{kl} - 2\nabla_k v_{0l} \right)\ , 
\nonumber \\
\B &=& D - \tfrac{16}{3}v_{0 i}v^{0i} - \tfrac{4}{3}v^2_{(+)} - 4v^2_{(-)}  + 
2\nabla_iv^{0i}\ , \nonumber \\
\B^i &=& 2\nabla_0v^{0i}  + 2\nabla_j v^{ji} + 
\tfrac{8}{3}\epsilon^{ijkl}v_{0j}v_{kl} +2\epsilon^{ijkl}
\nabla_j v_{kl}\ . \label{defs-aux}
\end{eqnarray}
\eqref{lin-sys-time-aux'} may be expanded in the oscillator basis. However it 
is simpler to substitute the
conditions arising from the gravitino and gaugino equations into the system as 
is discussed in the text.

\section{Killing spinor identities}\label{ksis}

\subsection{In a time-like basis} \label{ksitime-like}
We will first expand
\begin{equation}
\mathcal{E}(A)^\mu_I \gamma_\mu \epsilon^{\mathbf{i}} - \mathcal{E}(M)_I \epsilon^{\mathbf{i}} = 0 \ ,
\end{equation}
in a time-like basis acting on a generic spinor 
\begin{equation}
\ep = \la 1 + \mu_1 e^1 +\mu_2 e^2 + \sigma e^{12} \ , 
\end{equation}
from which we obtain
\begin{eqnarray}
\la(\mathcal{E}(A^I)_0 -\mathcal{E}(M^I))
-\sqrt{2}\mu_1(\mathcal{E}(A^I)_1 ) 
-\sqrt{2}\mu_2(\mathcal{E}(A^I)_2 ) 
&=&0 \ ,\nonumber\\
\la(\mathcal{E}(A^I)_{\bar1} )
 + \tfrac{\mu_1}{\sqrt2}(\mathcal{E}(A^I)_0 + \mathcal{E}(M^I)) 
-\si(\mathcal{E}(A^I)_2 ) 
&=&0 \ ,\nonumber\\
\la(\mathcal{E}(A^I)_{\bar2} )
+ \tfrac{\mu_2}{\sqrt2}(\mathcal{E}(A^I)_0 + \mathcal{E}(M^I)) 
+\si(\mathcal{E}(A^I)_1 ) 
&=&0 \ ,\nonumber\\
\tfrac{\mu_1}{\sqrt2}(\mathcal{E}(A^I)_{\bar2} ) 
- \tfrac{\mu_2}{\sqrt2}(\mathcal{E}(A^I)_{\bar1} ) 
+\tfrac{\si}{2}(\mathcal{E}(A^I)_0  - \mathcal{E}(M^I))
&=&0\ .
\end{eqnarray}
Whilst for
\begin{eqnarray}
\left[ \tfrac{1}{8} \mathcal{E}(v)^{ab} + \tfrac{1}{2} \mathcal{E}(D) v^{ab}
\right] \gamma_{ab} \epsilon^{\mathbf{i}} + \nabla^a \mathcal{E}(D) \gamma_a 
\epsilon^{\mathbf{i}} 
&=&0 \; ,
\end{eqnarray}
we obtain 
\begin{eqnarray}
 \la\left[ \tfrac{1}{4} \mathcal{E}(v)_\alpha^{\phantom{\alpha}\alpha} +
\mathcal{E}(D) v_\alpha^{\phantom{\alpha}\alpha} +\nabla_0 \mathcal{E}(D) 
\right]
-\tfrac{\mu_1}{\sqrt2}\left[ \tfrac{1}{2} \mathcal{E}(v)^{0\bar1} + 2
\mathcal{E}(D) v^{0\bar1} +2\nabla_{1} \mathcal{E}(D) 
 \right] &&\nonumber\\
-\tfrac{\mu_2}{\sqrt2}\left[ \tfrac{1}{2} \mathcal{E}(v)^{0\bar2} + 2
\mathcal{E}(D) v^{0\bar1} +2\nabla_{2} \mathcal{E}(D) 
 \right]
-\si\left[ \tfrac{1}{2} \mathcal{E}(v)^{\bar1 \bar2} + 2 \mathcal{E}(D) v^{\bar1
\bar2} \right] 
&=&0\ ,\nonumber\\
\la\left[ -\tfrac{1}{4} \mathcal{E}(v)^{01} -  \mathcal{E}(D) v^{01} +
\nabla_{\bar1} \mathcal{E}(D)  \right] &&\nonumber\\
-\tfrac{\mu_1}{\sqrt2}\left[ \tfrac{1}{4} \mathcal{E}(v)^{1\bar1} -\tfrac{1}{4}
\mathcal{E}(v)^{2\bar2} + \mathcal{E}(D) (v^{1\bar1} -v^{2\bar2}) -
\nabla_0 \mathcal{E}(D) \right] &&\nonumber\\
+\tfrac{\mu_2}{\sqrt2}\left[ -\tfrac{1}{2} \mathcal{E}(v)^{1\bar2}  -2
\mathcal{E}(D) v^{1\bar2} \right]
+\tfrac{\si}{2}\left[ \tfrac{1}{2} \mathcal{E}(v)^{0\bar2} +2 \mathcal{E}(D)
v^{0\bar2} -2 \nabla_{2} \mathcal{E}(D) \right]
&=&0\ ,\nonumber\\
\la\left[ -\tfrac{1}{4} \mathcal{E}(v)^{02} -  \mathcal{E}(D) v^{02} +
\nabla_{\bar2} \mathcal{E}(D)  \right]
+\tfrac{\mu_1}{\sqrt2}\left[ \tfrac{1}{2} \mathcal{E}(v)^{\bar1 2}  +2
\mathcal{E}(D) v^{\bar1 2} \right] &&\nonumber\\
+\tfrac{\mu_2}{\sqrt2}\left[ \tfrac{1}{4} \mathcal{E}(v)^{1\bar1} -\tfrac{1}{4}
\mathcal{E}(v)^{2\bar2} + \mathcal{E}(D)( v^{1\bar1} - v^{2\bar2}) +
\nabla_0 \mathcal{E}(D)\right]&&\nonumber\\
+\tfrac{\si}{2}\left[ -\tfrac{1}{2} \mathcal{E}(v)^{0\bar1} -2 \mathcal{E}(D)
v^{0\bar1} +2 \nabla_{1} \mathcal{E}(D)\right]
&=&0\ ,\nonumber\\
\la\left[ \tfrac{1}{4} \mathcal{E}(v)^{12} +  \mathcal{E}(D) v^{12} \right]
+\tfrac{\mu_1}{\sqrt2}\left[ \tfrac{1}{4} \mathcal{E}(v)^{02} +  \mathcal{E}(D)
v^{02} + \nabla_{\bar2} \mathcal{E}(D)  \right] &&\nonumber\\
-\tfrac{\mu_2}{\sqrt2}\left[ \tfrac{1}{4} \mathcal{E}(v)^{01} +  \mathcal{E}(D)
v^{01} + \nabla_{\bar1} \mathcal{E}(D)  \right]
+\tfrac{\si}{2}\left[ \tfrac{1}{4} \mathcal{E}{(v)^{\alpha}}_\alpha + 
\mathcal{E}(D) {v^{\alpha}}_\alpha + \nabla_0 
\mathcal{E}(D)\right]
&=&0\ .\nonumber\\
\end{eqnarray}
Finally for 
\begin{equation}
\left. \mathcal{E}(e)_a^\mu \gamma^a \epsilon^{\mathbf{i}} \right|_{\text{other bosons
on-shell}} = 0 \ ,
\end{equation}
we obtain
\begin{eqnarray}
 \la\mathcal{E}(e)_0^\mu - \sqrt2 \mu_1 \mathcal{E}(e)_{1}^\mu - \sqrt2 \mu_2
\mathcal{E}(e)_{2}^\mu &=&0\ , \nonumber\\
\la\mathcal{E}(e)_{\bar1}^\mu + \tfrac{1}{\sqrt2} \mu_1 \mathcal{E}(e)_0^\mu -
\si
\mathcal{E}(e)_2^\mu &=&0 \ ,\nonumber\\
\la\mathcal{E}(e)_{\bar2}^\mu + \tfrac{1}{\sqrt2} \mu_2 \mathcal{E}(e)_0^\mu +
\si
\mathcal{E}(e)_1^\mu &=&0 \ , \nonumber\\
\tfrac{\mu_1}{\sqrt2}\mathcal{E}(e)_{\bar2}^\mu -
\tfrac{\mu_2}{\sqrt2}\mathcal{E}(e)_{\bar1}^\mu +
\tfrac{\si}{2}\mathcal{E}(e)_0^\mu &=&0\ .
\end{eqnarray}

\subsection{In a null basis} \label{ksinull}
We will first expand
\begin{equation}
\mathcal{E}(A)^\mu_I \gamma_\mu \epsilon^{\mathbf{i}} - \mathcal{E}(M)_I \epsilon^{\mathbf{i}} = 0
\end{equation}
in the null basis acting on a generic spinor with first component 
\begin{equation}
\ep^{\mathbf{1}} = z_1 (1+e^1) + z_2(e^{12}-e^2) +z_3 (1-e^1) + z_4( e^{12}+e^2) \ .
\end{equation}
Dropping the $I$ index for clarity we get
\begin{eqnarray}
-z_1(i\mathcal{E}(A)^3 +\mathcal{E}(M)) + 
z_2(\mathcal{E}(A)^{1}+i\mathcal{E}(A)^{2}) + z_3\sqrt2\mathcal{E}(A)^{-} &=&0 \ ,\nonumber 
\\
-z_1(\mathcal{E}(A)^1 - i\mathcal{E}(A)^2) + z_2(i\mathcal{E}(A)^3 - 
\mathcal{E}(M)) + z_4\sqrt2 \mathcal{E}(A)^{-} &=&0 \ ,\nonumber\\
z_1\sqrt2 \mathcal{E}(A)^+ +  z_3(i\mathcal{E}(A)^3 - \mathcal{E}(M)) - 
z_4(\mathcal{E}(A)^{1}+i\mathcal{E}(A)^{2}) &=&0 \ ,\nonumber\\
z_2\sqrt2 \mathcal{E}(A)^+ + z_3(\mathcal{E}(A)^1 -i\mathcal{E}(A)^2) - 
z_4(i\mathcal{E}(A)^3 + \mathcal{E}(M))&=&0 \ .
\end{eqnarray}
Whilst for
\begin{equation}
\left[ \tfrac{1}{8} \mathcal{E}(v)^{ab} + \tfrac{1}{2} \mathcal{E}(D) v^{ab} 
\right] \gamma_{ab} \epsilon^{\mathbf{i}} + \nabla^a \mathcal{E}(D) \gamma_a \epsilon^{\mathbf{i}} =0 \ ,
\end{equation}
we obtain
\begin{eqnarray}
 -z_1\left[ \tfrac{1}{4} (\mathcal{E}(v)_{-+} -i\mathcal{E}(v)_{12}) + 
 \mathcal{E}(D)( v_{-+}-i v_{12}) -i\nabla_3 \mathcal{E}(D)\right] &&\nonumber\\
-z_2\left[ \tfrac{1}{4} (\mathcal{E}(v)_{23}-i\mathcal{E}(v)_{13}) + 
\mathcal{E}(D) (v_{23}-iv_{13}) +(\nabla_{1}+i\nabla_2) \mathcal{E}(D)\right] 
&&\nonumber\\
-z_3 \sqrt2 \left[ \tfrac{i}{4} \mathcal{E}(v)_{+3} + i \mathcal{E}(D) v_{+3} 
-\nabla_{+} \mathcal{E}(D)\right] &&\nonumber\\
+z_4\sqrt2 \left[ \tfrac{1}{4} (\mathcal{E}(v)_{+1}+i\mathcal{E}(v)_{+2}) +  
\mathcal{E}(D) (v_{+1}+iv_{+2}) \right] &=&0\ ,\nonumber\\
z_1\left[ \tfrac{1}{4} (\mathcal{E}(v)_{23}+i\mathcal{E}(v)_{13}) +  
\mathcal{E}(D) (v_{23}+iv_{13}) + (\nabla_{1}-i\nabla_{2}) 
\mathcal{E}(D)\right] &&\nonumber\\
-z_2\left[ \tfrac{1}{4} (\mathcal{E}(v)_{-+} + i\mathcal{E}(v)_{12}) + 
\mathcal{E}(D) (v_{-+} +iv_{12}) + i\nabla_3 \mathcal{E}(D)\right] &&\nonumber\\
-z_3\sqrt2 \left[ \tfrac{1}{4} (\mathcal{E}(v)_{+1}-i\mathcal{E}(v)_{+2})  + 
\mathcal{E}(D) (v_{+1}-iv_{+2}) \right] &&\nonumber\\
+z_4 \sqrt2 \left[ \tfrac{i}{4} \mathcal{E}(v)_{+3} +i \mathcal{E}(D) v_{+3} + 
\nabla_{+} \mathcal{E}(D)\right]&=&0 \ ,\nonumber\\
z_1\sqrt2 \left[ \tfrac{i}{4} \mathcal{E}(v)_{-3} +  i\mathcal{E}(D) v_{-3} + 
\nabla_{-} \mathcal{E}(D)\right]&&\nonumber\\
-z_2\sqrt2 \left[ \tfrac{1}{4} (\mathcal{E}(v)_{-1}+i\mathcal{E}(v)_{-2})  + 
\mathcal{E}(D) (v_{-1}+iv_{-2}) \right] &&\nonumber\\
+ z_3 \left[ \tfrac{1}{4} (\mathcal{E}(v)_{-+} + i\mathcal{E}(v)_{12}) + 
\mathcal{E}(D) (v_{-+} + iv_{12}) - i\nabla_3 \mathcal{E}(D)\right]&& 
\nonumber\\
+z_4\left[ -\tfrac{1}{4} (\mathcal{E}(v)_{23}-i\mathcal{E}(v)_{13}) - 
\mathcal{E}(D) (v_{23}-iv_{13}) + (\nabla_{1}-i\nabla_2) 
\mathcal{E}(D)\right]&=&0 \ ,\nonumber\\
z_1\sqrt2 \left[ \tfrac{1}{4} (\mathcal{E}(v)_{-1}-i\mathcal{E}(v)_{-2}) +  
\mathcal{E}(D) (v_{-1}-iv_{-2}) \right]&&\nonumber\\
+z_2 \sqrt2 \left[ -\tfrac{i}{4} \mathcal{E}(v)_{-3} -i  \mathcal{E}(D) v_{-3} 
+ \nabla_- \mathcal{E}(D) \right] &&\\
+z_3\left[ \tfrac{1}{4} (\mathcal{E}(v)_{23}+i\mathcal{E}(v)_{13}) +  
\mathcal{E}(D) (v_{23}+iv_{13}) - (\nabla_1-i\nabla_2) \mathcal{E}(D) \right] 
&&\nonumber\\
+z_4\left[ \tfrac{1}{4} (\mathcal{E}(v)_{-+}-i\mathcal{E}(v)_{12})+ 
\mathcal{E}(D) (v_{-+}-iv_{12}) + i\nabla_3 \mathcal{E}(D) \right]&=&0\ .\nonumber
\end{eqnarray}
Finally for 
\begin{equation}
\left. \mathcal{E}(e)_a^\mu \gamma^a \epsilon^{\mathbf{i}} \right|_{\text{other bosons 
on-shell}} = 0 \ ,
\end{equation}
we obtain
\begin{eqnarray}
 iz_1\mathcal{E}(e)_3^\mu -  z_2 
(\mathcal{E}(e)_{1}^\mu+i\mathcal{E}(e)_{2}^\mu) + \sqrt2 z_3 
\mathcal{E}(e)_{+}^\mu &=&0 \ ,\nonumber\\
z_1(\mathcal{E}(e)_{1}^\mu-i\mathcal{E}(e)_{2}^\mu) -  iz_2 
\mathcal{E}(e)_3^\mu +\sqrt2 z_4 \mathcal{E}(e)_+^\mu &=&0 \ ,\nonumber\\
\sqrt2 z_1\mathcal{E}(e)_{-}^\mu - iz_3 \mathcal{E}(e)_3^\mu + z_4 
(\mathcal{E}(e)_1^\mu+i\mathcal{E}(e)_2^\mu) &=&0\ ,\nonumber \\
\sqrt2 z_2\mathcal{E}(e)_{-}^\mu - z_3(\mathcal{E}(e)_{1}^\mu 
-i\mathcal{E}(e)_{2}^\mu) +iz_4\mathcal{E}(e)_3^\mu &=&0\ .
\end{eqnarray}

\section{Some useful identities for simplifying the E.o.M.s}
\label{tlidentities}
We briefly decribe the identities used to simplify the equations of motion that 
are not implied by supersymmetry, in the case of the first orbit. Similar 
identities can be derived in the case of the second orbit.
Firstly  we discuss some of the consequences of (anti)selfduality for terms 
that appear in the equations of motion.
Let $A,B,C$ be three antisymmetric tensors with Euclidean indices and that 
$A,C$ satisfy the (anti)self-duality conditions
\begin{equation}
 \tfrac 12 \epsilon_{ijkl} A^{kl} = \sigma_A A_{ij} \ , \qquad
 \tfrac 12 \epsilon_{ijkl} C^{kl} = \sigma_C C_{ij} \ ,
\end{equation}
where $\sigma_A, \sigma_C$ take values $\pm 1$.
Making use of these identities, together with
\begin{equation}
 \epsilon_{i_1 i_2 i_3 i_4} \epsilon^{j_1 j_2 j_3 j_4} = 4! \delta^{[j_1}_{i_1} 
\delta^{j_2}_{i_2} \delta^{j_3}_{i_3} \delta^{j_4]}_{i_4} \ ,
\end{equation}
one can prove the following formula
\begin{equation}
 \sigma_A \sigma_C (ABC)_{ij} = (CBA)_{ij} - (CAB)_{ij} - (BCA)_{ij} - \tfrac 
12 (AC)B_{ij} + \delta_{ij} {\rm tr}(ABC) \ .
\end{equation}
We make use of the shorthand notation
\begin{equation}
 (ABC)_{ij} = A_{ih} B^{hk} C_{kj} \ , \qquad  (AB) = A_{ij} B^{ij} \ , \qquad 
 {\rm tr}(ABC) = A_{ih} B^{hk} C_{k}^{\phantom A i} \ .
\end{equation}
Note that from antisymmetry of $A,B,C$ we get
\begin{equation}
 {\rm tr}(ABC) = -  {\rm tr}(ACB) \ . \label{trace}
\end{equation}
We adopt the shorthand notation 
\begin{equation}
 G_{ij}^{(\pm)} \equiv (\pm)_{ij} \ .
\end{equation}
Let us first consider $(+++)_{ij}$. Using the identity (\ref{trace}) we can 
immediately see ${\rm tr}(+++)=0$. Therefore, 
the general formula in this case boils down to
\begin{equation}
 (+++)_{ij} = -\tfrac 14 (++)(+)_{ij} \ .
\end{equation}
The $(---)_{ij}$ case is completely analogous:
\begin{equation}
 (---)_{ij} = -\tfrac 14 (--)(-)_{ij} \ .
\end{equation}
 We then turn to $(++-)_{ij}$, for which the general formula gives
\begin{equation}
 (++-)_{ij} = (+-+)_{ij} \ .
\end{equation}
Note that the matrix on the RHS is manifestly antisymmetric. If we consider the 
ordering $(+-+)_{ij}$ the general formula 
reads instead
\begin{equation}
 (++-)_{ij} + (-++)_{ij} = -\tfrac 12 (++)(-)_{ij} \ .
\end{equation}
Combining the last two equations we find
\begin{equation}
 (++-)_{ij} = (+-+)_{ij} = (-++)_{ij} = -\tfrac 14 (++)(-)_{ij} \ .
\end{equation}
With the same strategy the $(--+)$ form yields
\begin{equation}
 (--+)_{ij} = (-+-)_{ij} = (+--)_{ij} = -\tfrac 14 (--)(+)_{ij} \ .
\end{equation}

Next let us consider terms that include a $\Theta$.
Let us first consider $(\Theta + +)$ where $\Theta$ is self-dual in the first 
time-like orbit.
The trace argument applies and we have thus ${\rm tr}(\Theta ++) = 0$. 
From the general formula applied to $(\Theta ++)_{ij}$ we get
\begin{equation}
 (\Theta ++)_{ij} + (+\Theta +)_{ij} = -\tfrac12 (\Theta +)(+)_{ij} \ . 
\label{firsttheta}
\end{equation}
If we use $(+\Theta +)_{ij}$ instead we find
\begin{equation}
 (\Theta ++)_{ij} + (++\Theta)_{ij} = -\tfrac 12 (++)\Theta_{ij} \ . 
\label{secondtheta}
\end{equation}
Note that the  $(++\Theta)_{ij}$ equation gives us nothing new.
From (\ref{firsttheta}) we can infer that $(\Theta ++)$ is antisymmetric, since 
the other two terms are manifestly antisymmetric.
(\ref{secondtheta}) then gives us
\begin{equation}
 (\Theta ++ )_{ij} = (++\Theta)_{ij} = -\tfrac 14 (++)\Theta_{ij} \ .
\end{equation}
Plugging it back into (\ref{firsttheta}), we find
\begin{equation}
 (+\Theta +)_{ij} = -\tfrac 12 (\Theta+)(+)_{ij} + \tfrac 14 (++)\Theta_{ij} \ .
\end{equation}

Let us now turn to the $(\Theta --)$ terms. Once again the trace is zero. From 
the $(\Theta --)$ formula
we read off
\begin{equation}
 (\Theta -- )_{ij} = (-\Theta -)_{ij} \ .
\end{equation}
From $(-\Theta-)$ we get instead
\begin{equation}
 (\Theta --)_{ij} + (--\Theta)_{ij} = -\tfrac 12 (--)\Theta_{ij} \ .
\end{equation}
The same logic applies as before: the first equation ensures antisymmetry of 
$(\Theta --)$, so that the second equation gives
the answer for $(\Theta --)_{ij}$; plugging it back into the first equation we 
also find $(-\Theta -)_{ij}$. In the end,
\begin{equation}
 (\Theta --)_{ij} = (--\Theta)_{ij} =( - \Theta -)_{ij}= -\tfrac 14 
(--)\Theta_{ij} \ .
\end{equation}

Finally, let us discuss the $(\Theta + -)$ terms. This time the trace arguments 
fail. Let us adopt the following parameterization:
\begin{align}
 (\Theta +-) &\equiv A \ ,&  (-+\Theta ) &\equiv -A^T \ ,\nonumber\\
 (\Theta -+) &\equiv B \ ,&  (+-\Theta ) &\equiv -B^T \ ,\nonumber\\
 (+\Theta -) &\equiv C \ ,&  (-\Theta+ ) &\equiv -C^T \ .
\end{align}
As far as traces are concerned,
\begin{equation}
 {\rm tr}A = -{\rm tr} B = -{\rm tr} C \ .
\end{equation}
The three equations for orderings $(\Theta + -), (\Theta - +)$ give 
respectively ($(+ \Theta -)$ is redundant)
\begin{align}
 &-A = -A^T + C^T + B^T + \mathbb I \; {\rm tr} A \ ,\nonumber\\
& B = -B^T -C +A^T -\tfrac 12 (\Theta +)(-) - \mathbb I\; {\rm tr}A \ . 
\end{align}
It is convenient to analyse these relations decomposing every matrix in 
symmetric and antisymmetric part. Doing this, we find that the $(\Theta + -)$ matrices are determined up 
to an arbitrary symmetric matrix $X$. More precisely,
\begin{align}
& (\Theta + - ) = (\Theta - +) = (-\Theta +) = X-\tfrac 14 (\Theta +)(-) \ ,\nonumber\\
& (+ \Theta  - ) = (-+\Theta) = (+-\Theta) =- X-\tfrac 14 (\Theta +)(-) \ . 
\end{align}
However, the first line just states $A=B$, and since they must have opposite 
traces, we get
\begin{equation}
{\rm tr} X = 0 \ .
\end{equation}

We also make use of some differential identities. First let us define 
$T_{ij}=e^{-2\phi}G_{ij}$, which is a closed two form on the base space, and we omit the hats on the base space quantities.
Using the identity in four dimensions for a two-form
\begin{equation}
{\nabla}_{j}  T^{ji} =\left( *d*T \right)^i\ ,
\end{equation}
we have that since $dT^{(+)} + dT^{(-)}=0$ that
\begin{equation}
J^i:={\nabla}_{j}  T^{(+)ji}={\nabla}_{j}  T^{(-)ji} \ , \label{Jis}
\end{equation}
and this is conserved $\nabla_i J^i=0$ by Ricci flatness.
Note that we are using the conventions for the Hodge dual of a p-form $\alpha$ such that
\begin{equation}
 \star \alpha_{j_1\cdots j_{4-p}}=\tfrac{1}{p!}{\epsilon_{j_{1}\cdots j_{4-p}}}^{i_1\cdots i_p}\alpha_{i_1\cdots i_p}.
\end{equation}

The Bianchi identity can be written
\begin{equation}
\nabla_i T_{j k} + 2\nabla_{[j}T_{k] i}=0 \ .
\end{equation}
Splitting $T$ into (anti)selfdual parts and operating with $\nabla^i$ gives
\begin{equation}
\nabla^2 T^{(+)}_{j k} +  \nabla^2 T^{(-)}_{j k} + 
2\nabla^i\nabla_{[j}T^{(+)}_{k] i} + 2\nabla^i\nabla_{[j}T^{(-)}_{k] i}=0 \ .
\end{equation}
Finally commuting the covariant derivative, using the selfduality of the 
curvature tensor and using (\ref{Jis}), we get an expression for the exterior 
derivative of $J$
\begin{equation}
dJ_{ij} = \tfrac 12 \nabla^2 T^{(+)}_{i j} + \tfrac 12 \nabla^2 T^{(-)}_{ij} + 
\tfrac 12 R_{ij}^{\phantom{ij}kl}T^{(+)}_{kl}\ , \label{dJ}
\end{equation}
In the same way there is a simpler identity for $\Theta^I$, namely
\begin{equation}
\nabla^2\Theta^I_{ij} = -R_{ij}^{\phantom{ij}kl}\Theta^I_{kl} \ . 
\label{laplaciantheta}
\end{equation}
In particular it is important to remember that whilst the $\Theta^I$ are 
harmonic with respect to the form Laplacian, they are not (necessarily) 
harmonic with respect to the connection Laplacian.
Note that apart from the identification of ${\nabla}_{j}  T^{(+)ji}$ with ${\nabla}_{j}  T^{(-)ji}$ and setting the right hand side of the Bianchi identity to zero in equation \eqref{dJ} we have not used the closure of $T$, so for an arbitrary two form $\alpha$ one can also derive the relation
\begin{equation}
(d\star d\star \alpha)_{ij}=  \nabla^2 \alpha^{(+)}_{i j} +  \nabla^2 \alpha^{(-)}_{ij} + 
 R_{ij}^{\phantom{ij}kl}\alpha^{(+)}_{kl}\ -  \nabla^k(d\alpha)_{ijk}, \label{dfJ}
\end{equation}
But this is equally valid for $\star \alpha$ so taking linear combinations we obtain
\begin{eqnarray}
-2\nabla_{[i}\nabla^{k}  \alpha^{(+)}_{j]k} + \nabla^k (df T^{(+)})_{ijk}&=& \nabla^2 \alpha^{(+)}_{i j} +  R_{ij}^{\phantom{ij}kl}\alpha^{(+)}_{kl} \ ,\\
-2\nabla_{[i}\nabla^{k}  \alpha^{(-)}_{j]k} + \nabla^k (df T^{(-)})_{ijk}&=&  \nabla^2 \alpha^{(-)}_{ij} \ .
\end{eqnarray}
Defining $K^{\pm}_i={\nabla}^{j} \alpha^{(\pm)}_{ji}$ we have that $\star K^{\pm}=\star \star d \star \alpha^{\pm}=\mp d \alpha^{(\pm)}$, so we can write the above as 
\begin{eqnarray}
 dK^+_{ij} -  \nabla^k (\star K^+)_{ijk} &=& \nabla^2 \alpha^{(+)}_{i j} +  R_{ij}^{\phantom{ij}kl}\alpha^{(+)}_{kl} \ ,\\
 dK^-_{ij} +  \nabla^k (\star K^-)_{ijk}&=&  \nabla^2 \alpha^{(-)}_{ij} \ ,
\end{eqnarray}
but we have that $\nabla^k (\star K^{\pm})_{ijk}=-(\star d K^{\pm})_{ij}$, thus
\begin{eqnarray}
(dK^+)^{(+)}_{ij}=\tfrac 12 (dK^+_{ij} + (\star dK^+)_{ij}) &=& \tfrac 12 \nabla^2 \alpha^{(+)}_{i j} + \tfrac 12 R_{ij}^{\phantom{ij}kl}\alpha^{(+)}_{kl} \ ,\\
(dK^-)^{(-)}_{ij}=\tfrac 12 (dK^-_{ij} - (\star dK^-)_{ij}) &=& \tfrac 12 \nabla^2 \alpha^{(-)}_{ij} \ , \label{dK23}
\end{eqnarray}
and $(dK^\pm)^{(\mp)}$ are unconstrained by these arguments.

\bibliographystyle{JHEP.bst}
\bibliography{osksis.bib}
\end{document}